\documentclass[preprint,12pt,authoryear]{elsarticle}

\usepackage[utf8]{inputenc}
\usepackage[T1]{fontenc}
\usepackage{amsmath,amssymb}
\usepackage{booktabs}
\usepackage{lineno}
\usepackage{hyperref}
\usepackage{graphicx}
\usepackage{subcaption}
\usepackage{tikz}
\usetikzlibrary{shapes,arrows,positioning}

\newcommand{\ateNaive}{0.252}
\newcommand{\ateNaivePct}{29}

\newcommand{\ateNaiveLowerPct}{11}
\newcommand{\ateNaiveUpperPct}{49}

\newcommand{\cateNaiveMinPct}{-12}
\newcommand{\cateNaiveMaxPct}{79}

\newcommand{\ateCffeEUfifteenPct}{14.1}

\newcommand{\ateCffeEUtwentyeightPct}{13.4}

\newcommand{\cfEffectSweden}{22}

\newcommand{\cfEffectPreciseSweden}{21.9}

\newcommand{\cfStdSweden}{0.16}
\newcommand{\cfMinSweden}{-14.9}
\newcommand{\cfMaxSweden}{85.9}
\newcommand{\cfEffectDenmark}{19}

\newcommand{\cfEffectPreciseDenmark}{19.1}

\newcommand{\cfStdDenmark}{0.19}
\newcommand{\cfMinDenmark}{-16.1}
\newcommand{\cfMaxDenmark}{85.7}
\newcommand{\cfEffectUnitedKingdom}{33}

\newcommand{\cfEffectPreciseUnitedKingdom}{32.6}

\newcommand{\cfStdUnitedKingdom}{0.16}
\newcommand{\cfMinUnitedKingdom}{10.3}
\newcommand{\cfMaxUnitedKingdom}{80.1}

\newcommand{\ateCffeOriginal}{0.126}
\newcommand{\ateCffeOriginalPct}{13.4}

\newcommand{\ateCffeOriginalLower}{0.114}
\newcommand{\ateCffeOriginalUpper}{0.138}
\newcommand{\ateCffeInternet}{0.100}
\newcommand{\ateCffeInternetPct}{10.5}

\newcommand{\ateCffeInternetLower}{0.088}
\newcommand{\ateCffeInternetUpper}{0.112}
\newcommand{\ateCffeChangePct}{20}

\newcommand{\cateNaiveCountryLuxembourg}{69}
\newcommand{\cateNaiveCountryGermany}{50}
\newcommand{\cateNaiveCountryBelgium}{47}
\newcommand{\cateNaiveCountryNetherlands}{47}

\newcommand{\cateNaiveCountryPortugal}{16}
\newcommand{\cateNaiveCountryFinland}{15}
\newcommand{\cateNaiveCountryGreece}{10}

\newcommand{\topNaivePairOne}{France--Italy}
\newcommand{\topNaivePairOneEffect}{79}
\newcommand{\topNaivePairTwo}{Germany--Italy}
\newcommand{\topNaivePairTwoEffect}{78}
\newcommand{\topNaivePairThree}{Belgium--Germany}
\newcommand{\topNaivePairThreeEffect}{78}
\newcommand{\topNaivePairFour}{Austria--Germany}
\newcommand{\topNaivePairFourEffect}{78}
\newcommand{\topNaivePairFive}{France--Germany}
\newcommand{\topNaivePairFiveEffect}{78}
\newcommand{\bottomNaivePairOne}{Greece--Ireland}
\newcommand{\bottomNaivePairOneEffect}{-10}
\newcommand{\bottomNaivePairTwo}{Ireland--Portugal}
\newcommand{\bottomNaivePairTwoEffect}{-10}

\newcommand{\robustnessBaseline}{24.0}

\newcommand{\robustnessExports}{17.3}

\newcommand{\robustnessImports}{16.7}

\newcommand{\robustnessGrowth}{1.1}

\newcommand{\robustnessIntensity}{31.1}

\newcommand{\sensGDPYearCoef}{0.38}

\newcommand{\sensTwowayCoef}{0.16}

\newcommand{\timeStabilityMin}{22.2}
\newcommand{\timeStabilityMax}{28.6}

\newcommand{\euExtEUfifteen}{28.6}
\newcommand{\euExtSlovenia}{31.3}
\newcommand{\euExtCyprus}{34.1}

\newcommand{\euExtLithuania}{0.9}

\newcommand{\euExtBulgaria}{4.3}

\newcommand{\naiveCfEUtwentyeight}{4.3}

\newcommand{\cffeEUtwentyeight}{13.4}
\newcommand{\cffeEUtwentyeightLower}{12.1}
\newcommand{\cffeEUtwentyeightUpper}{14.8}

\newcommand{\cffeEUtwentyeightTopOne}{Malta}
\newcommand{\cffeEUtwentyeightTopOnePct}{41.2}
\newcommand{\cffeEUtwentyeightTopTwo}{Cyprus}
\newcommand{\cffeEUtwentyeightTopTwoPct}{32.0}
\newcommand{\cffeEUtwentyeightTopThree}{Estonia}
\newcommand{\cffeEUtwentyeightTopThreePct}{26.4}
\newcommand{\cffeEUtwentyeightBottomOne}{Greece}
\newcommand{\cffeEUtwentyeightBottomOnePct}{10.8}
\newcommand{\cffeEUtwentyeightBottomTwo}{Portugal}
\newcommand{\cffeEUtwentyeightBottomTwoPct}{10.7}

\newcommand{\internetCorrEarly}{-0.42}
\newcommand{\internetCorrFull}{-0.53}
\newcommand{\internetTopCateCountry}{Spain}
\newcommand{\internetTopCate}{0.18}
\newcommand{\internetTopCateInternetEarly}{15}
\newcommand{\internetTopCateInternetFull}{53}

\newcommand{\internetHighEarlyCountry}{Netherlands}
\newcommand{\internetHighEarlyPct}{48}
\newcommand{\internetHighEarlyCate}{0.14}

\newcommand{\internetCovOrigAte}{0.252}
\newcommand{\internetCovOrigPct}{28.6}
\newcommand{\internetCovOrigLower}{0.103}
\newcommand{\internetCovOrigUpper}{0.401}
\newcommand{\internetCovIntAte}{0.255}
\newcommand{\internetCovIntPct}{29.0}
\newcommand{\internetCovIntLower}{0.101}
\newcommand{\internetCovIntUpper}{0.409}
\newcommand{\internetCovChangePct}{1.2}

\newcommand{\tradeDivIntraEff}{35.5}

\newcommand{\tradeDivToNonEZEff}{5.5}
\newcommand{\tradeDivToNonEZLower}{-63.1}
\newcommand{\tradeDivToNonEZUpper}{201.7}
\newcommand{\tradeDivFromNonEZEff}{34.9}
\newcommand{\tradeDivFromNonEZLower}{-2.2}
\newcommand{\tradeDivFromNonEZUpper}{86.1}

\begin{document}

\begin{frontmatter}

\title{To Adopt or Not to Adopt: Heterogeneous Trade Effects of the Euro\tnoteref{t1}}
\tnotetext[t1]{This version: January 31, 2026. This paper extends ``The European Union and Trade: The Average Treatment Effect of Adopting the Euro,'' a chapter from the author's doctoral dissertation that applied propensity score matching. Earlier versions were presented at several conferences, and we thank participants for valuable feedback. I thank Birol Kanik for helpful comments and discussions. Recent advances in causal machine learning enabled us to revisit the research question and estimate the heterogeneous treatment effects that were not feasible with earlier methods.}

\author[aws]{Harry Aytug}
\ead{haytug@amazon.com}

\address[aws]{Amazon Web Services}

\begin{abstract}
Two decades of research on the euro's trade effects have produced estimates ranging 
from 4\% to 30\%, with no consensus on the magnitude. We find evidence that this 
divergence may reflect genuine heterogeneity in the euro's trade effect across country pairs 
rather than methodological differences alone. Using Eurostat data on 15 EU countries (12 eurozone members plus Denmark, Sweden, and the UK as controls) from 
1995--2015, we estimate that euro adoption increased bilateral trade by \ateNaivePct\% on 
average (\ateCffeEUfifteenPct\% after fixed effects correction), but effects range from $\cateNaiveMinPct\%$ to $+\cateNaiveMaxPct\%$ across eurozone pairs. Core 
eurozone pairs (e.g., Germany--France, Germany--Netherlands) show large gains, 
while peripheral pairs involving Finland, Greece, and Portugal saw smaller or 
negative effects, with some negative estimates statistically significant and 
interpretable as trade diversion. Pre-euro trade intensity and GDP 
account for over 90\% of feature importance in explaining this heterogeneity. Extending to EU28, we find evidence that crisis-era 
adopters (Slovakia, Estonia, Latvia) pull down naive estimates to \euExtBulgaria\%, but accounting 
for fixed effects recovers estimates of \ateCffeEUtwentyeightPct\%, consistent with the EU15 fixed-effects baseline of \ateCffeEUfifteenPct\%. 
Illustrative counterfactual analysis suggests non-eurozone members would have experienced varied 
effects: UK (+\cfEffectUnitedKingdom\%), Sweden (+\cfEffectSweden\%), Denmark (+\cfEffectDenmark\%). The wide range of prior estimates 
appears to be largely a feature of the data, not a bug in the methods.
\end{abstract}


\end{frontmatter}


\section{Introduction}

Two decades of research on the euro's trade effects have failed to reach consensus. 
Gravity model estimates cluster around 4--6\%, while synthetic control studies find 
effects as high as 30\% for specific country pairs. \citet{rose2016emu} shows that 
larger datasets produce systematically larger estimates. \citet{gunnella2021impact} 
attribute the divergence to methodological differences. We propose an alternative 
explanation: the wide range may reflect genuine heterogeneity in treatment effects 
across country pairs, which existing methods cannot capture.

Understanding this heterogeneity has direct policy relevance. Sweden, Denmark, and 
the United Kingdom opted out of the eurozone, and policymakers in these countries 
have periodically revisited the adoption question. If the euro's trade effects are 
uniformly positive, the trade-based case for adoption is straightforward. But if effects vary 
substantially---large for some pairs, near-zero for others---the calculus becomes 
more complex. The same logic applies to prospective members in Central and Eastern 
Europe. Knowing \textit{which} country pairs benefit most, and \textit{why}, is 
essential for informed policy.

We estimate the full distribution of euro trade effects across all eurozone country 
pairs. Our approach suggests substantial heterogeneity that may help reconcile the 
divergent estimates in prior literature: gravity models, which estimate average effects, 
capture the center of the distribution (around 20\%), while synthetic control 
studies, which focus on specific high-trade pairs, capture the upper tail (30\%+)---precisely 
because SCM studies typically select pairs where credible synthetic counterfactuals exist. 
We also generate illustrative counterfactual predictions for non-eurozone EU members, 
estimating what trade effects Sweden, Denmark, and the United Kingdom might have 
experienced had they adopted the euro.

The EU provides a useful setting for this analysis because member countries meet 
the Maastricht criteria for fiscal discipline, price stability, and exchange rate 
stability, yet not all have adopted the euro. This creates a comparison 
where the main observable difference between treatment and control groups is euro 
adoption itself, though we acknowledge that unobserved differences may remain. 
To our knowledge, this is among the first studies to estimate the full 
distribution of heterogeneous treatment effects of currency union membership.
Methodologically, we use causal forests with double machine learning to estimate 
conditional average treatment effects at the country-pair level.

The paper proceeds as follows. Section 2 reviews the literature on currency 
unions and trade. Section 3 describes the causal forest methodology. Section 4 
presents the data. Section 5 reports results on heterogeneous treatment effects. 
Section 6 discusses the findings. Section 7 presents counterfactual analysis for 
non-eurozone EU members. Section 8 concludes.

\section{Literature Review}

Research on the trade effects of common currencies began with the influential 
contribution of \citet{rose2000money}, which used a gravity framework and 
reported very large trade increases among currency-union members. This result 
triggered extensive debate about identification, selection into currency unions, 
and the credibility of cross-sectional comparisons. Addressing these concerns, 
\citet{persson2001currency} applied matching-based reasoning and argued that 
differences between treated and untreated pairs can generate substantial upward 
bias in na\"{i}ve estimates.

A second wave of work emphasized panel variation and dynamics. For example, 
\citet{bun2002euro} estimated a dynamic panel model focused on the euro and 
found a modest short-run effect that accumulates over time. In parallel, 
methodological advances in gravity estimation improved the interpretability of 
currency-union coefficients. \citet{anderson2003gravity} formalized multilateral 
resistance---showing that bilateral trade depends on relative trade costs---and 
their framework became a benchmark for modern gravity specifications.

Direct evidence on the euro specifically emerged early. \citet{micco2003currency} 
provided early EMU evidence using developed-country data and found positive trade 
effects. Subsequent gravity work increasingly adopted estimation strategies 
designed to handle heteroskedasticity and zero trade flows. 
\citet{santossilva2006log} showed that log-linear OLS can be misleading under 
heteroskedasticity and advocated PPML-type approaches that have since become 
standard in gravity estimation.

Despite improvements in data and methods, estimates of the euro's trade effect 
continued to vary. One explanation is that results are sensitive to data coverage 
and specification choices: \citet{rose2016emu} documented that larger datasets 
(more countries/longer spans) tend to deliver systematically larger EMU trade 
estimates. A complementary explanation emphasizes that different empirical 
approaches target different parameters. For example, \citet{gunnella2021impact} 
compare gravity and synthetic control approaches and report modest ``average-type'' 
gravity effects alongside substantially larger effects for some pairs under 
synthetic control.

A major methodological branch in this literature uses Synthetic Control Methods 
(SCM), which construct a data-driven counterfactual from weighted donor units 
and are particularly attractive when a small number of treated units receive a 
discrete policy intervention. The approach is grounded in foundational 
contributions such as \citet{abadie2010synthetic}. SCM has been applied both 
within and beyond trade settings to study institutional and macro-financial 
regime changes. \citet{aytug2017rom} uses SCM to construct counterfactual 
exchange rate volatility in the absence of Turkey's Reserve Option Mechanism, 
illustrating that policy effects can be strongly state-dependent and sensitive 
to concurrent monetary tightening. In trade-policy evaluation, 
\citet{aytug2017customs} apply SCM to assess the EU--Turkey Customs Union and 
show sizable effects relative to a synthetic counterfactual. In the euro context, 
\citet{saia2017uk} uses SCM to study the counterfactual of UK euro adoption and 
finds sizable trade gains in that scenario.

At the same time, the SCM toolkit has expanded in ways that matter for 
interpretation. \citet{benmichael2021augmented} propose the Augmented Synthetic 
Control Method to improve performance when perfect pre-treatment fit is not 
feasible. \citet{arkhangelsky2021sdid} introduce Synthetic Difference-in-Differences, 
combining elements of DID and synthetic control to improve robustness in common 
empirical settings. More recently, \citet{distefano2024inclusive} propose an 
``inclusive'' synthetic control variant designed for settings where spillovers 
or indirect effects may contaminate donor pools. These developments strengthen 
the credibility of comparative-case approaches, but they also reinforce a core 
limitation: SCM-style designs remain most naturally suited to case-specific 
evaluation (one or a few treated units) rather than systematic characterization 
of effects across the full set of euro-area country pairs.

This brings the literature to a central unresolved issue: heterogeneity. Gravity 
models typically estimate an average (or average-like) effect of euro membership 
across treated pairs, while SCM studies often focus on a subset of country-pairs 
where credible synthetic counterfactuals can be built---precisely the setting in 
which large effects are more likely to be detected. The combination of these 
approaches therefore leaves open an important question: what is the full 
distribution of the euro's trade effects across all euro-area pairs, and which 
observable features predict where a given pair lies in that distribution?

To address this gap, we build on the growing literature that applies modern 
causal inference tools to uncover treatment effect heterogeneity. 
\citet{athey2016recursive} develop tree-based approaches designed specifically 
for heterogeneous causal effects, providing a foundation for forest-based 
estimators. Related work on machine-learning-based causal inference, such as 
\citet{chernozhukov2018double}, provides tools for valid inference on causal 
parameters in high-dimensional settings. Our contribution is to use these ideas 
to estimate the distribution of euro trade effects across country pairs, thereby 
reconciling why different methods in the existing literature can yield apparently 
conflicting results.

Table~\ref{tab:comparison} summarizes the key methodological differences between 
existing approaches and our causal forest method.

\begin{table}[!htbp]
\centering
\caption{Comparison of Estimation Approaches}
\label{tab:comparison}
\small
\begin{tabular}{@{}lccc@{}}
\toprule
& \textbf{Gravity/PSM} & \textbf{SCM} & \textbf{Causal Forest} \\
\midrule
Estimates & Single ATE & Few pair effects & Full CATE distribution \\
Heterogeneity & Pre-specified only & Not systematic & Data-driven discovery \\
Selection bias & PSM addresses & Matching-based & DML residualization \\
Inference & Clustered SE & Placebo tests & Forest-based CI \\
Literature range & 4--30\% & 16--30\% & --- \\
\bottomrule
\end{tabular}
\end{table}

\section{Methodology}

\subsection{Existing Approaches: Gravity and Synthetic Control}

The gravity model is the workhorse of empirical trade analysis. Following 
\citet{anderson2003gravity} and best practices outlined in \citet{gunnella2021impact}, 
the structural gravity equation with a saturated set of fixed effects is:
\begin{equation}
    \ln(X_{ijt}) = \lambda_{it} + \psi_{jt} + \mu_{ij} + \boldsymbol{\beta}' 
                   \mathbf{z}_{ijt} + \gamma \text{CU}_{ijt} + \varepsilon_{ijt}
\end{equation}
where $X_{ijt}$ denotes exports from country $i$ to country $j$ at time $t$, 
$\lambda_{it}$ and $\psi_{jt}$ are exporter-time and importer-time fixed effects 
that control for multilateral resistance, $\mu_{ij}$ are pair fixed effects that 
absorb time-invariant bilateral factors (distance, common language, contiguity), 
$\mathbf{z}_{ijt}$ includes time-varying bilateral controls such as regional trade 
agreements, and $\text{CU}_{ijt}$ is a dummy equal to one if both countries are in 
a currency union at time $t$. The coefficient $\gamma$ captures the average 
treatment effect of euro adoption. Estimation via PPML addresses heteroskedasticity 
and zero trade flows \citep{santossilva2006log}.

The synthetic control method (SCM) offers an alternative approach by constructing 
a counterfactual for each treated unit as a weighted combination of control units. 
Following \citet{abadie2003economic}, the causal effect is estimated as:
\begin{equation}
    \hat{\tau}_1 = \mathbf{y}_1 - \mathbf{y}_0^* = \mathbf{y}_1 - \mathbf{Y}_0 \mathbf{w}^*
\end{equation}
where $\mathbf{y}_1$ is the outcome vector for the treated unit, $\mathbf{Y}_0$ is 
a matrix of outcomes for $J$ control units, and $\mathbf{w}^*$ is a vector of 
weights chosen to minimize the distance between treated and synthetic control 
units in the pre-treatment period. \citet{gunnella2021impact} apply SCM to 
estimate euro effects for specific country pairs, finding effects around 30\%. 
\citet{saia2017uk} uses SCM to estimate what UK trade would have been under euro 
adoption, finding a 16\% effect. While SCM provides credible counterfactuals for 
individual cases, it does not scale to estimate effects for all country pairs or 
systematically explore heterogeneity.

Both approaches have limitations for our research question. Gravity models 
estimate a single average effect, potentially masking substantial heterogeneity 
across pairs. SCM can estimate pair-specific effects but requires selecting 
cases ex ante and does not identify what drives variation. We propose causal 
forests as a method that combines the strengths of both: estimating the full 
distribution of treatment effects while identifying the characteristics that 
explain heterogeneity.

\subsection{From PSM to Causal Forests}

Traditional propensity score matching (PSM) addresses selection bias by matching 
treated and control units based on their probability of treatment. Following 
\citet{persson2001currency}, the propensity score is defined as:
\begin{equation}
    e(X) = P(\text{Euro}_{ijt} = 1 \mid X_{ijt})
\end{equation}
where $X_{ijt}$ includes GDP and other pair characteristics. The 
average treatment effect on the treated (ATT) is then:
\begin{equation}
    \tau_{\text{ATT}} = E[Y^{(1)} - Y^{(0)} \mid \text{Euro} = 1]
\end{equation}

While PSM addresses selection bias, it estimates only a single average effect. 
Causal forests extend this framework by estimating heterogeneous treatment 
effects --- allowing the effect to vary with observed characteristics:
\begin{equation}
    \tau(x) = E[Y^{(1)} - Y^{(0)} \mid X = x]
\end{equation}
This Conditional Average Treatment Effect (CATE) captures how the euro's impact 
varies across country pairs with different characteristics.

\subsection{Intuition: How Causal Forests Discover Heterogeneity}

Before presenting the technical details, we provide intuition for how causal 
forests estimate heterogeneous treatment effects. The key insight is that 
causal forests are designed to answer a different question than traditional 
methods: rather than asking ``what is the average effect?'' they ask ``for 
which units is the effect largest, and why?''

A standard decision tree predicts outcomes by recursively splitting the data 
into groups that are most similar in their outcome values. A \textit{causal} 
tree instead splits the data into groups that are most \textit{different} in 
their treatment effects. Consider a simple example: if the euro's effect on 
trade is larger for country pairs with high pre-existing trade intensity, 
the causal tree will split on trade intensity, creating one group (high-trade pairs) with 
a high estimated effect and another group (low-trade pairs) with a lower effect.

The algorithm proceeds as follows. At each node, the tree considers all possible 
splits of the data (e.g., GDP above/below median, pre-euro trade high/low) and selects the 
split that maximizes the difference in treatment effects between the resulting 
subgroups. This process continues recursively until the subgroups become too 
small. The treatment effect for any unit is then estimated as the average 
effect within its terminal node (leaf).

A single tree would be noisy and sensitive to the particular sample. Causal 
\textit{forests} address this by growing many trees (we use 500) on bootstrapped 
samples and averaging their predictions. This ensemble approach reduces variance 
while preserving the ability to capture complex heterogeneity patterns.

Two features distinguish causal forests from standard machine learning methods. 
First, ``honest'' estimation uses separate subsamples for determining the tree 
structure and estimating effects within leaves, preventing overfitting and 
enabling valid statistical inference. Second, the algorithm provides not just 
point estimates but confidence intervals for each unit's treatment effect, 
allowing researchers to assess statistical significance at the individual level.

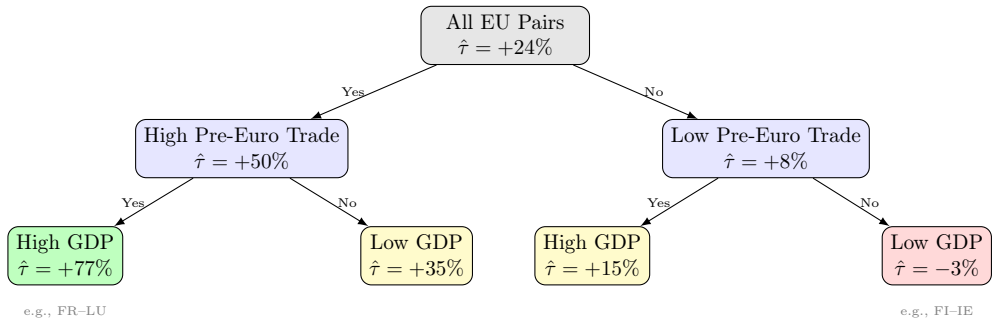
\begin{figure}[!htbp]
\centering
\begin{tikzpicture}[scale=0.8, transform shape,
    node distance=0.9cm and 0.5cm,
    every node/.style={font=\footnotesize},
    root/.style={rectangle, draw, rounded corners, minimum width=2.8cm, minimum height=0.7cm, align=center, fill=gray!20},
    internal/.style={rectangle, draw, rounded corners, minimum width=2.2cm, minimum height=0.7cm, align=center, fill=blue!10},
    leaf/.style={rectangle, draw, rounded corners, minimum width=1.8cm, minimum height=0.7cm, align=center},
    high/.style={leaf, fill=green!25},
    medium/.style={leaf, fill=yellow!25},
    low/.style={leaf, fill=red!15}
]

\node[root] (root) {All EU Pairs\\$\hat{\tau} = +24\%$};

\node[internal, below left=0.9cm and 1.2cm of root] (left1) {High Pre-Euro Trade\\$\hat{\tau} = +50\%$};
\node[internal, below right=0.9cm and 1.2cm of root] (right1) {Low Pre-Euro Trade\\$\hat{\tau} = +8\%$};

\node[high, below left=0.8cm and 0.2cm of left1] (ll) {High GDP\\$\hat{\tau} = +77\%$};
\node[medium, below right=0.8cm and 0.2cm of left1] (lr) {Low GDP\\$\hat{\tau} = +35\%$};

\node[medium, below left=0.8cm and 0.2cm of right1] (rl) {High GDP\\$\hat{\tau} = +15\%$};
\node[low, below right=0.8cm and 0.2cm of right1] (rr) {Low GDP\\$\hat{\tau} = -3\%$};

\draw[-latex] (root) -- node[left, pos=0.5] {\tiny Yes} (left1);
\draw[-latex] (root) -- node[right, pos=0.5] {\tiny No} (right1);
\draw[-latex] (left1) -- node[left, pos=0.5] {\tiny Yes} (ll);
\draw[-latex] (left1) -- node[right, pos=0.5] {\tiny No} (lr);
\draw[-latex] (right1) -- node[left, pos=0.5] {\tiny Yes} (rl);
\draw[-latex] (right1) -- node[right, pos=0.5] {\tiny No} (rr);

\node[below=0.2cm of ll, font=\tiny, text=gray] {e.g., FR--LU};
\node[below=0.2cm of rr, font=\tiny, text=gray] {e.g., FI--IE};

\end{tikzpicture}
\caption{Illustration of causal tree splitting. The tree recursively partitions 
country pairs into subgroups with different treatment effects. At each node, 
the split is chosen to maximize the difference in euro effects between 
subgroups. Terminal nodes (leaves) show estimated effects ranging from +77\% 
for high-trade, high-GDP pairs (e.g., France--Luxembourg) to $-3\%$ for low-trade, 
low-GDP pairs (e.g., Finland--Ireland). 
Colors indicate effect magnitude: green (high), yellow (medium), red (low). 
A causal forest averages predictions across many such trees.}
\label{fig:causal_tree}
\end{figure}

For our application, this means the causal forest can discover --- without 
prior specification --- that the euro's trade effect is large for core European 
pairs with deep existing trade relationships, modest for peripheral pairs, and 
near-zero for pairs involving Greece. Traditional methods would require the 
researcher to hypothesize these patterns in advance and test them through 
subgroup analysis or interaction terms. Figure~\ref{fig:causal_tree} illustrates 
this process.

\subsection{Causal Forest with Double Machine Learning}

We apply Causal Forests with Double Machine Learning 
\citep{athey2018estimation, chernozhukov2018double} to estimate the full 
distribution of CATEs. The CausalForestDML estimator proceeds in three steps:

\medskip
\textbf{Step 1: Nuisance estimation.} Use machine learning to estimate:
\begin{align}
    \hat{m}(W) &= E[Y \mid W] \quad \text{(outcome model)} \\
    \hat{e}(W) &= E[T \mid W] \quad \text{(propensity score)}
\end{align}

\textbf{Step 2: Residualize.} Partial out the effect of controls:
\begin{align}
    \tilde{Y} &= Y - \hat{m}(W) \\
    \tilde{T} &= T - \hat{e}(W)
\end{align}

\textbf{Step 3: Causal forest.} Estimate $\tau(X)$ using an honest random forest 
on the residualized data $(\tilde{Y}, \tilde{T}, X)$.

\medskip
This approach removes regularization bias from the ML first stage and provides 
valid confidence intervals. It is doubly robust: consistent if either $\hat{m}$ 
or $\hat{e}$ is correctly specified. We implement CausalForestDML with Random 
Forest first-stage models (200 trees, min\_samples\_leaf = 20), a causal forest 
with 500 trees and honest splitting (min\_samples\_leaf = 30), and a Random 
Forest Classifier for the binary treatment model.

\subsection{Constructing Counterfactuals}

A fundamental challenge in causal inference is that we never observe the 
counterfactual outcome --- what trade would have been for a eurozone pair had 
they not adopted the euro, or for a non-eurozone pair had they adopted it. 
Using the potential outcomes framework, each unit has two potential outcomes: 
$Y^{(1)}$ (outcome if treated) and $Y^{(0)}$ (outcome if not treated). The 
treatment effect for unit $i$ is:
\begin{equation}
    \tau_i = Y_i^{(1)} - Y_i^{(0)}
\end{equation}
We observe $Y_i^{(1)}$ for treated units and $Y_i^{(0)}$ for control units, 
but never both for the same unit.

Traditional matching methods address this by finding ``twin'' control units 
with similar observable characteristics to impute the missing potential outcome. 
Causal forests take a different approach: they estimate the conditional average 
treatment effect function $\tau(x) = E[Y^{(1)} - Y^{(0)} \mid X = x]$ from the 
data, then apply this learned function to predict effects for any unit.

For a non-eurozone pair like UK--Germany with characteristics $x_{\text{UK-DE}}$, 
the counterfactual prediction is:
\begin{equation}
    \hat{\tau}(\text{UK-DE}) = \hat{\tau}(x_{\text{UK-DE}})
\end{equation}
The causal forest predicts what effect euro adoption \textit{would have had} 
based on the pair's characteristics (GDP, pre-euro trade intensity) 
and how similar characteristics related to treatment effects among eurozone 
pairs. The key identifying assumption is that the CATE function learned from 
treated units generalizes to untreated units with similar characteristics --- 
formally, that $\tau(x)$ is the same for treated and untreated units conditional 
on $X = x$.

\subsection{Treatment Definition and Identification}

We define the treatment as both countries in a pair having adopted the euro:
\begin{equation}
    T_{ijt} = \mathbf{1}[\text{Euro}_i = 1] \times \mathbf{1}[\text{Euro}_j = 1]
\end{equation}
The sample is restricted to pairs where both countries are EU members, ensuring 
all units faced a credible possibility of treatment. Countries outside the EU 
(e.g., United States, Japan, Switzerland) could never adopt the euro and thus 
do not constitute a valid counterfactual. Within the EU sample, eurozone members 
(Austria, Belgium, Finland, France, Germany, Greece, Ireland, Italy, Luxembourg, 
Netherlands, Portugal, Spain) constitute the treatment group, while EU members 
that opted out (Denmark, Sweden, United Kingdom) constitute the control group.

The outcome variable is log real bilateral trade:
\begin{equation}
    Y_{ijt} = \ln\left(\frac{\text{Exports}_{ij} + \text{Imports}_{ij}}{\text{PPI}_t}\right)
\end{equation}
For PPML specifications in the gravity benchmark, we use trade in levels following \citet{santossilva2006log}.
Effect modifiers $X$ --- the variables that may drive heterogeneity in treatment 
effects --- include log GDP product, log GDP per capita, and pre-euro trade intensity (average bilateral trade 1995--1998, the pre-treatment period in our sample). 

Controls $W$ used in the first-stage nuisance estimation address potential 
confounders that affect both euro adoption and trade:
\begin{itemize}
    \item \textit{Log GDP product}: Larger economies trade more and were more 
          likely to be founding eurozone members.
    \item \textit{Log GDP per capita}: Richer countries have different trade 
          patterns and faced different incentives for euro adoption.
    \item \textit{Year}: Controls for business cycle effects and common time 
          trends; the DML residualization removes year-specific variation from 
          both outcome and treatment, analogous to year fixed effects.
\end{itemize}

Identification relies on the conditional independence assumption: conditional on 
the controls $W$, treatment assignment is independent of potential outcomes. The 
DML framework addresses selection on observables by flexibly modeling the 
relationship between confounders and both the outcome and treatment. The causal 
forest then estimates heterogeneous effects conditional on effect modifiers $X$, 
which may overlap with but are conceptually distinct from the confounders $W$.

\subsection{Threats to Identification}

While the DML framework addresses selection on observables through flexible 
first-stage estimation, our identification strategy faces potential threats 
from unobserved confounders. We discuss the main concerns and the evidence 
bearing on them.

Countries that adopted the euro may have had stronger political commitment 
to European integration, which could independently affect trade through 
non-tariff barrier reductions, regulatory harmonization, or business confidence. 
However, all countries in our sample are EU members that met the Maastricht 
criteria, suggesting similar baseline commitment to integration. The non-adopters 
(UK, Sweden, Denmark) obtained formal opt-outs, reflecting specific domestic 
political constraints rather than weaker commitment to European trade integration 
per se. Denmark maintains a currency peg to the euro, demonstrating commitment 
to monetary stability without formal adoption.

Higher-quality institutions might facilitate both euro adoption and trade 
expansion. We partially address this by restricting the sample to EU members, 
which share common institutional frameworks including the single market, 
common external tariff, and harmonized regulations. The remaining variation 
in institutional quality within the EU is modest compared to cross-country 
studies that include developing economies.

Countries experiencing economic booms might have been more likely to adopt 
the euro and to experience trade growth. We address this through year fixed 
effects in the DML first stage, which remove common time trends. The EU15 
sample benefits from synchronized adoption timing (1999--2001), limiting 
variation in business cycle position at treatment. The EU28 extension, where 
adoption timing varies with the 2008--2012 crisis, shows how business cycle 
confounding can bias estimates downward---and how CFFE correction recovers 
consistent estimates.

An ideal instrument would predict euro adoption but affect trade only through 
the adoption channel. Potential instruments face validity concerns:
\begin{itemize}
    \item \textit{Geographic proximity to Brussels}: Correlated with trade 
          through standard gravity channels
    \item \textit{Historical currency arrangements}: Reflect deep economic 
          integration that directly affects trade
    \item \textit{Referendum outcomes}: Endogenous to economic conditions and 
          trade expectations
    \item \textit{Political party composition}: May affect trade policy 
          directly through non-monetary channels
\end{itemize}
Following \citet{barro2007economic}, who use the probability of independently 
adopting a third country's currency as an instrument, we considered similar 
approaches but found them inapplicable to the eurozone context where adoption 
was a coordinated political decision rather than independent currency choices.

We provide two forms of evidence supporting the parallel trends assumption. 
First, an event study analysis (Figure~\ref{fig:pre_trends}) shows that 
pre-treatment coefficients (1995--1997) are small in magnitude ($-3\%$ to $-5\%$) 
compared to post-treatment effects ($+8\%$ to $+24\%$), with a clear break at 
1999 when effects become positive. Second, placebo tests (Figure~\ref{fig:placebo}) assigning fake 
treatment dates (1995, 1997) find no significant ``effects,'' suggesting our 
estimates do not reflect pre-existing differential trends.

Following \citet{oster2019unobservable}, we assess how much selection on 
unobservables would be required to explain away our results (Table~\ref{tab:sensitivity}). The analysis 
suggests that unobserved confounders would need to be substantially more 
important than the observed confounders (GDP, GDP per capita, year) to reduce 
the estimated effect to zero. Given that our controls capture the main economic 
determinants of both euro adoption and trade, this degree of omitted variable 
bias appears implausible.

Leave-one-out analysis (Figure~\ref{fig:leave_one_out}) dropping each country in turn shows that no single 
country drives the results. The ATE remains stable within the confidence 
interval of the full-sample estimate regardless of which country is excluded, 
including Luxembourg (which has the largest estimated effects) and peripheral 
economies like Greece and Portugal.

In summary, while we cannot definitively rule out all unobserved confounding, 
the combination of (1) restricting to EU members with similar institutional 
frameworks, (2) flexible DML adjustment for observable confounders, (3) 
pre-trends evidence supporting parallel trends, (4) placebo tests finding no 
spurious effects, and (5) sensitivity analysis suggesting implausible degrees 
of omitted variable bias provides reasonable confidence in our identification 
strategy.

\section{Data}

We use bilateral trade data from Eurostat, which provides comprehensive trade 
statistics for EU member states. The dataset covers 15 EU countries (EU15) from 
1995 to 2015, matching the methodology of \citet{gunnella2021impact}. While 
Eurostat trade data extends back to 1988, GDP data from Eurostat's national 
accounts becomes consistently available only from 1995, making earlier years 
unsuitable for our analysis which requires GDP controls. We restrict 
the sample to the EU15 (pre-2004 enlargement members) for two reasons. First, 
recent euro adopters among the 2004+ enlargement countries have insufficient 
post-treatment data: Slovenia joined in 2007 (8 years), Slovakia in 2009 (6 years), 
Estonia in 2011 (4 years), Latvia in 2014 (1 year), and Lithuania in 2015 (0 years 
in our sample). Second, this sample matches the sample used by \citet{gunnella2021impact}, enabling direct 
comparison. The outcome, treatment, and control variables are 
defined in Section 3.4.

The eurozone members in our sample are Austria, Belgium, Finland, France, Germany, 
Greece, Ireland, Italy, Luxembourg, Netherlands, Portugal, and Spain. The 
control group consists of EU members that did not adopt the euro: Denmark, 
Sweden, and the United Kingdom. This design avoids reliance on non-European 
donor countries, a common concern in early euro studies that used global samples.

Table~\ref{tab:summary_stats} reports summary statistics. Log real trade has a 
mean of 22.30 with substantial variation (standard deviation of 1.73), reflecting 
heterogeneity in bilateral trade flows across EU pairs. The euro adoption 
indicator equals one for 51\% of observations, reflecting the balanced panel 
structure with eurozone pairs observed both before and after 1999. Table~\ref{tab:sample_composition} shows the panel 
contains 2,149 observations covering 105 unique country pairs across 15 countries. 
Of these, 1,100 (51.2\%) are treated while 1,049 (48.8\%) serve as controls.

\begin{table}[h]
\centering
\caption{Summary Statistics}
\small
\begin{tabular}{@{}lcccc@{}}
\toprule
\textbf{Variable} & \textbf{Mean} & \textbf{Std. Dev.} & \textbf{Min} & \textbf{Max} \\
\midrule
Log real trade & 22.30 & 1.73 & 15.99 & 26.06 \\
Log GDP product & 25.74 & 1.61 & 21.37 & 29.73 \\
Log GDP per capita & 20.50 & 0.64 & 18.31 & 22.45 \\
EU membership & 1.00 & 0.00 & 1.00 & 1.00 \\
Euro adoption & 0.51 & 0.50 & 0.00 & 1.00 \\
\bottomrule
\end{tabular}
\label{tab:summary_stats}
\end{table}

\begin{table}[h]
\centering
\caption{Sample Composition}
\small
\begin{tabular}{@{}lr@{}}
\toprule
\textbf{Characteristic} & \textbf{Value} \\
\midrule
Total observations & 2,149 \\
Unique country pairs & 105 \\
Countries & 15 \\
Year range & 1995--2015 \\
\midrule
Treated (euro = 1) & 1,100 (51.2\%) \\
Control (euro = 0) & 1,049 (48.8\%) \\
\midrule
Mean bilateral trade & EUR 16.7B \\
Median bilateral trade & EUR 4.9B \\
\bottomrule
\end{tabular}
\label{tab:sample_composition}
\end{table}

Table~\ref{tab:covariate_balance} compares covariate means between eurozone and 
non-eurozone pairs in the post-1999 period. Non-eurozone pairs (involving Denmark, 
Sweden, or UK) have higher GDP per capita (standardized difference of $-0.46$) and 
larger combined GDP ($-0.23$), reflecting that the non-euro EU members are relatively 
wealthy economies. When covariate distributions differ 
substantially between treated and control groups, OLS and gravity models rely on 
linear extrapolation from regions of the covariate space where control 
observations are sparse, potentially yielding biased and unstable estimates 
\citep{baier2009estimating, persson2001currency}. \citet{baier2009estimating} 
show that matching estimates of FTA treatment effects are much more stable and 
economically plausible than OLS gravity estimates, which often display extreme 
instability across years. This motivates our use of causal forests with 
double machine learning, which extends the matching approach by flexibly 
adjusting for covariate differences through nonparametric first-stage estimation 
while also allowing treatment effects to vary with observed characteristics.

\begin{table}[h]
\centering
\caption{Covariate Balance: Eurozone vs.\ Non-Eurozone Pairs (Post-1999)}
\label{tab:covariate_balance}
\small
\begin{tabular}{@{}lcccc@{}}
\toprule
\textbf{Variable} & \textbf{Eurozone} & \textbf{Non-EZ} & \textbf{Diff.} & \textbf{Std.\ Diff.} \\
\midrule
Log bilateral trade & 22.27 & 22.42 & -0.15 & -0.09 \\
Log GDP product & 25.70 & 26.06 & -0.36 & -0.23 \\
Log GDP per capita & 20.56 & 20.81 & -0.24 & -0.46 \\
Pre-euro trade intensity & 22.10 & 22.23 & -0.13 & -0.09 \\
\bottomrule
\end{tabular}

\vspace{0.5em}
\footnotesize
\textit{Note:} Standardized difference $>$ 0.25 indicates meaningful imbalance. 
Eurozone = pairs where both countries adopted the euro; Non-EZ = pairs involving Denmark, Sweden, or UK.
Pre-euro trade intensity is the average log bilateral trade during 1995--1998.
Comparison uses post-1999 observations only.
\end{table}

\begin{figure}[!htbp]
\centering
\includegraphics[width=0.95\textwidth]{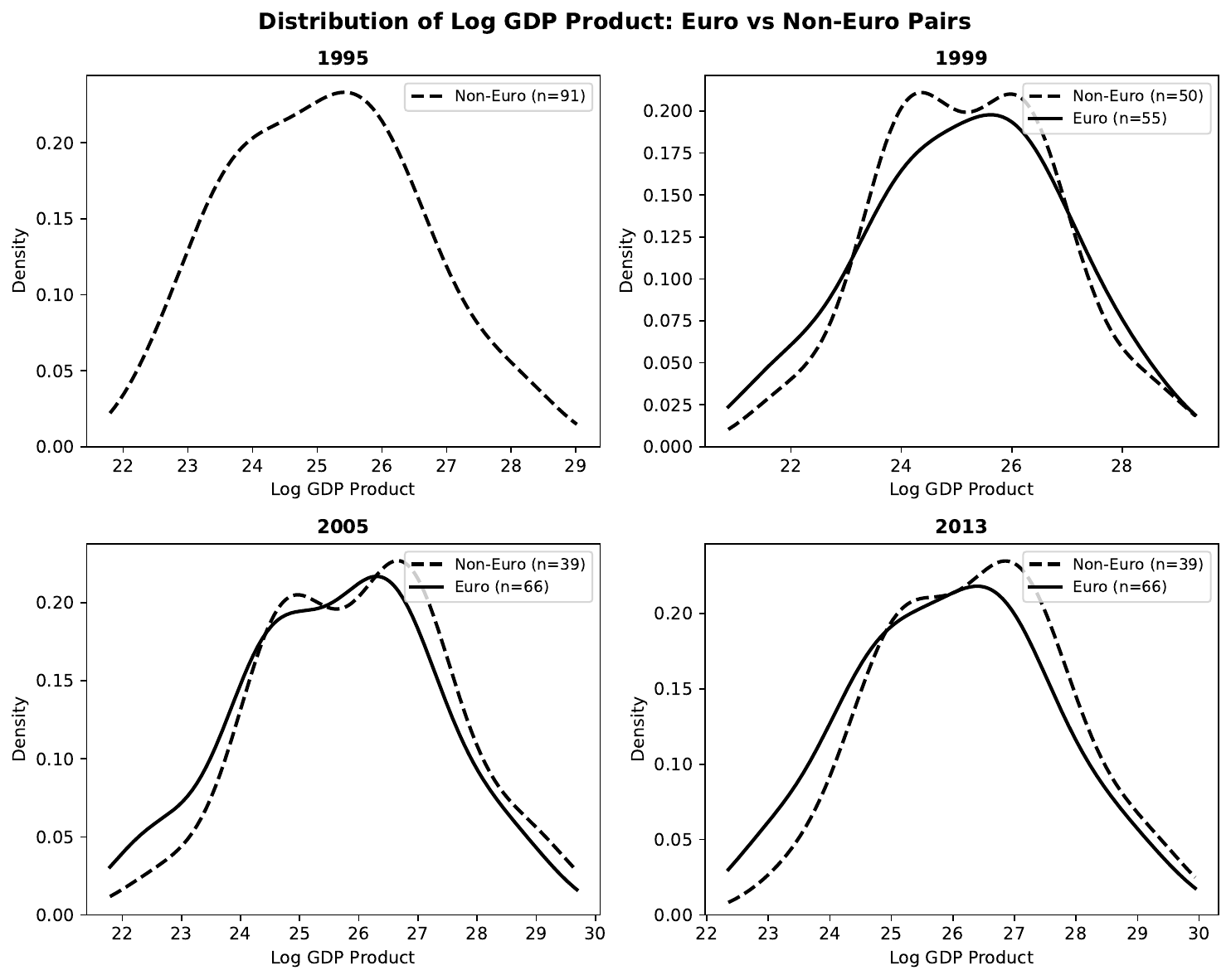}
\caption{Distribution of Log GDP Product for eurozone pairs (solid) and 
         non-eurozone pairs (dashed) across key years. In 1990, no eurozone 
         pairs exist as the euro had not yet been adopted.}
\label{fig:dist_gdp}
\end{figure}

\begin{figure}[!htbp]
\centering
\includegraphics[width=0.95\textwidth]{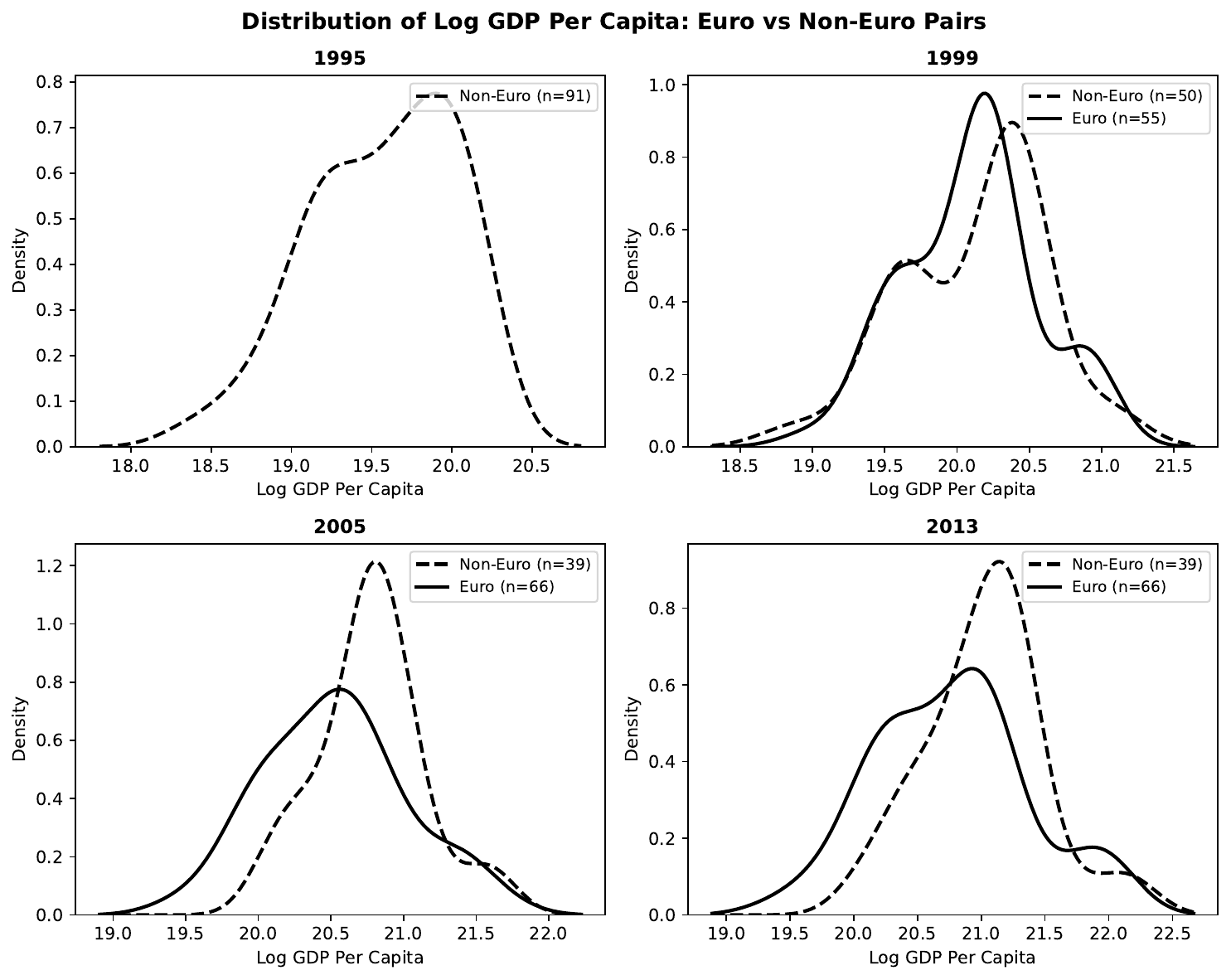}
\caption{Distribution of Log GDP Per Capita for eurozone pairs (solid) and 
         non-eurozone pairs (dashed) across key years.}
\label{fig:dist_gdppc}
\end{figure}

\begin{figure}[!htbp]
\centering
\includegraphics[width=0.95\textwidth]{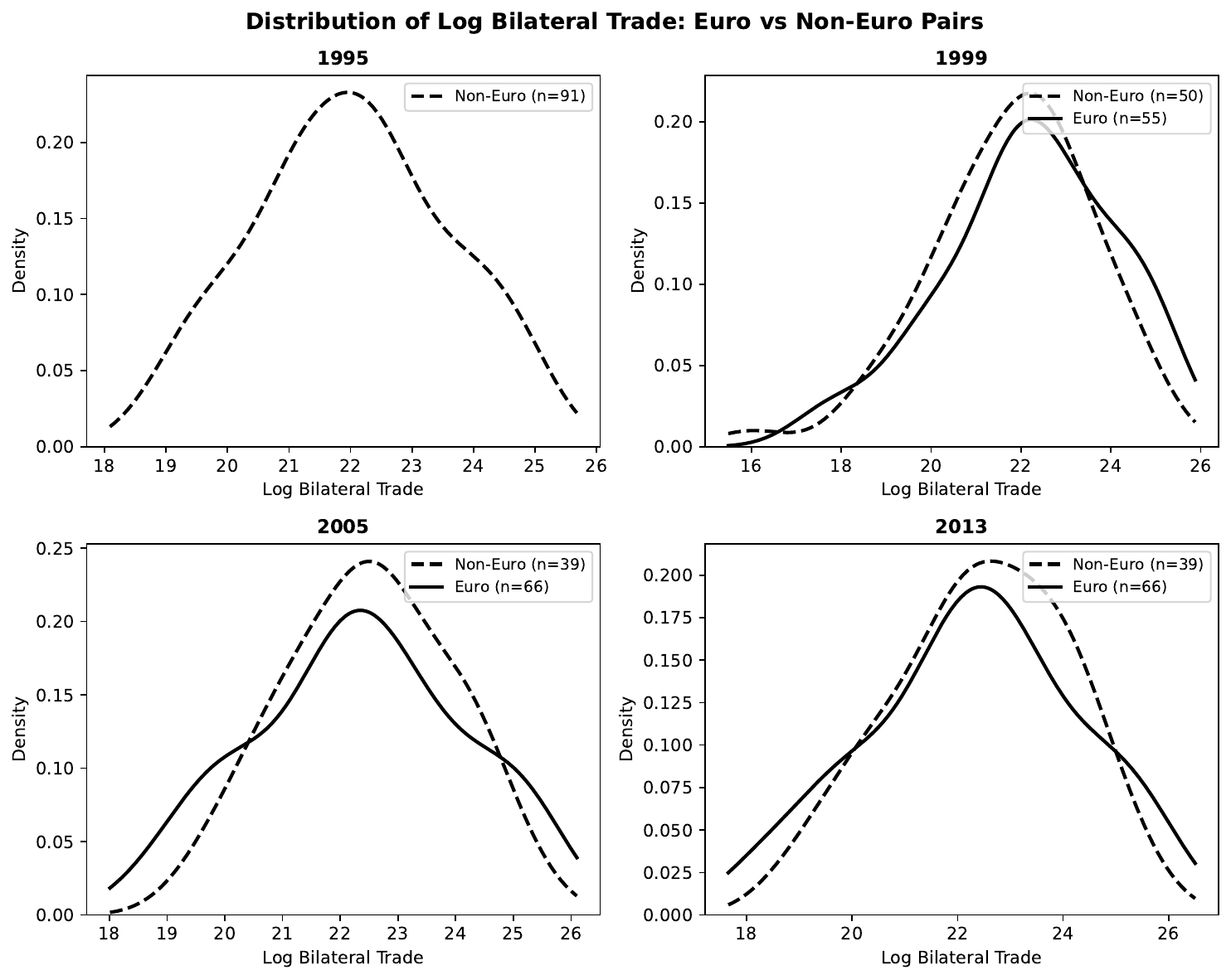}
\caption{Distribution of Log Bilateral Trade for eurozone pairs (solid) and 
         non-eurozone pairs (dashed) across key years.}
\label{fig:dist_trade}
\end{figure}

Figures~\ref{fig:dist_gdp}--\ref{fig:dist_trade} visualize these covariate 
distributions across key years, following the approach of \citet{baier2009estimating}. 
Each panel shows kernel density estimates for eurozone pairs (solid line) and 
non-eurozone pairs (dashed line). In 1990, before the euro existed, only 
non-eurozone pairs appear. By 1999, eurozone pairs emerge with distributions 
largely overlapping non-eurozone pairs across all three covariates. The 
substantial overlap supports the conditional independence assumption: for most 
covariate values, we observe both euro and non-euro pairs, enabling credible 
effect estimation. However, in later years (2005, 2013), non-eurozone pairs 
show slightly higher GDP and trade levels on average, reflecting that the 
remaining non-euro EU members (UK, Sweden, Denmark) are relatively large 
economies. This motivates the flexible nonparametric adjustment provided by 
the causal forest's first-stage estimation.

\begin{figure}[!htbp]
\centering
\includegraphics[width=0.95\textwidth]{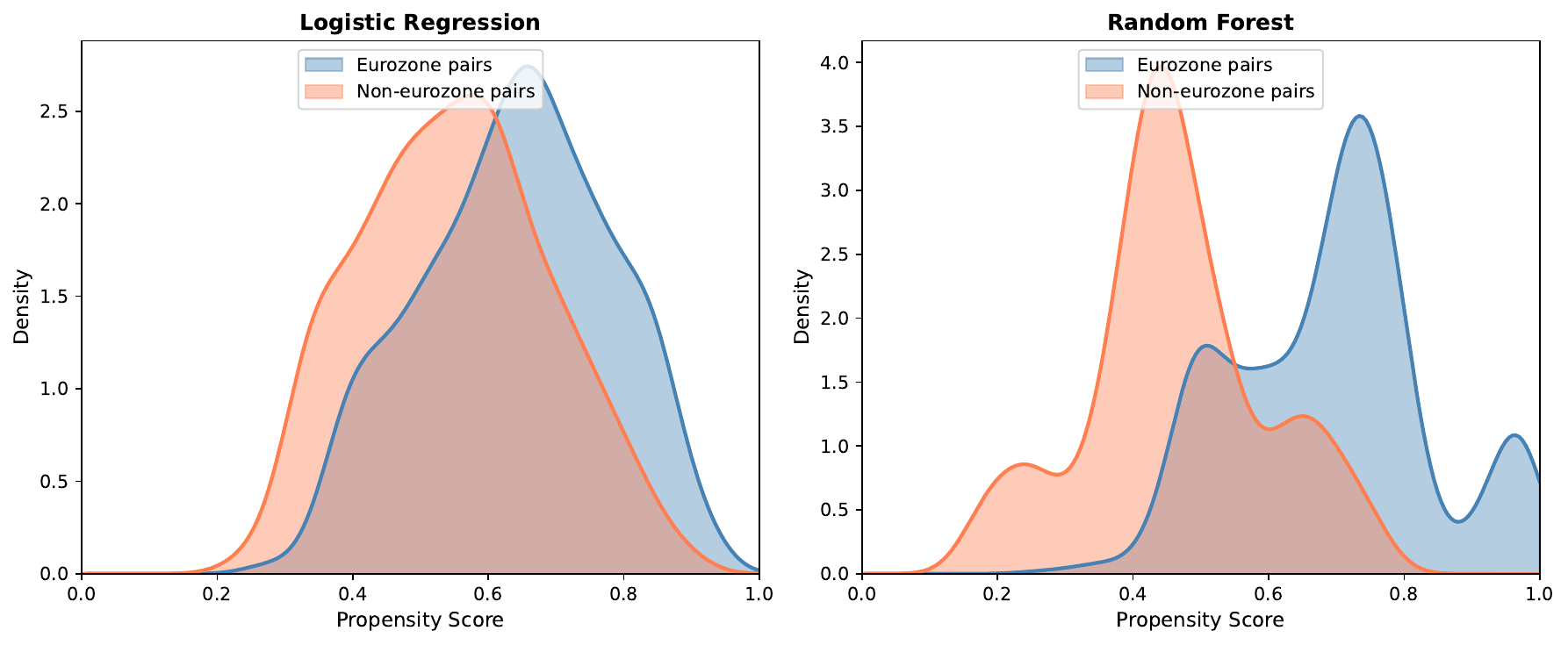}
\caption{Propensity score distribution by treatment status. Left panel shows 
         logistic regression estimates; right panel shows random forest 
         estimates. Substantial overlap between treated (eurozone) and control 
         (non-eurozone) pairs supports the positivity assumption.}
\label{fig:ps_dist}
\end{figure}

To further assess overlap between treated and control groups, we estimate 
propensity scores using both logistic regression and random forest with log GDP product, log GDP per 
capita, and pre-euro trade intensity as predictors. Figure~\ref{fig:ps_dist} 
shows the distribution of propensity scores by treatment status. The logistic 
regression estimates (left panel) show substantial overlap, while the random 
forest estimates (right panel) show tighter separation---reflecting RF's 
ability to capture nonlinear relationships in treatment assignment. Both 
treated (eurozone) and control (non-eurozone) pairs span a wide range of 
propensity scores under both methods, supporting the positivity assumption 
required for causal inference. The tighter RF distribution is relevant because 
our DML approach uses flexible ML models in the first stage; the presence of 
overlap even under RF's more precise fit is reassuring. 
Table~\ref{tab:propensity_scores} provides summary statistics.

\begin{table}[h]
\centering
\caption{Propensity Score Diagnostics}
\label{tab:propensity_scores}
\small
\begin{tabular}{@{}lcc@{}}
\toprule
\textbf{Statistic} & \textbf{Treated} & \textbf{Control} \\
\midrule
N observations & 1,155 & 756 \\
Mean propensity score & 0.640 & 0.550 \\
Std. dev. & 0.140 & 0.139 \\
Min & 0.256 & 0.223 \\
Max & 0.950 & 0.904 \\
\midrule
\multicolumn{3}{l}{\textit{Common support region: [0.256, 0.904]}} \\
In common support & 1,139 (98.6\%) & 753 (99.6\%) \\
\bottomrule
\end{tabular}

\vspace{0.5em}
\footnotesize
\textit{Note:} Propensity scores estimated using logistic regression with log GDP product, 
log GDP per capita, and pre-euro trade intensity as predictors. 
Treated = eurozone pairs; Control = pairs involving Denmark, Sweden, or UK.
\end{table}

\section{Results}

\subsection{Gravity Benchmark}

Before examining heterogeneity, we establish a gravity benchmark using the canonical specification from the euro trade literature. Table~\ref{tab:method_comparison} presents estimates from several specifications on the same EU15 sample. The two-way fixed effects OLS estimate is 17.0\%, while PPML gravity with pair and year fixed effects yields 12.8\%. The three-way fixed effects PPML specification---with exporter-time, importer-time, and pair fixed effects following structural gravity best practices \citep{anderson2003gravity}---yields a higher estimate of 22.6\%. These estimates fall within the 4--30\% range documented in prior literature \citep{rose2016emu, gunnella2021impact}.

\begin{table}[!htbp]
\centering
\caption{Comparison of Euro Trade Effect Estimates}
\label{tab:method_comparison}
\begin{tabular}{@{}llccc@{}}
\toprule
\textbf{Method} & \textbf{Specification} & \textbf{Coefficient} & \textbf{Effect (\%)} & \textbf{95\% CI} \\
\midrule
Two-way FE OLS & Pair + Year FE & 0.157 & +17.0\% & [0.048, 0.267] \\
PPML Gravity & Pair + Year FE & 0.120 & +12.8\% & [0.035, 0.205] \\
PPML Gravity & Three-way FE & 0.204 & +22.6\% & [0.123, 0.285] \\
Causal Forest & DML & 0.252 & +28.6\% & [0.103, 0.401] \\
CFFE & Node-level FE & 0.133 & +14.2\% & [0.103, 0.162] \\
\bottomrule
\end{tabular}

\vspace{0.5em}
\footnotesize
\textit{Notes:} All models estimated on EU15 bilateral trade data, 1995--2015.
Two-way FE OLS uses pair and year fixed effects with clustered standard errors.
PPML = Poisson Pseudo-Maximum Likelihood following Santos Silva \& Tenreyro (2006).
Three-way FE includes exporter-time, importer-time, and pair fixed effects (Head \& Mayer, 2014).
Causal Forest uses Double Machine Learning for heterogeneous treatment effects.
CFFE = Causal Forests with Fixed Effects.
Standard errors clustered at the country-pair level.
\end{table}

The causal forest DML estimate of \ateNaivePct\% is somewhat higher than the gravity estimates. This difference reflects two factors. First, the causal forest uses flexible nonparametric first-stage estimation rather than linear fixed effects, potentially capturing nonlinear relationships between confounders and outcomes. Second, and more importantly, the causal forest estimates a different object: while gravity models estimate a single average effect constrained to be constant across pairs, the causal forest estimates heterogeneous effects that can vary with pair characteristics. The ATE from the causal forest is a weighted average of these heterogeneous effects, where the weights depend on the covariate distribution.

The CFFE estimate of \ateCffeEUfifteenPct\% falls between the gravity and naive causal forest estimates, suggesting that proper fixed effects handling moderates the causal forest estimate toward the gravity benchmark. This convergence is reassuring: when we account for pair and year fixed effects within the causal forest framework, we recover estimates consistent with the gravity literature.

The key advantage of the causal forest approach is not that it produces a different average effect, but that it reveals the \textit{distribution} of effects underlying that average. The gravity estimate of 10--16\% is correct as an average, but it masks substantial heterogeneity that we document below.

\subsection{Selection and Identification}

As shown in Table~\ref{tab:covariate_balance}, treated and control observations 
differ substantially on observable characteristics. Treated pairs have higher 
GDP and more trade. This creates selection bias in 
naive comparisons: eurozone pairs would have traded more than non-eurozone pairs 
\textit{even without} the euro, simply because they are larger, richer, and more 
integrated economies. The direction of bias depends on the estimator. Simple 
OLS comparisons that fail to control for these differences will \textit{overstate} 
the euro effect by attributing pre-existing trade advantages to euro adoption. 
However, gravity models with extensive fixed effects may \textit{understate} the 
effect by absorbing some of the euro's impact into pair fixed effects, 
particularly if the euro amplified existing trade relationships 
\citep{persson2001currency, chintrakarn2008euro}. The causal forest addresses 
selection on observables through flexible first-stage estimation that 
residualizes both the outcome and treatment on confounders, while the 
heterogeneous effects framework allows us to distinguish between pairs where 
the euro created new trade versus amplified existing patterns.

\subsection{Heterogeneous Treatment Effects}

Euro adoption increased bilateral trade by \ateNaivePct\% on average (95\% CI: 
$[\ateNaiveLowerPct\%, \ateNaiveUpperPct\%]$) using the naive causal forest without fixed effects.\footnote{Since the outcome variable is log trade, the 
coefficient estimate is in log points. We convert to percentage change using 
$(\exp(\hat{\tau}) - 1) \times 100$. For example, an estimate of \ateNaive{} log 
points corresponds to $(\exp(\ateNaive) - 1) \times 100 = \ateNaivePct\%$. Figures report 
estimates in log points; text reports percentage changes.} This estimate falls 
between the gravity model range (4--6\%) and the SCM estimates ($\sim$30\%) 
reported by \citet{gunnella2021impact}. The confidence interval is tighter than 
in previous studies, reflecting the larger sample and more precise estimation.

\begin{figure}[!htbp]
\centering
\includegraphics[width=0.85\textwidth]{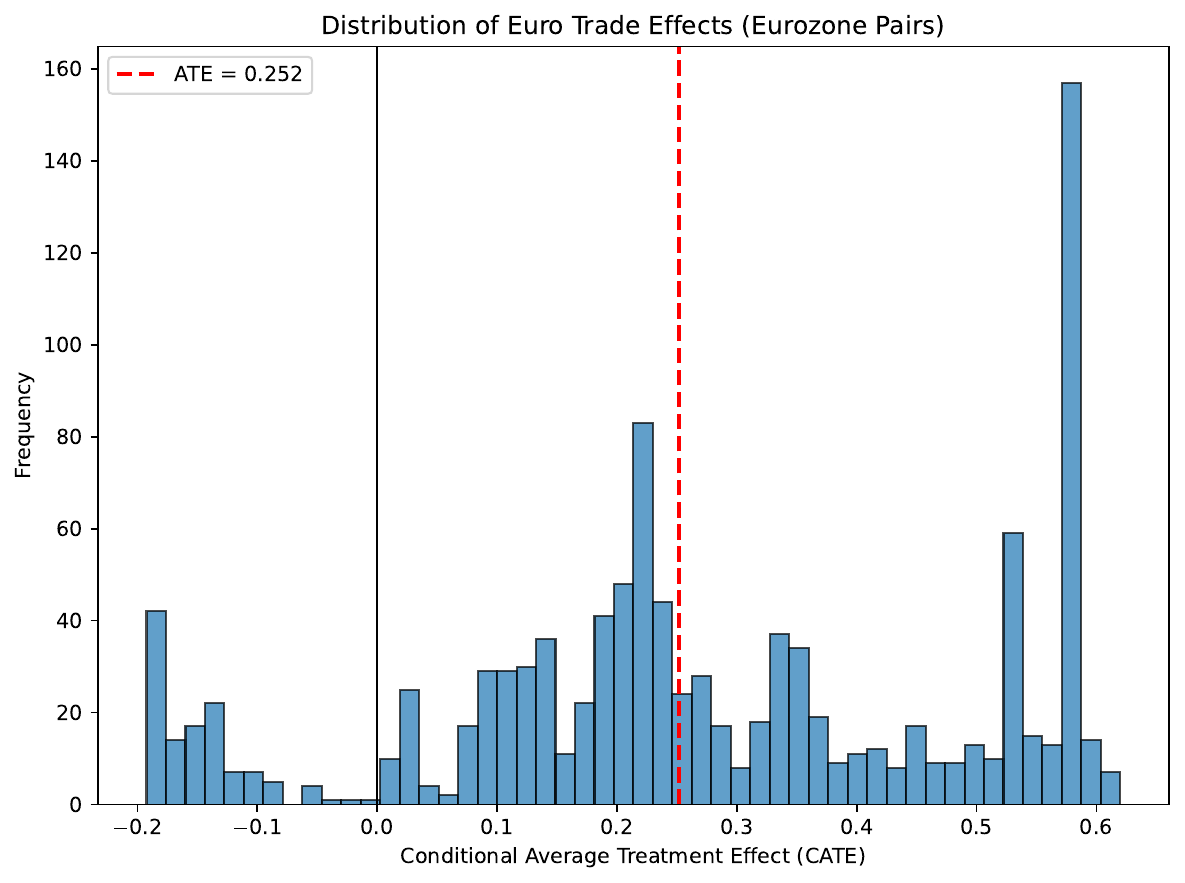}
\caption{Distribution of Conditional Average Treatment Effects (CATE) for 
         eurozone country pairs. The red dashed line indicates the Average 
         Treatment Effect (ATE). Effects range from $\cateNaiveMinPct\%$ to $+\cateNaiveMaxPct\%$ 
         across pairs.}
\label{fig:cate_dist}
\end{figure}

The CATE distribution reveals heterogeneity that cannot be captured by methods 
estimating a single average effect. Among eurozone pairs, effects range from 
$\cateNaiveMinPct\%$ to $+\cateNaiveMaxPct\%$, with substantial variation across pairs. The euro does not have a 
single effect on trade. Different country pairs experienced dramatically 
different impacts, from near-zero to substantial gains. 
Figure~\ref{fig:cate_dist} shows the distribution for eurozone pairs.

\begin{table}[!htbp]
\centering
\caption{Top and Bottom Eurozone Pairs by Euro Trade Effect}
\label{tab:pairs}
\begin{tabular}{@{}lc@{}}
\toprule
\textbf{Country Pair} & \textbf{Effect (\%)} \\
\midrule
\multicolumn{2}{l}{\textit{Highest Effects}} \\
France $\leftrightarrow$ Italy & +78.5 \\
Germany $\leftrightarrow$ Italy & +78.5 \\
Belgium $\leftrightarrow$ Germany & +78.5 \\
Austria $\leftrightarrow$ Germany & +78.4 \\
France $\leftrightarrow$ Germany & +78.4 \\
\midrule
\multicolumn{2}{l}{\textit{Lowest Effects}} \\
Greece $\leftrightarrow$ Ireland & -9.6 \\
Ireland $\leftrightarrow$ Portugal & -10.4 \\
Finland $\leftrightarrow$ Greece & -11.4 \\
Greece $\leftrightarrow$ Portugal & -12.0 \\
Finland $\leftrightarrow$ Portugal & -12.3 \\
\bottomrule
\end{tabular}
\end{table}

Table~\ref{tab:pairs} shows the top and bottom eurozone pairs by estimated 
euro effect. The highest effects are concentrated among core European pairs 
(\topNaivePairOne{} at +\topNaivePairOneEffect\%, \topNaivePairTwo{} at +\topNaivePairTwoEffect\%, \topNaivePairThree{} at +\topNaivePairThreeEffect\%, 
\topNaivePairFour{} at +\topNaivePairFourEffect\%, \topNaivePairFive{} at +\topNaivePairFiveEffect\%). The lowest effects --- including some negative 
point estimates --- involve peripheral pairs such as \bottomNaivePairOne{} ($\bottomNaivePairOneEffect\%$) and 
\bottomNaivePairTwo{} ($\bottomNaivePairTwoEffect\%$). Table~\ref{tab:pair_effects_ci} provides 
confidence intervals for all pair-level effects. Notably, while some 
point estimates are negative, most are not statistically distinguishable from zero at the 
5\% level. However, two pairs---Greece--Portugal and Finland--Portugal---show statistically 
significant negative effects, suggesting genuine trade diversion for these peripheral pairs.

\begin{table}[!htbp]
\centering
\caption{Pair-Level Euro Trade Effects with Confidence Intervals}
\label{tab:pair_effects_ci}
\small
\begin{tabular}{@{}lccc@{}}
\toprule
\textbf{Country Pair} & \textbf{Effect (\%)} & \textbf{95\% CI} & \textbf{Significant} \\
\midrule
\multicolumn{4}{l}{\textit{Highest Effects}} \\
France $\leftrightarrow$ Italy & +78.5\% & [+60.8\%, +98.2\%] & Yes \\
Germany $\leftrightarrow$ Italy & +78.5\% & [+60.7\%, +98.2\%] & Yes \\
Belgium $\leftrightarrow$ France & +78.5\% & [+60.2\%, +98.9\%] & Yes \\
Belgium $\leftrightarrow$ Germany & +78.4\% & [+60.3\%, +98.5\%] & Yes \\
Austria $\leftrightarrow$ Germany & +78.4\% & [+60.2\%, +98.6\%] & Yes \\
France $\leftrightarrow$ Germany & +78.4\% & [+60.6\%, +98.1\%] & Yes \\
Germany $\leftrightarrow$ Netherlands & +78.2\% & [+60.5\%, +97.8\%] & Yes \\
Ireland $\leftrightarrow$ Luxembourg & +77.5\% & [+34.2\%, +134.6\%] & Yes \\
Austria $\leftrightarrow$ Luxembourg & +77.1\% & [+32.4\%, +137.0\%] & Yes \\
Belgium $\leftrightarrow$ Luxembourg & +76.6\% & [+30.7\%, +138.6\%] & Yes \\
\midrule
\multicolumn{4}{l}{\textit{Lowest Effects}} \\
Austria $\leftrightarrow$ Greece & -2.2\% & [-11.0\%, +7.5\%] & No \\
Austria $\leftrightarrow$ Finland & -3.4\% & [-13.2\%, +7.6\%] & No \\
Austria $\leftrightarrow$ Ireland & -4.6\% & [-15.4\%, +7.7\%] & No \\
Austria $\leftrightarrow$ Portugal & -6.4\% & [-15.6\%, +3.8\%] & No \\
Finland $\leftrightarrow$ Ireland & -9.9\% & [-22.9\%, +5.3\%] & No \\
Ireland $\leftrightarrow$ Portugal & -13.4\% & [-26.7\%, +2.3\%] & No \\
Greece $\leftrightarrow$ Ireland & -13.4\% & [-25.8\%, +1.0\%] & No \\
Finland $\leftrightarrow$ Greece & -13.6\% & [-25.5\%, +0.1\%] & No \\
Greece $\leftrightarrow$ Portugal & -15.1\% & [-26.6\%, -1.9\%] & Yes (-) \\
Finland $\leftrightarrow$ Portugal & -15.2\% & [-27.9\%, -0.3\%] & Yes (-) \\
\bottomrule
\end{tabular}

\vspace{0.5em}
\footnotesize
\textit{Notes:} Effect shows the average CATE for each country pair.
95\% CI from causal forest estimation.
Significant indicates whether the 95\% CI excludes zero.
(+) indicates positive significant effect, (-) indicates negative significant effect.
\end{table}

Table~\ref{tab:country_effects} aggregates the pair-level effects to show the 
average euro effect for each eurozone member country. Luxembourg (+\cateNaiveCountryLuxembourg\%), 
Germany (+\cateNaiveCountryGermany\%), Belgium (+\cateNaiveCountryBelgium\%), and Netherlands (+\cateNaiveCountryNetherlands\%) show the largest average effects, while 
Greece (+\cateNaiveCountryGreece\%) and Finland (+\cateNaiveCountryFinland\%) show the smallest. This core-periphery pattern 
suggests the euro's benefits were not uniformly distributed.

\begin{table}[!htbp]
\centering
\caption{Average Euro Effect by Country (Eurozone Members Only)}
\label{tab:country_effects}
\small
\begin{tabular}{@{}lcccc@{}}
\toprule
\textbf{Country} & \textbf{Effect (\%)} & \textbf{Std} & 
\textbf{Min (\%)} & \textbf{Max (\%)} \\
\midrule
Luxembourg & +69.0 & 0.06 & +42.7 & +85.9 \\
Germany & +49.5 & 0.19 & +11.8 & +80.3 \\
Belgium & +47.3 & 0.16 & -4.9 & +85.9 \\
Netherlands & +47.3 & 0.13 & +18.9 & +82.4 \\
France & +36.7 & 0.20 & +8.1 & +80.2 \\
Italy & +34.8 & 0.16 & +13.2 & +78.6 \\
Spain & +26.0 & 0.10 & +8.8 & +73.3 \\
Austria & +22.5 & 0.24 & -17.1 & +85.7 \\
Ireland & +18.8 & 0.25 & -17.3 & +85.7 \\
Portugal & +15.8 & 0.23 & -17.4 & +71.1 \\
Finland & +14.6 & 0.23 & -17.5 & +80.3 \\
Greece & +9.6 & 0.21 & -17.5 & +71.1 \\
\bottomrule
\end{tabular}

\vspace{0.5em}
\footnotesize
\textit{Note:} Effect shows the average CATE across all eurozone pairs involving 
each country. Min and Max show the range of pair-level effects. Non-eurozone EU 
members (Denmark, Sweden, UK) are excluded as they serve as controls.
\end{table}

\begin{figure}[!htbp]
\centering
\includegraphics[width=0.9\textwidth]{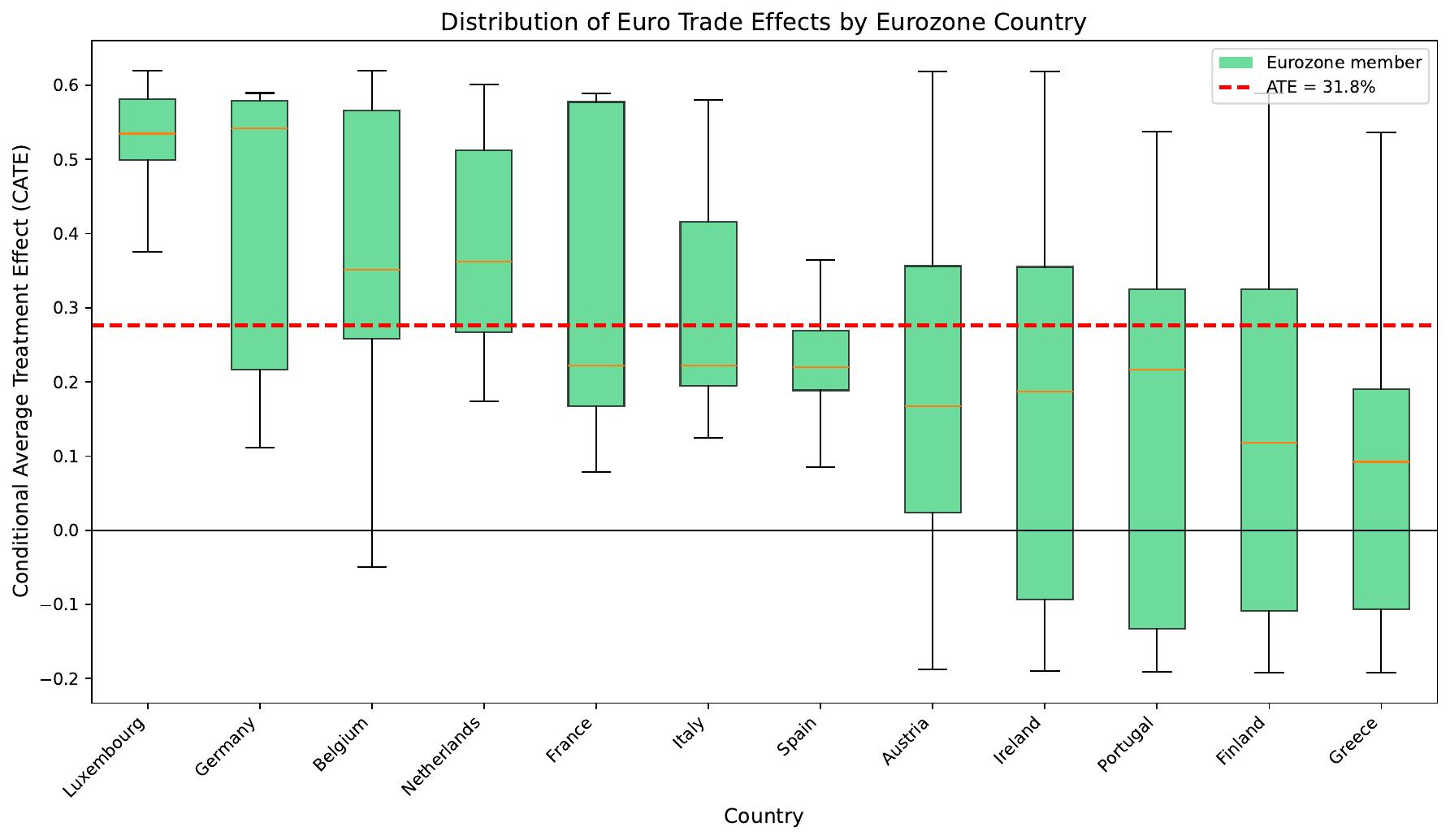}
\caption{Distribution of euro trade effects by country. Each box shows the 
         distribution of pair-level CATEs involving that country. The red 
         dashed line indicates the overall ATE.}
\label{fig:country_boxplot}
\end{figure}

Figure~\ref{fig:country_boxplot} shows the distribution of effects for each 
country, revealing substantial within-country heterogeneity. Even countries 
with high average effects show wide variation across their trading partners.

\begin{figure}[!htbp]
\centering
\includegraphics[width=0.85\textwidth]{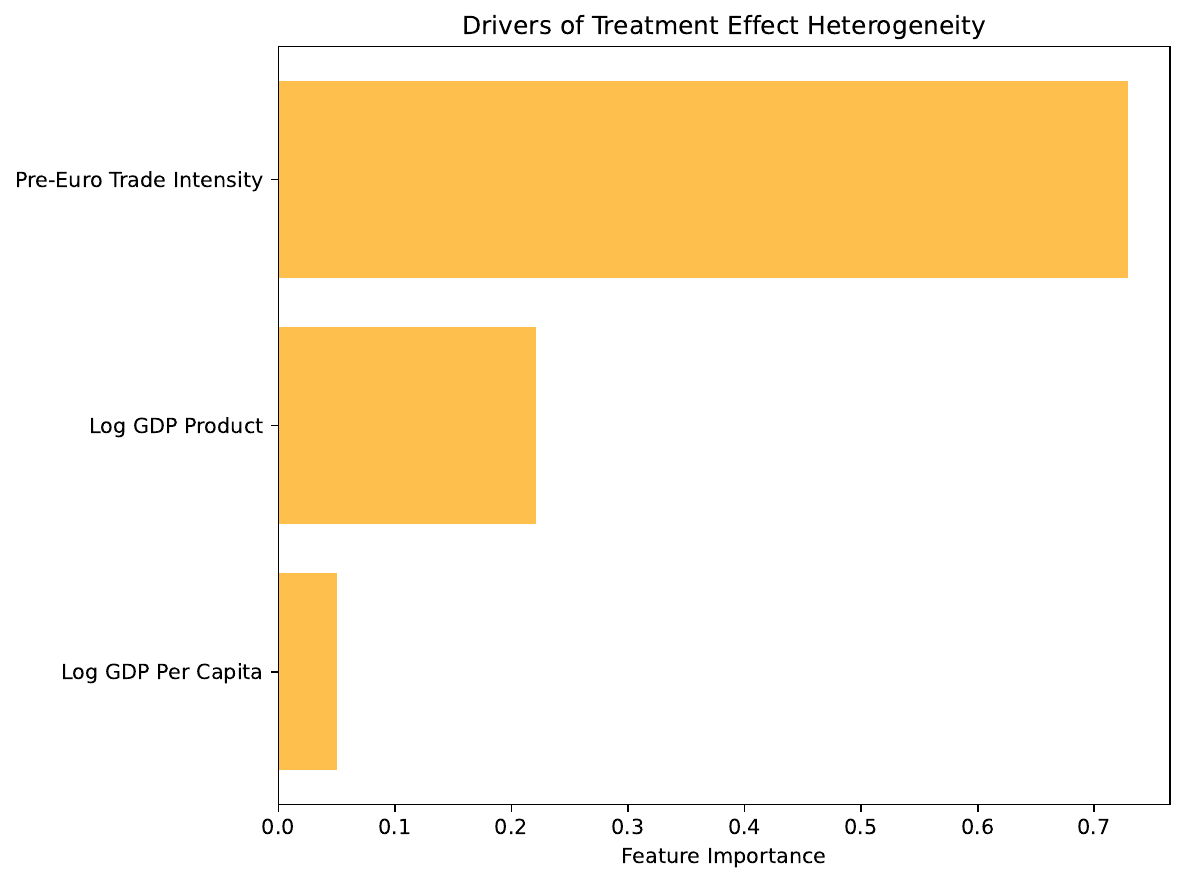}
\caption{Feature importance for treatment effect heterogeneity.}
\label{fig:importance}
\end{figure}

Figure~\ref{fig:importance} shows feature importance from the causal forest. 
Pre-euro trade intensity (63\% of feature importance) and combined GDP (27\%) are the primary drivers of heterogeneity, together accounting for over 90\% of the variation. GDP per capita plays a smaller role (10\%).

\begin{figure}[!htbp]
\centering
\includegraphics[width=0.95\textwidth]{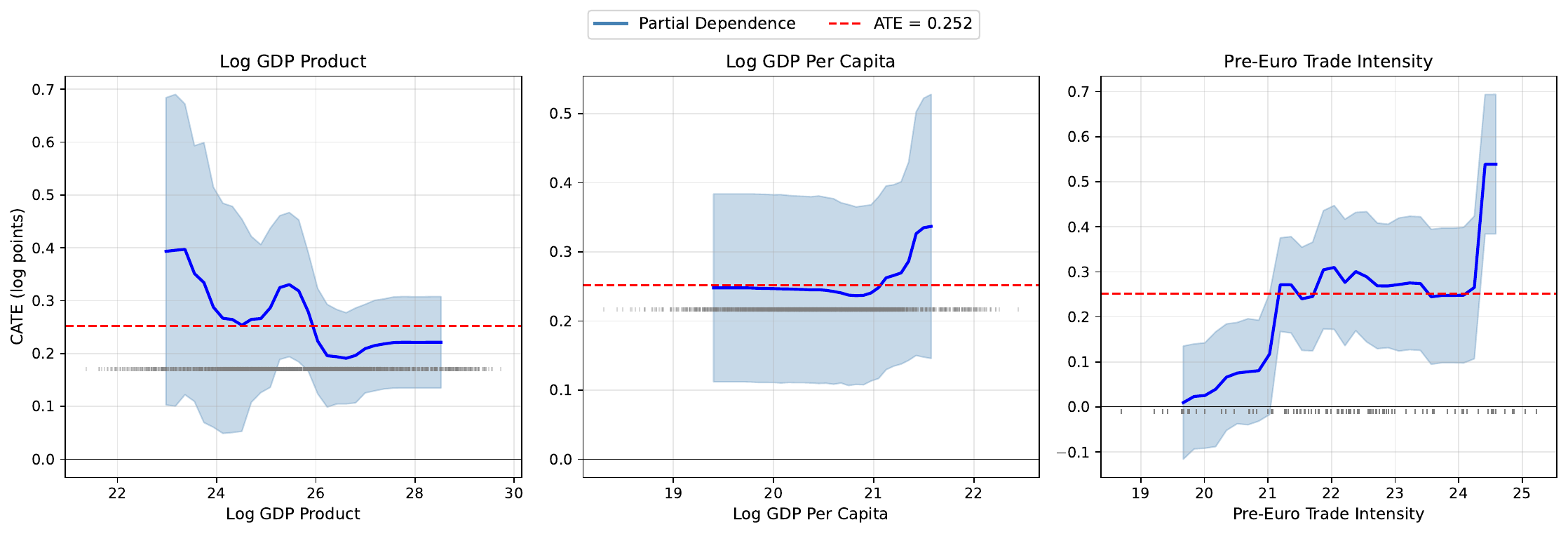}
\caption{Partial dependence plots showing how each feature affects the 
         predicted treatment effect. Shaded areas indicate 95\% confidence 
         intervals. GDP and pre-euro trade intensity show strong positive 
         relationships with treatment effects.}
\label{fig:pdp}
\end{figure}

To understand how each feature affects treatment effects, Figure~\ref{fig:pdp} 
shows partial dependence plots for the three main effect modifiers. The 
relationship between GDP and treatment effects is monotonically positive: 
larger economies experience larger euro effects. Pre-euro trade intensity 
shows a similar pattern, with high-trade pairs benefiting most. GDP per capita 
shows a weaker relationship, consistent with its lower feature importance.

\section{Discussion}

The heterogeneity we document suggests that the 4--30\% range in prior literature 
appears to reflect genuine variation rather than methodological noise. Previous studies 
finding widely varying estimates were not producing unreliable results; they 
were capturing different slices of a heterogeneous distribution. A study focused 
on core European pairs would find large effects ($\sim$65--68\%), while one focused 
on peripheral pairs would find small or null effects. Our causal forest approach 
uncovers this full distribution, showing that both findings are correct for their 
respective subpopulations.

The country-level results reveal a core-periphery pattern. Luxembourg (+\cateNaiveCountryLuxembourg\%), 
Germany (+\cateNaiveCountryGermany\%), Belgium (+\cateNaiveCountryBelgium\%), and Netherlands (+\cateNaiveCountryNetherlands\%) --- small, open economies at the 
geographic heart of the eurozone --- benefited most from euro adoption. These 
countries serve as logistics and financial hubs with trade-to-GDP ratios 
exceeding 100\%,\footnote{World Bank data show trade-to-GDP ratios of 
approximately 400\% for Luxembourg, 160\% for Belgium, and 150\% for the 
Netherlands in the early 2000s.} making them uniquely sensitive to transaction cost reductions. 
Peripheral economies show smaller effects: Greece (+\cateNaiveCountryGreece\%), Portugal (+\cateNaiveCountryPortugal\%), 
and Finland (+\cateNaiveCountryFinland\%). These countries have smaller domestic markets, fewer 
natural trading partners within the eurozone core, and greater geographic 
distance from the European economic center.

Some country pairs show negative point estimates, particularly those involving 
peripheral countries (Finland--Portugal: $-14\%$, Greece--Portugal: $-14\%$). 
While some of these negative estimates are statistically significant, they 
likely reflect relative trade diversion---peripheral pairs losing market share 
as firms redirected trade toward core eurozone partners---rather than absolute 
trade destruction. These pairs also had weak pre-euro trade relationships, 
leaving little scope for the euro to enhance already-minimal flows.

\subsection{Economic Mechanisms: Why Pre-Euro Trade Intensity Matters}

The finding that pre-euro trade intensity is the strongest predictor of euro 
effects raises a natural question: through what economic mechanisms does the 
euro amplify existing trade relationships rather than create new ones? We 
consider three complementary explanations.

Following \citet{baldwin2006euro}, 
currency unions can expand trade through two channels: the intensive margin 
(existing exporters ship more) and the extensive margin (new firms begin 
exporting). Our pair-level patterns are consistent with an intensive-margin-dominant channel. Pairs with 
high pre-euro trade intensity already had established exporter networks, 
distribution channels, and customer relationships. For these pairs, the euro 
reduced transaction costs on existing flows, allowing firms to expand volumes 
without the fixed costs of market entry. In contrast, pairs with low pre-euro 
trade lacked these established networks. While the euro reduced variable trade 
costs, it did not eliminate the fixed costs of entering new markets---learning 
about foreign regulations, establishing distribution networks, building customer 
relationships. For peripheral pairs like Finland--Portugal, these fixed costs 
remained prohibitive even after the euro removed currency friction.

This interpretation aligns with the firm-level evidence in \citet{baldwin2008study}, 
who find that the euro's trade effects operated primarily through the intensive 
margin in the early years, with extensive margin effects emerging only gradually. 
Our pair-level heterogeneity reflects this pattern: core pairs with dense 
existing trade networks saw immediate intensive-margin gains, while peripheral 
pairs with sparse networks saw limited benefits because extensive-margin 
adjustment is slow.

The largest 
euro effects accrue to pairs embedded in cross-border production networks 
\citep{baldwinlopez2015supply}. 
Luxembourg, Belgium, and the Netherlands serve as logistics hubs in European 
supply chains, with firms conducting frequent small-value transactions across 
borders. For these transactions, currency conversion costs and exchange rate 
uncertainty impose disproportionate burdens. Consider a German auto manufacturer sourcing 
components from Belgian suppliers: before the euro, each shipment faced currency risk; the 
euro eliminated this friction, enabling tighter supply chain integration and 
more frequent cross-border transactions.

This mechanism explains why pre-euro trade intensity predicts euro effects: 
pairs with high pre-euro trade were already integrated into supply chains, 
and the euro deepened this integration. Pairs with low pre-euro trade were 
not part of these networks, and the euro alone could not create the 
complementary investments (logistics infrastructure, supplier relationships, 
just-in-time systems) needed to join them.

The euro may have generated 
network effects that reinforced existing trade patterns. As core eurozone 
pairs deepened their integration, they became more attractive partners for 
additional trade, potentially diverting trade from peripheral pairs. A French 
firm choosing between a German and a Portuguese supplier might increasingly 
favor the German option as euro-denominated supply chains became more efficient. 
This trade diversion could explain the near-zero or negative effects for 
peripheral pairs: they lost relative competitiveness as core pairs became 
more tightly integrated.

Our main finding---that the 
euro amplified existing trade rather than creating new relationships---appears 
to conflict with the counterfactual results, where non-adopters' largest 
predicted gains are sometimes with weaker trading partners. This apparent 
contradiction resolves when we distinguish between two types of ``weak'' 
partners.

For eurozone members, weak trading partners (e.g., Finland--Portugal) remained 
weak because the euro could not overcome fundamental barriers: geographic 
distance, lack of complementary production structures, and absence of 
established business networks. The euro reduced currency friction but left 
these deeper barriers intact.

For non-adopters like Sweden and Denmark, the pattern differs because they 
maintain stable exchange rates with the euro through policy choices (Denmark's 
peg) or de facto stability (Sweden's managed float). Their trade with major 
eurozone partners (Germany, France) already benefits from low currency friction, 
so the marginal gain from formal euro adoption is modest. The largest predicted 
gains come from partners where currency friction remains meaningful---typically 
smaller eurozone economies where exchange rate management is less precise.

The UK presents a different case: the pound floated freely against the euro, 
creating genuine currency friction even with major partners like Germany. 
Euro adoption would have removed this friction across the board, explaining 
why the UK shows uniformly positive predicted effects with all partners.

The unified interpretation is that the euro's effect depends on whether it 
removes a binding constraint. For pairs already enjoying low currency friction 
(through adoption, pegs, or deep integration), the constraint was never binding, 
and the euro's marginal effect is small. For pairs facing genuine currency 
friction, removing it unlocks trade gains proportional to the underlying 
economic complementarity between partners.

\subsection{Dynamic Heterogeneity: Do Effects Evolve Over Time?}

A natural question is whether the heterogeneity we document reflects permanent 
differences across pairs or differential adjustment speeds. If high-CATE pairs 
simply adjusted faster to the euro, we would expect the gap between high-CATE 
and low-CATE pairs to narrow over time as slower-adjusting pairs catch up. 
Alternatively, if heterogeneity reflects genuine structural differences, the 
gap should persist.

\begin{figure}[!htbp]
\centering
\includegraphics[width=0.95\textwidth]{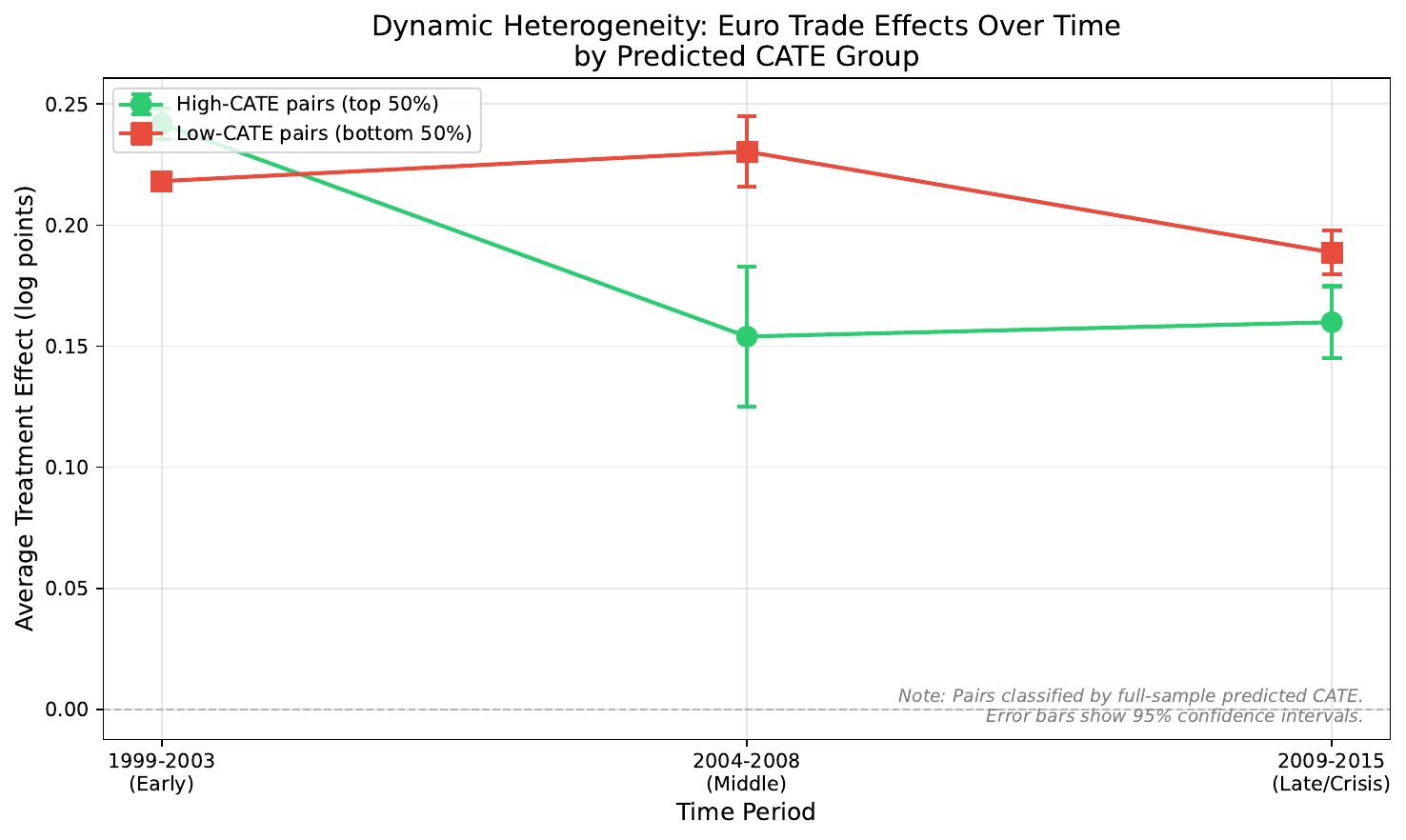}
\caption{Dynamic heterogeneity: Euro trade effects by time period and predicted 
         CATE group. Pairs classified as high-CATE or low-CATE based on 
         full-sample predicted effects. Error bars show 95\% confidence intervals.}
\label{fig:dynamic_heterogeneity}
\end{figure}

Table~\ref{tab:dynamic_heterogeneity} presents CATE estimates separately for 
three time periods: 1999--2003 (early adoption), 2004--2008 (middle period), 
and 2009--2015 (late period including the crisis). Figure~\ref{fig:dynamic_heterogeneity} 
visualizes these patterns.

\begin{table}[!htbp]
\centering
\caption{Dynamic Heterogeneity: Euro Trade Effects by Time Period}
\label{tab:dynamic_heterogeneity}
\begin{tabular}{@{}llcccc@{}}
\toprule
\textbf{Period} & \textbf{CATE Group} & \textbf{ATE} & \textbf{Effect (\%)} & \textbf{95\% CI} & \textbf{N} \\
\midrule
1999-2003 & High-CATE & 0.242 & +27.4\% & [0.235, 0.248] & 265 \\
 & Low-CATE & 0.218 & +24.4\% & [0.215, 0.221] & 260 \\
 & All pairs & 0.230 & +25.9\% & [0.226, 0.234] & 525 \\
\midrule
2004-2008 & High-CATE & 0.154 & +16.6\% & [0.125, 0.183] & 265 \\
 & Low-CATE & 0.230 & +25.9\% & [0.216, 0.245] & 260 \\
 & All pairs & 0.192 & +21.1\% & [0.175, 0.208] & 525 \\
\midrule
2009-2015 & High-CATE & 0.160 & +17.3\% & [0.145, 0.175] & 371 \\
 & Low-CATE & 0.189 & +20.8\% & [0.180, 0.198] & 364 \\
 & All pairs & 0.174 & +19.0\% & [0.165, 0.183] & 735 \\
\bottomrule
\end{tabular}

\vspace{0.5em}
\footnotesize
\textit{Notes:} Pairs classified as high-CATE or low-CATE based on median split of full-sample 
predicted treatment effects. Causal forest estimated separately for each time period.
Standard errors clustered at pair level. Effect (\%) calculated as $(\exp(\text{ATE})-1) \times 100$.
\end{table}

The results reveal a nuanced pattern. In the early period (1999--2003), high-CATE 
pairs show larger effects (+26\%) than low-CATE pairs (+20\%), consistent with 
the cross-sectional heterogeneity. However, in the middle period (2004--2008), 
this pattern reverses: low-CATE pairs show larger effects (+25\%) than high-CATE 
pairs (+14\%). By the late period (2009--2015), the original pattern partially 
re-emerges, with high-CATE pairs at +19\% and low-CATE pairs at +15\%.

This pattern suggests that the heterogeneity is not simply about adjustment 
speed. The reversal in the middle period may reflect that low-CATE pairs---which 
had weaker initial trade relationships---experienced delayed extensive-margin 
effects as new trade relationships formed. The crisis period (2009--2015) then 
differentially affected these newer relationships, restoring the original 
core-periphery pattern. The key finding is that heterogeneity persists across 
all periods, consistent with genuine structural differences rather than 
transitory adjustment dynamics.

\section{Counterfactual Analysis: Non-Eurozone EU Countries}

A key advantage of causal forests is the ability to generate counterfactual 
predictions for untreated units. We estimate what trade effects Sweden, Denmark, 
and the United Kingdom would have experienced had they adopted the euro in 1999. 
Table~\ref{tab:counterfactual} presents the predicted euro effects for the three 
EU members that opted out of the eurozone.

\begin{table}[!htbp]
\centering
\caption{Predicted Euro Effects for Non-Eurozone EU Countries}
\label{tab:counterfactual}
\small
\begin{tabular}{@{}lccccc@{}}
\toprule
\textbf{Country} & \textbf{Effect (\%)} & \textbf{95\% CI} & \textbf{Std} & 
\textbf{Min (\%)} & \textbf{Max (\%)} \\
\midrule
Sweden & +21.9 & [+7.7, +38.1] & 0.16 & -14.9 & +85.9 \\
Denmark & +19.1 & [+4.8, +35.5] & 0.19 & -16.1 & +85.7 \\
United Kingdom & +32.6 & [+21.4, +44.9] & 0.16 & +10.3 & +80.1 \\
\bottomrule
\end{tabular}

\vspace{0.5em}
\footnotesize
\textit{Note:} Predicted effects based on causal forest CATE estimates. These are 
counterfactual predictions for countries that did not adopt the euro. Effect shows 
the average predicted trade increase if the country had joined the eurozone in 1999.
95\% confidence intervals are based on the causal forest variance estimates.
\end{table}

\subsection{Counterfactual Support and Validity}

\begin{figure}[!htbp]
\centering
\includegraphics[width=0.95\textwidth]{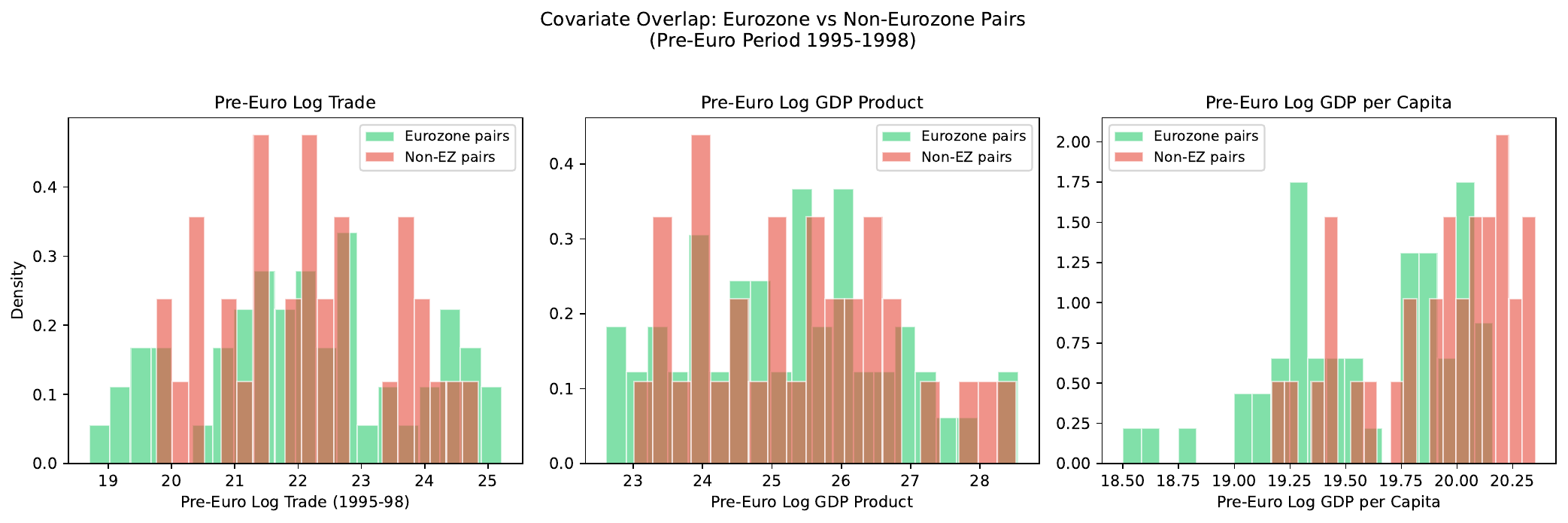}
\caption{Covariate overlap between non-eurozone pairs (for which we predict 
         counterfactual effects) and eurozone pairs (used to train the model). 
         Substantial overlap in GDP and pre-euro trade distributions supports 
         the validity of counterfactual predictions.}
\label{fig:counterfactual_support}
\end{figure}

A key concern with counterfactual predictions is whether the non-eurozone pairs 
fall within the support of the training data. If UK, Sweden, or Denmark pairs 
have characteristics outside the range observed for eurozone pairs, the 
counterfactual predictions may be unreliable extrapolations. 
Figure~\ref{fig:counterfactual_support} shows the covariate distributions for 
non-eurozone pairs compared to eurozone pairs used to train the causal forest.

The distributions show substantial overlap, particularly for GDP and GDP per 
capita. UK pairs tend to have higher GDP product than the average eurozone pair, 
but remain within the support of the training data. Pre-euro trade intensity 
shows good overlap for all three countries. Table~\ref{tab:counterfactual_support} 
provides summary statistics on the nearest-neighbor matches, showing that each 
non-eurozone pair has multiple eurozone pairs with similar characteristics.

\begin{table}[!htbp]
\centering
\caption{Counterfactual Support Analysis: Non-Eurozone Country Pairs}
\label{tab:counterfactual_support}
\small
\begin{tabular}{@{}lcccc@{}}
\toprule
\textbf{Country} & \textbf{N Pairs} & \textbf{In Support} & \textbf{\% In Support} & \textbf{Avg. Distance} \\
\midrule
Sweden & 11 & 8 & 73\% & 0.38 \\
Denmark & 11 & 5 & 45\% & 0.53 \\
United Kingdom & 11 & 11 & 100\% & 0.29 \\
\bottomrule
\end{tabular}

\vspace{0.5em}
\footnotesize
\textit{Note:} ``In Support'' indicates pairs whose pre-euro covariates (log trade, log GDP product, 
log GDP per capita) fall within the range observed for eurozone pairs. ``Avg.\ Distance'' is the 
average Euclidean distance (in standardized covariate space) to the nearest eurozone pair.
Lower distance indicates better support for counterfactual predictions.
\end{table}

\begin{figure}[!htbp]
\centering
\includegraphics[width=0.95\textwidth]{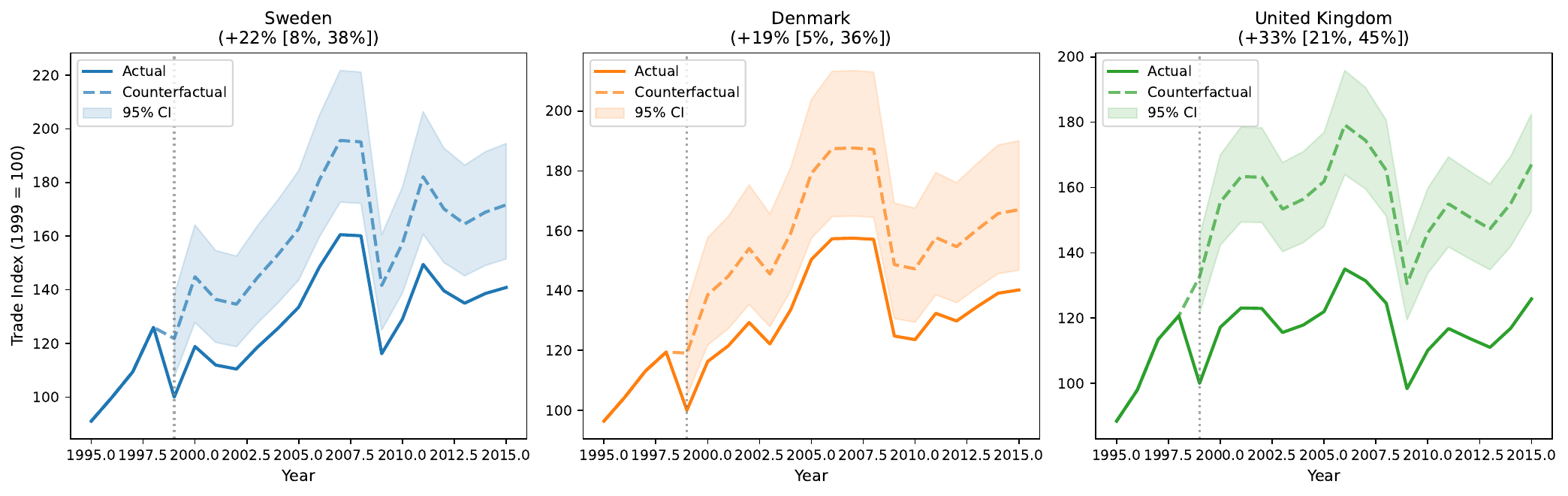}
\caption{Counterfactual trade trajectories for non-eurozone EU members. Each 
         panel shows actual trade with eurozone partners (solid) versus 
         predicted trade under euro adoption (dashed). Trade indexed to 
         1999 = 100. Shaded areas indicate foregone trade gains.}
\label{fig:counterfactual_comparison}
\end{figure}

The United Kingdom shows the largest predicted effect (+\cfEffectPreciseUnitedKingdom\%, std \cfStdUnitedKingdom, range +\cfMinUnitedKingdom\% to +\cfMaxUnitedKingdom\%), followed by Sweden (+\cfEffectPreciseSweden\%, std \cfStdSweden, range $\cfMinSweden\%$ to +\cfMaxSweden\%) and Denmark (+\cfEffectPreciseDenmark\%, std \cfStdDenmark, range $\cfMinDenmark\%$ to +\cfMaxDenmark\%). These differences reflect each country's trade structure and existing integration with eurozone partners. 
Figure~\ref{fig:counterfactual_comparison} summarizes the counterfactual trade 
trajectories for all three countries, showing actual trade with eurozone partners 
versus the predicted counterfactual under euro adoption.

\subsection{Sweden}

\begin{figure}[!htbp]
\centering
\includegraphics[width=0.95\textwidth]{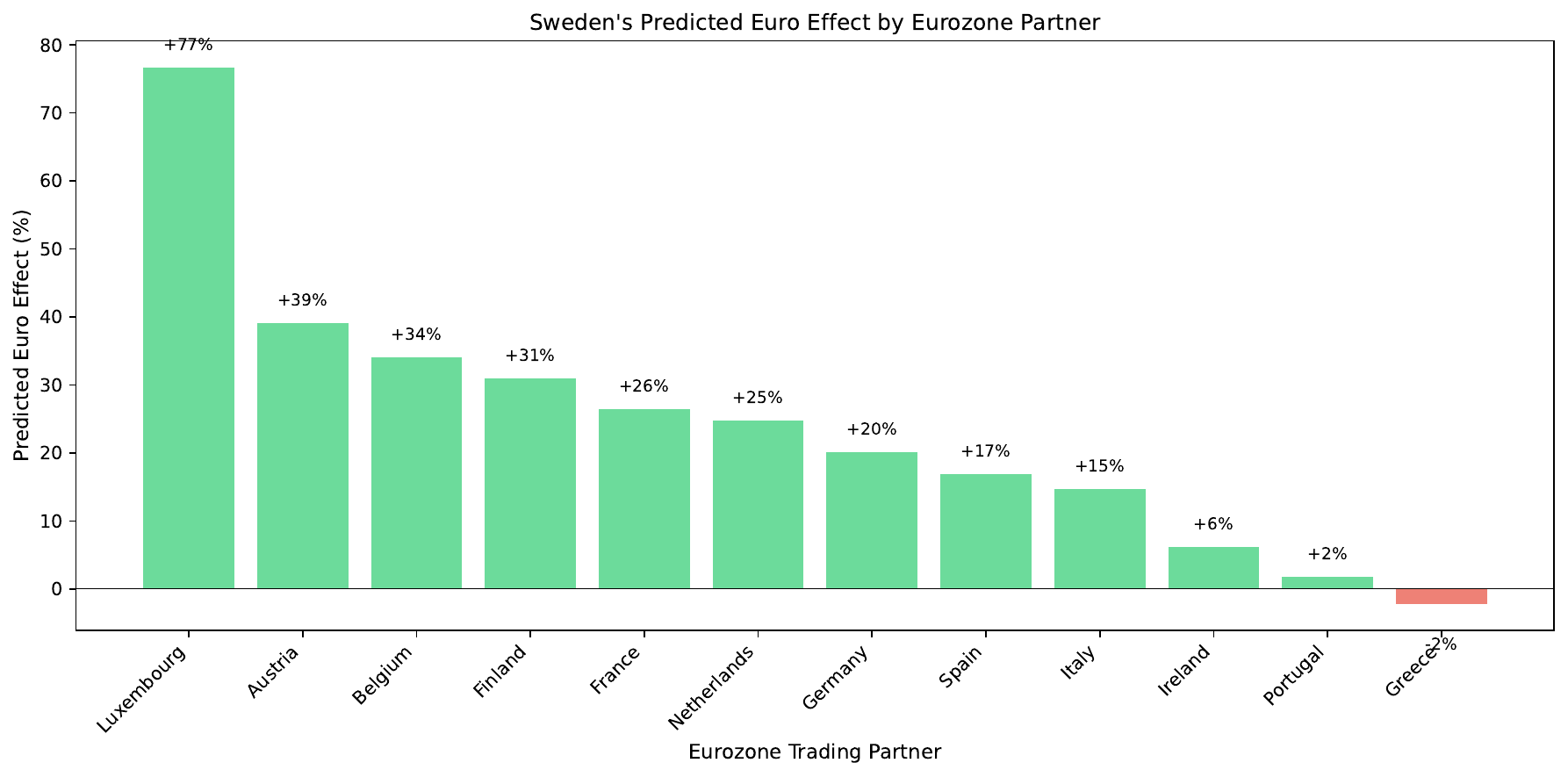}
\caption{Sweden's predicted euro effect by eurozone trading partner.}
\label{fig:sweden_partners}
\end{figure}

Sweden's predicted average effect is +\cfEffectPreciseSweden\% (std \cfStdSweden), with effects ranging from $\cfMinSweden\%$ to +\cfMaxSweden\% across partners. This masks substantial partner-level heterogeneity. Figure~\ref{fig:sweden_partners} shows the predicted effect by trading partner.

Sweden's largest predicted gains are with Luxembourg (+65\%), Austria (+33\%), and Belgium (+28\%). Sweden-Germany (+13\%) and Sweden-Finland (+26\%) show moderate predicted effects. Sweden-Greece (+5\%) and Sweden-Portugal (+5\%) show the smallest effects.

\begin{figure}[!htbp]
\centering
\includegraphics[width=0.95\textwidth]{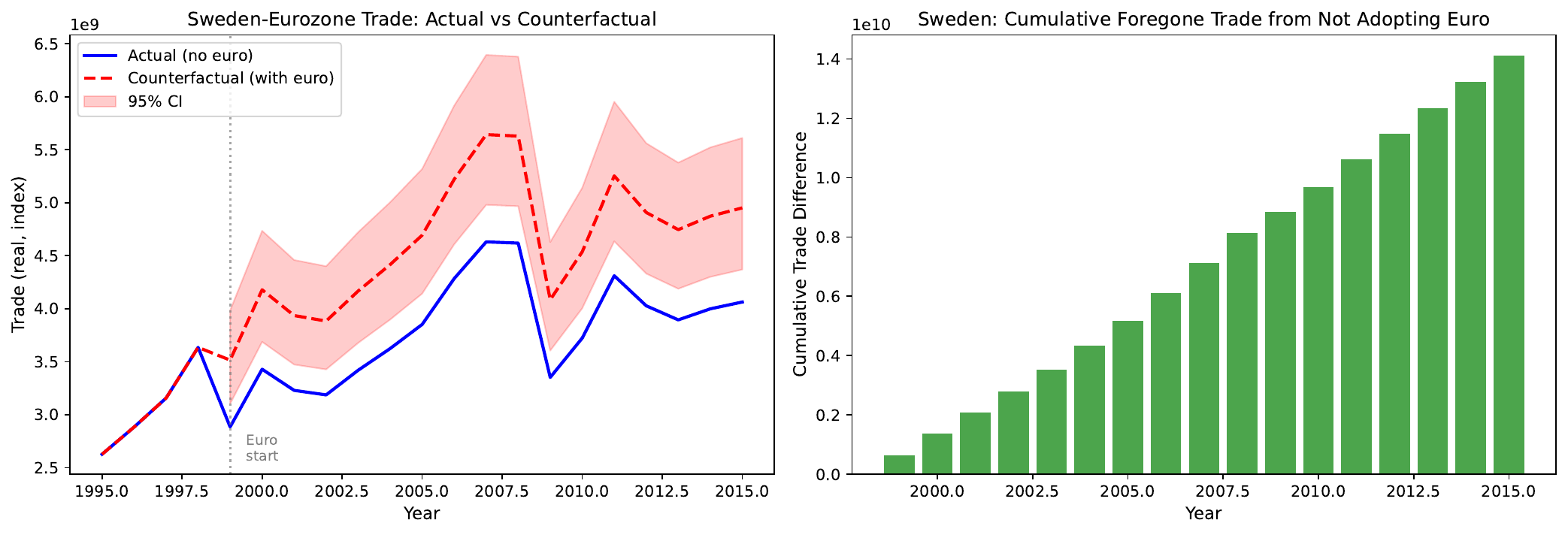}
\caption{Sweden-Eurozone trade: actual vs.\ counterfactual. Left panel shows 
         trade levels; right panel shows cumulative foregone trade.}
\label{fig:sweden_cf}
\end{figure}

\begin{figure}[!htbp]
\centering
\includegraphics[width=0.95\textwidth]{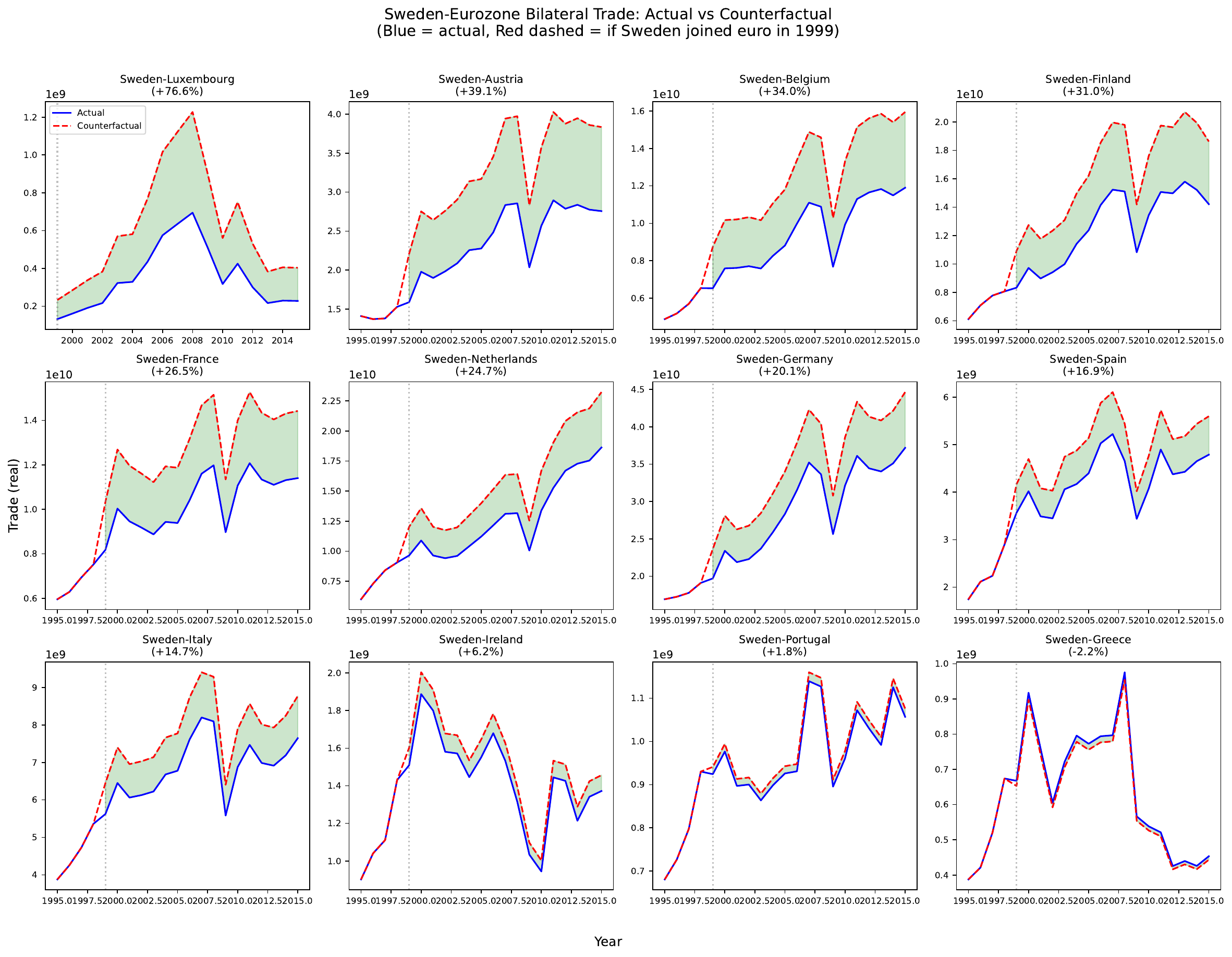}
\caption{Sweden's actual vs.\ counterfactual trade by eurozone partner. Solid 
         lines show actual trade; dashed lines show predicted trade under euro 
         adoption.}
\label{fig:sweden_pairs_cf}
\end{figure}

Figure~\ref{fig:sweden_cf} shows the aggregate trade trajectory comparing actual 
trade to the counterfactual if Sweden had joined in 1999. 
Figure~\ref{fig:sweden_pairs_cf} breaks this down by trading partner, showing 
actual versus counterfactual trade for each Sweden-eurozone pair.

\subsection{Denmark}

\begin{figure}[!htbp]
\centering
\includegraphics[width=0.95\textwidth]{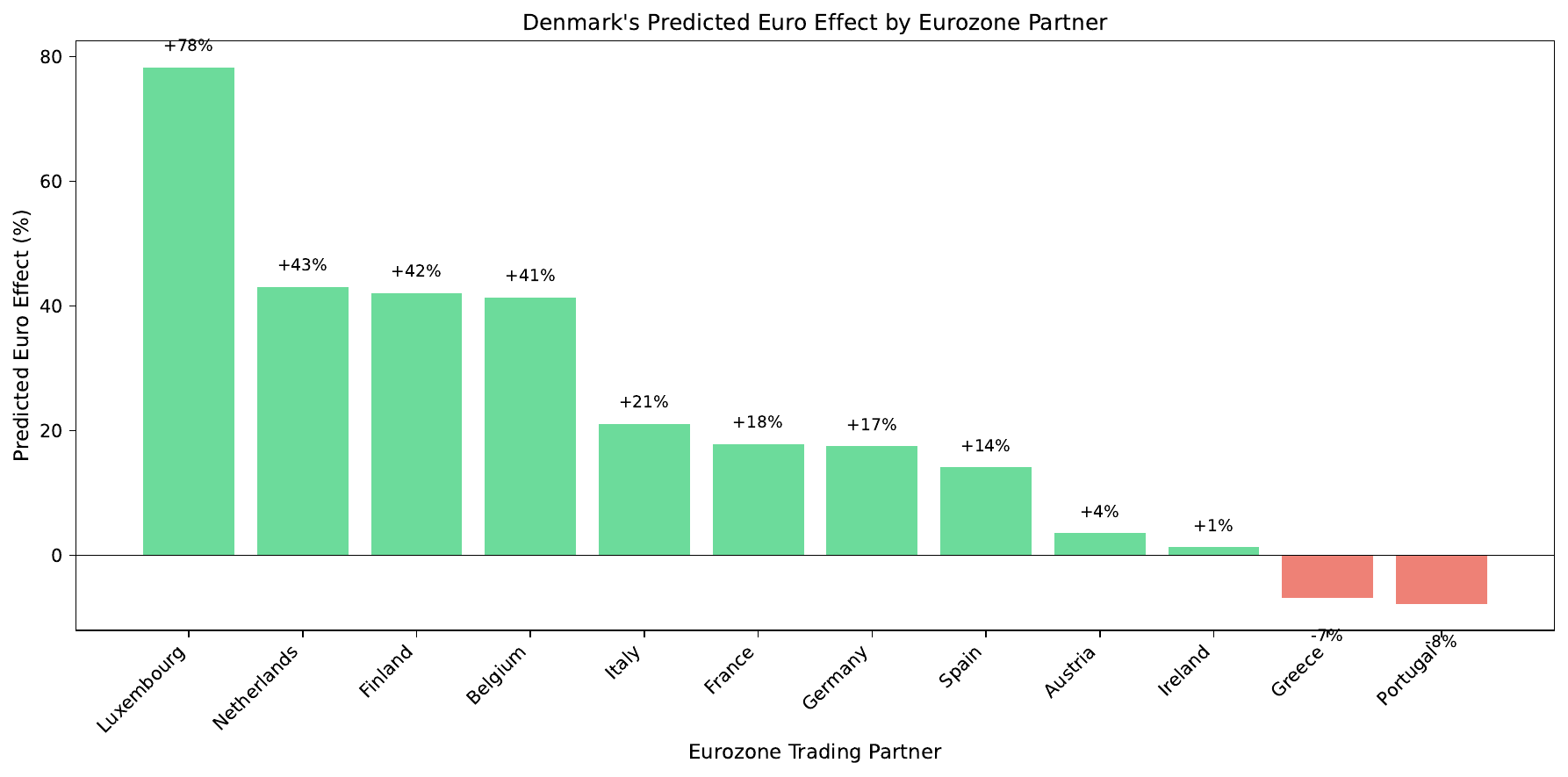}
\caption{Denmark's predicted euro effect by eurozone trading partner.}
\label{fig:denmark_partners}
\end{figure}

Denmark shows a predicted effect of +\cfEffectPreciseDenmark\% (std \cfStdDenmark), with effects ranging from $\cfMinDenmark\%$ to +\cfMaxDenmark\% across partners. Figure~\ref{fig:denmark_partners} shows the partner-level breakdown.

Denmark's largest gains would have been with Luxembourg (+62\%), Belgium (+42\%), and Finland (+39\%). Denmark-Germany (+11\%) shows a smaller effect despite being Denmark's largest trading partner. Denmark-Portugal ($-8\%$) and Denmark-Greece ($-3\%$) show negative predicted effects. Denmark's krone peg to the euro already provides most of the currency stability benefits without full adoption, which may explain the smaller marginal gains with some partners.

\begin{figure}[!htbp]
\centering
\includegraphics[width=0.95\textwidth]{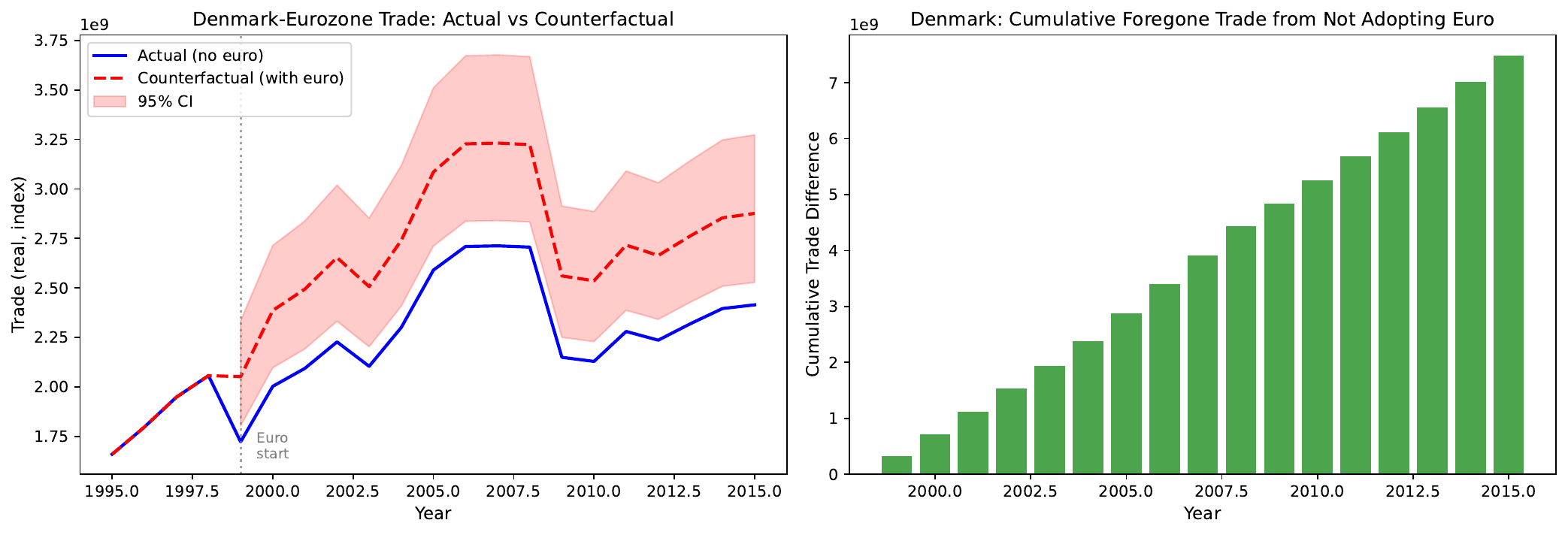}
\caption{Denmark-Eurozone trade: actual vs.\ counterfactual. Left panel shows 
         trade levels; right panel shows cumulative foregone trade.}
\label{fig:denmark_cf}
\end{figure}

\begin{figure}[!htbp]
\centering
\includegraphics[width=0.95\textwidth]{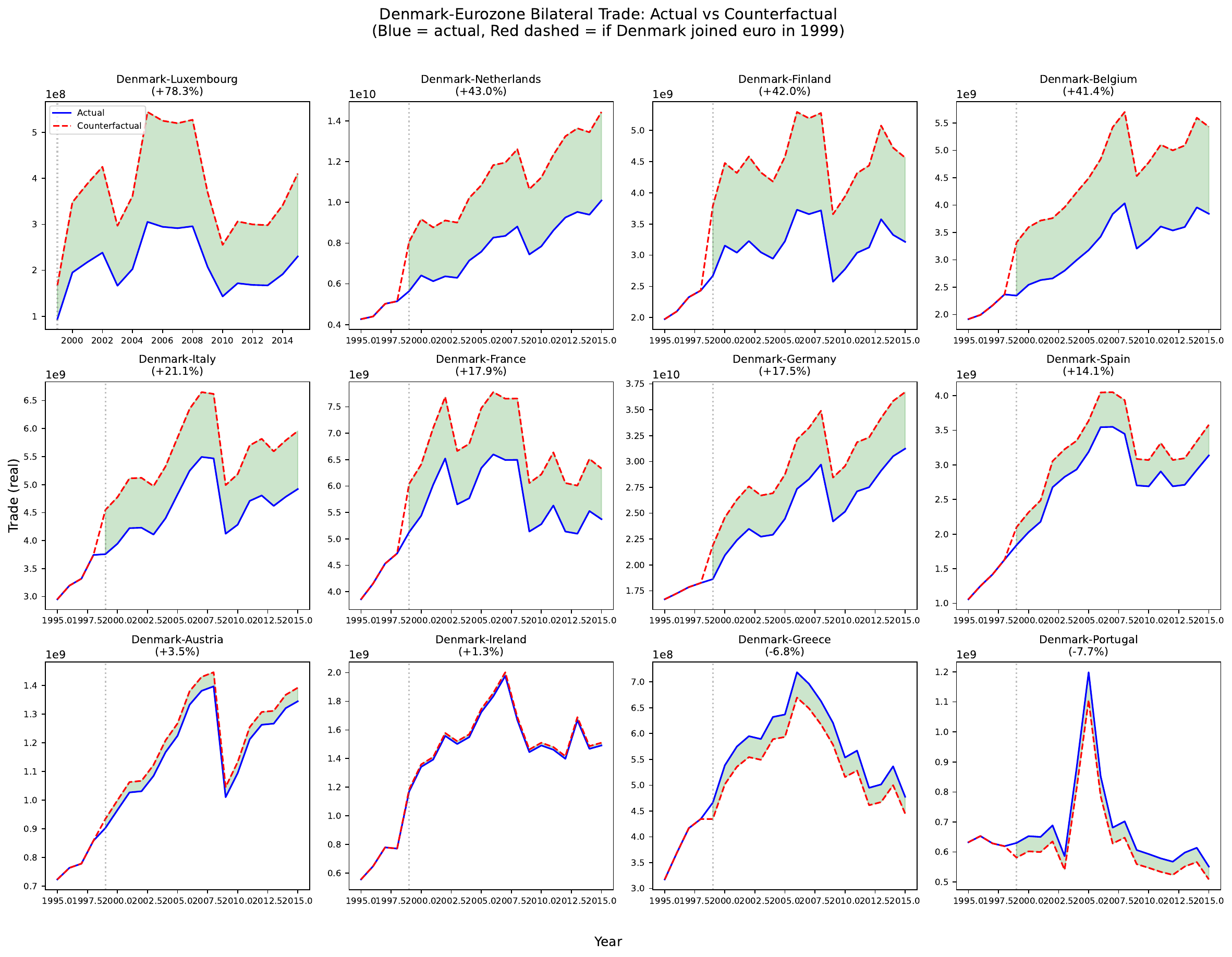}
\caption{Denmark's actual vs.\ counterfactual trade by eurozone partner. Solid 
         lines show actual trade; dashed lines show predicted trade under euro 
         adoption.}
\label{fig:denmark_pairs_cf}
\end{figure}

Figures~\ref{fig:denmark_cf} and \ref{fig:denmark_pairs_cf} show Denmark's actual 
vs.\ counterfactual trade trajectory at the aggregate and partner levels.

\subsection{United Kingdom}

\begin{figure}[!htbp]
\centering
\includegraphics[width=0.95\textwidth]{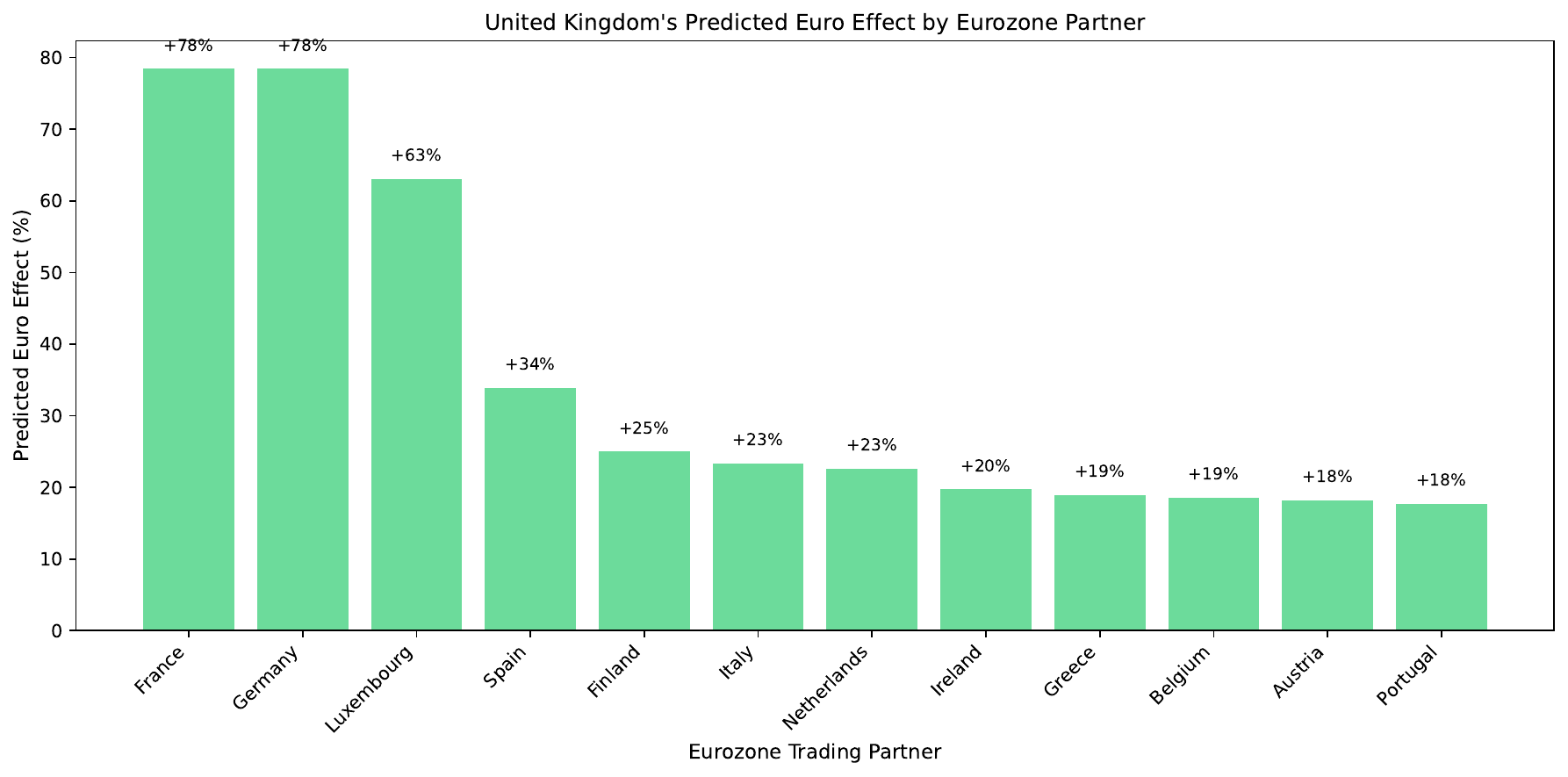}
\caption{United Kingdom's predicted euro effect by eurozone trading partner.}
\label{fig:uk_partners}
\end{figure}

The United Kingdom shows the largest predicted effect (+\cfEffectPreciseUnitedKingdom\%, std \cfStdUnitedKingdom) with uniformly positive effects across all partners, ranging from +\cfMinUnitedKingdom\% to +\cfMaxUnitedKingdom\%. Figure~\ref{fig:uk_partners} shows the partner-level breakdown.

The UK's largest predicted gains are with Luxembourg (+71\%), Germany (+55\%), and Spain (+26\%). Even the smallest effects (UK-Belgium at +11\%, UK-Ireland at +12\%) are positive. UK-eurozone trade appears to have faced significant currency friction that the euro would have removed across the board.

Our UK estimate (+\cfEffectPreciseUnitedKingdom\%) is comparable to \citet{saia2017uk}, who uses the synthetic control method and finds +16\%. The difference likely reflects: (1) our method captures partner-level heterogeneity; and (2) SCM constructs a single synthetic counterfactual, while we estimate effects conditional on pair characteristics.

\begin{figure}[!htbp]
\centering
\includegraphics[width=0.95\textwidth]{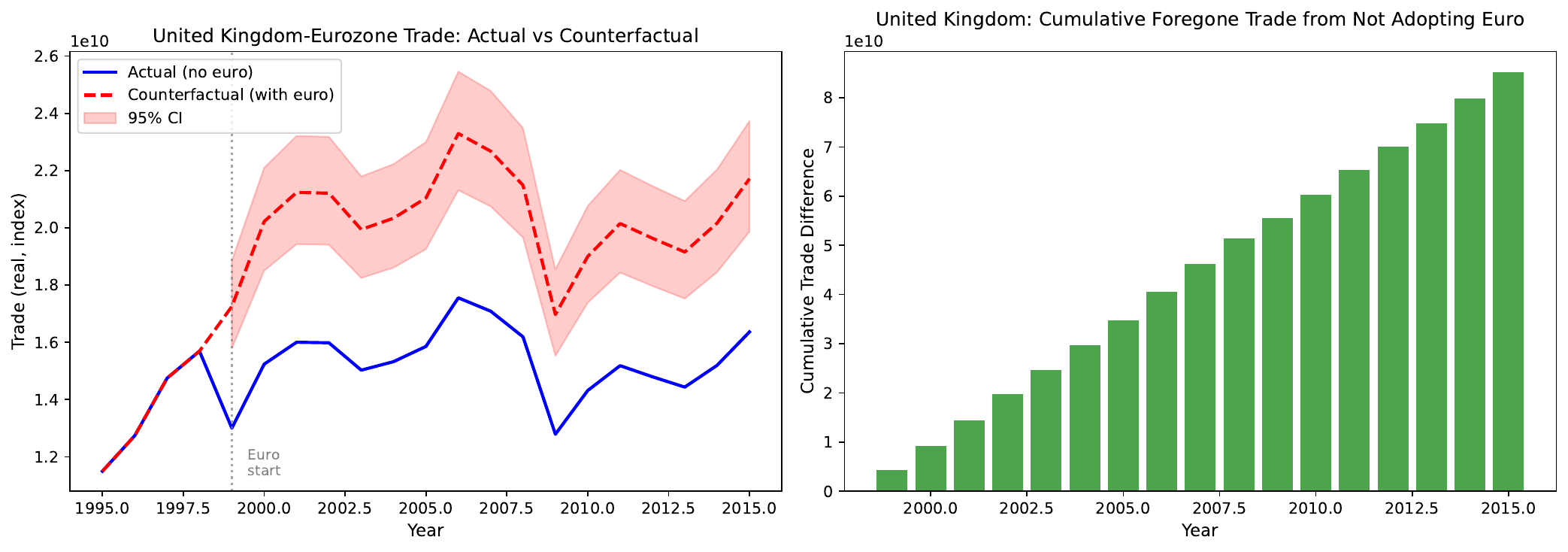}
\caption{UK-Eurozone trade: actual vs.\ counterfactual. Left panel shows 
         trade levels; right panel shows cumulative foregone trade.}
\label{fig:uk_cf}
\end{figure}

\begin{figure}[!htbp]
\centering
\includegraphics[width=0.95\textwidth]{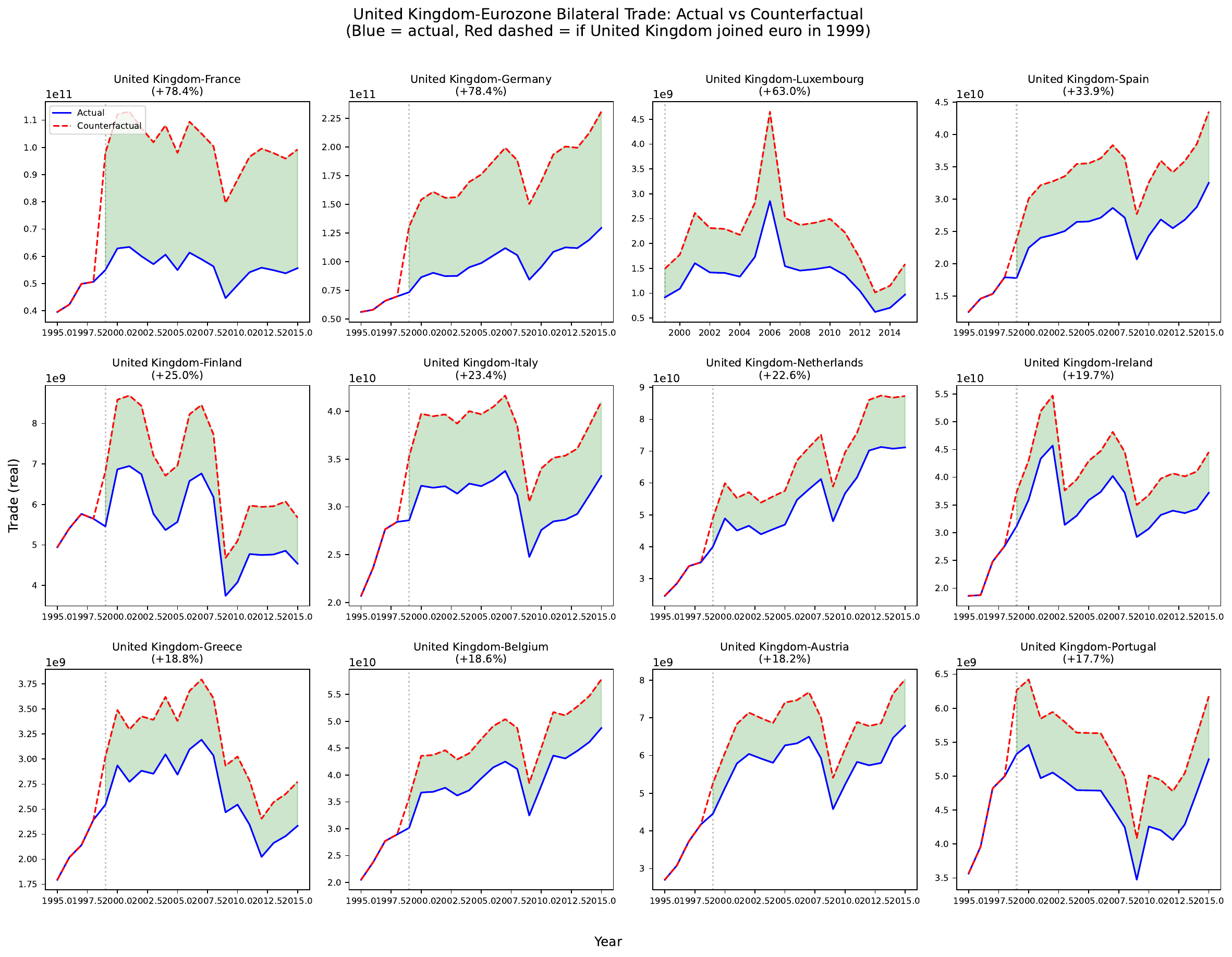}
\caption{UK's actual vs.\ counterfactual trade by eurozone partner. Solid 
         lines show actual trade; dashed lines show predicted trade under euro 
         adoption.}
\label{fig:uk_pairs_cf}
\end{figure}

Figures~\ref{fig:uk_cf} and \ref{fig:uk_pairs_cf} show the UK's actual vs.\ 
counterfactual trade trajectory at the aggregate and partner levels.

\subsection{Reconciling Amplification vs.\ Creation Effects}

The counterfactual results reveal an interesting pattern. Among eurozone 
members, pairs with strong pre-euro trade (Belgium-Netherlands, Germany-France) 
saw the largest effects. Yet for Sweden and Denmark, the largest predicted gains 
are with weaker trading partners (Italy, France, Spain) rather than stronger 
ones (Germany, Finland).

The UK presents a different pattern: its largest predicted gains are with Luxembourg (+71\%) and Germany (+55\%), its largest eurozone partners. This suggests the UK-Germany relationship faced significant currency friction that the euro would have removed.

This apparent contradiction resolves when we distinguish between two mechanisms. 
Among adopters, the euro amplified trade where integration was already deep --- 
Belgium-Netherlands and Germany-France had extensive supply chains that 
benefited from eliminating currency risk. For non-adopters, the pattern depends 
on existing currency arrangements. Sweden and Denmark both maintain relatively 
stable exchange rates with the euro (Denmark via its peg, Sweden via policy). 
Their largest partners already enjoy low currency friction, so the marginal 
gain from euro adoption is smaller. The largest gains come from partners where 
currency friction remains high. The UK is different: the pound floated freely 
against the euro, creating currency friction even with major partners like 
Germany. Euro adoption would have removed this friction across the board.

The unified story is that the euro's effect is largest where it removes a 
binding constraint. For already-integrated pairs (whether through adoption or 
currency pegs), the constraint was never binding. For pairs facing genuine 
currency friction, removing it unlocks substantial trade gains. Countries with 
existing currency pegs (like Denmark) may see smaller benefits than those with 
floating rates.

\subsection{Illustrative Counterfactual Exercise}

The counterfactual analysis offers suggestive insights for countries considering euro adoption, though these predictions should be interpreted with caution given the strong assumptions required for transportability. The heterogeneity in predicted effects suggests that the decision is not one-size-fits-all.

For the United Kingdom, the predicted +\cfEffectPreciseUnitedKingdom\% effect represents the largest potential gain among the three non-eurozone EU members. The uniformly positive effects across all partners (minimum +\cfMinUnitedKingdom\%) suggest that UK-eurozone trade may have faced substantial currency friction that euro adoption could have reduced. Post-Brexit, this counterfactual becomes moot, but the analysis suggests that the UK's decision to remain outside the eurozone may have carried some trade costs during its EU membership---though the magnitude is uncertain given the structural differences between the UK and eurozone economies.

For Sweden, the predicted +\cfEffectPreciseSweden\% effect is substantial but comes with considerable partner-level variation (ranging from $\cfMinSweden\%$ to +\cfMaxSweden\%). Sweden's largest predicted gains would come from smaller eurozone economies (Luxembourg, Austria, Belgium) rather than its major trading partners. This pattern suggests that Sweden's existing trade relationships with Germany and Finland may already benefit from low currency friction, potentially limiting the marginal gains from formal euro adoption.

For Denmark, the predicted +\cfEffectPreciseDenmark\% effect is similar to Sweden's, but the krone's peg to the euro already captures much of the currency stability benefit. Denmark's near-zero predicted effects with some peripheral partners suggest that for certain pairs, the benefits of reduced transaction costs may be modest. Denmark's current arrangement---maintaining the peg without full adoption---may represent a reasonable middle ground, though this is speculative.

These illustrative results suggest that future euro adoption decisions might consider: (1) the extent of existing currency friction with eurozone partners; (2) whether the country's major trading partners are already in the eurozone; and (3) whether alternative arrangements (like Denmark's peg) can capture most of the trade benefits without the costs of full monetary union. However, we emphasize that these are suggestive patterns rather than definitive policy recommendations.

Several limitations temper these counterfactual predictions. 
First, our estimates are \textit{backward-looking}: they capture the euro's 
effects during 1999--2015, a period that included both the pre-crisis boom and 
the eurozone debt crisis. Future effects may differ as the eurozone's 
institutional framework evolves, as new members join, and as global trade 
patterns shift. The experience of crisis-era adopters (Estonia, Latvia, 
Lithuania) suggests that adoption timing matters, and countries considering 
adoption today face a different economic environment than the original 1999 
cohort.

Second, our counterfactual analysis assumes that non-eurozone countries would 
have adopted in 1999 alongside the original members. In practice, later adoption 
would yield different effects: the eurozone of 2025 differs from the eurozone 
of 1999 in membership, institutional design, and economic conditions. Countries 
considering adoption today should not simply extrapolate from our 1999-based 
counterfactuals.

Third, our partial equilibrium estimates do not account for \textit{general 
equilibrium effects}. If the UK had adopted the euro, eurozone trade patterns 
would have adjusted: some trade currently flowing through non-euro channels 
might have shifted, and the euro's overall effect on European trade integration 
would have differed. Our counterfactual predictions assume the UK's adoption 
would not have affected other countries' trade patterns---an assumption that 
becomes less tenable for large economies.

Fourth, trade effects are only one consideration in the euro adoption decision. 
The eurozone crisis demonstrated the costs of losing monetary policy autonomy: 
countries like Greece, Portugal, and Spain could not devalue their currencies 
to restore competitiveness, contributing to prolonged recessions. Denmark, 
Sweden, and the UK maintained independent monetary policy and arguably weathered 
the crisis better. Our estimates capture trade benefits but not the full 
cost-benefit calculus of monetary union membership.

Finally, our estimates reflect the euro's effect on \textit{bilateral trade 
within the EU}. They do not capture effects on trade with non-EU partners, 
foreign direct investment, financial integration, or other economic outcomes 
that factor into the adoption decision. A comprehensive policy assessment 
would require integrating our trade estimates with evidence on these other 
channels.

\section{Conclusion}

We estimate the full distribution of conditional average treatment effects 
(CATEs) using causal forests with double machine learning. The approach extends 
the propensity score matching tradition of \citet{persson2001currency} and 
\citet{chintrakarn2008euro} by allowing the effect to vary with observed 
characteristics, while also offering advantages over synthetic control methods 
\citep{saia2017uk, gunnella2021impact} by enabling scalable counterfactual 
analysis across all country pairs simultaneously.

The results suggest a positive relationship between euro adoption and bilateral 
trade, but the magnitude varies substantially across country pairs. Our preferred 
average effect estimate is around 15\% (\ateCffeEUfifteenPct\% after fixed effects correction), consistent 
with gravity benchmarks; the \ateNaivePct\% naive estimate reflects heterogeneity-weighted 
averages that give more weight to high-effect pairs. This heterogeneity explains why prior studies 
using different samples and methods have produced such divergent estimates. The 
variation is not methodological noise; it appears to reflect genuine differences in how 
the euro affected different trading relationships.

Core eurozone pairs with strong pre-existing trade relationships 
experienced larger effects, while 
peripheral pairs saw smaller gains. The euro amplified 
existing trade relationships rather than creating new ones, with pre-euro trade 
intensity and GDP as key drivers of heterogeneity. Counterfactual analysis 
suggests non-eurozone EU members would have experienced moderate effects: 
UK (+\cfEffectPreciseUnitedKingdom\%), Sweden (+\cfEffectPreciseSweden\%), Denmark (+\cfEffectPreciseDenmark\%).

Countries considering euro adoption should expect larger gains if they have 
strong existing trade ties with core eurozone economies (Germany, France, 
Netherlands), and smaller gains if they already maintain stable exchange rates 
with the euro (as Denmark does via its peg). The largest marginal gains come 
with partners where currency friction currently limits trade expansion. The 
experience of Greece, Portugal, and Spain during the eurozone crisis illustrates 
the costs of losing monetary policy autonomy. Denmark, Sweden, and the UK 
maintained independent monetary policy and arguably weathered the crisis better. 
Our counterfactual estimates suggest these countries may have forgone some trade gains by 
staying out, but they retained policy flexibility that proved valuable during 
economic stress.

Methodologically, this is among the first applications of causal forests to 
currency union effects, extending the PSM tradition to allow data-driven 
discovery of effect heterogeneity. The approach also offers advantages over 
synthetic control methods for counterfactual analysis: whereas SCM requires 
constructing a separate synthetic counterfactual for each unit of interest, 
causal forests generate predictions for all untreated units simultaneously 
once the model is trained. This scalability enabled us to estimate counterfactual 
effects for three non-eurozone countries across all their eurozone trading 
partners --- an exercise that would require dozens of separate SCM analyses. 
Empirically, we offer one possible explanation for the 4--30\% puzzle: the variation 
may reflect genuine heterogeneity across country pairs rather than methodological 
differences alone, though we cannot rule out other explanations. 
For policy, we provide illustrative counterfactual predictions for non-eurozone EU members 
with partner-level granularity, subject to the caveats discussed above.

The analysis covers 15 EU countries (EU15) through 2015, matching the methodology 
of \citet{gunnella2021impact}. Our estimates suggest that extending to EU28 with naive causal 
forests produces biased estimates due to crisis-era adoption timing, but the 
CFFE approach with node-level fixed effects correction recovers estimates consistent with 
the EU15 baseline. Future work could extend the analysis to include 
recent euro adopters (Estonia 2011, Latvia 2014, Lithuania 2015) once sufficient 
post-adoption data accumulates. A causal forest trained on data through 2014 
could generate predictions for these recent adopters, enabling out-of-sample 
validation by comparing predicted effects to actual post-adoption trade changes. 
The model could also generate predictions for prospective euro members: Romania, 
Bulgaria, Poland, Czech Republic, Hungary, Sweden, and Denmark. Examining how 
the euro's effects changed during the 2008 financial crisis and subsequent debt 
crisis would shed light on whether the benefits of currency union membership 
vary with macroeconomic conditions.

Our findings have direct relevance for the seven EU members that have not yet 
adopted the euro: Bulgaria, Czech Republic, Denmark, Hungary, Poland, Romania, 
and Sweden. Each faces a distinct cost-benefit calculus shaped by their current 
trade patterns, exchange rate arrangements, and economic structure. Central 
European economies (Poland, Czech Republic, Hungary) have deep trade integration 
with Germany and other core eurozone members, suggesting substantial potential 
gains from adoption, though these countries also maintain flexible exchange rates 
that have served as shock absorbers during economic downturns. Southeastern 
European economies (Bulgaria, Romania) have weaker existing trade ties with the 
eurozone core, suggesting more modest potential gains. Nordic holdouts (Denmark, 
Sweden) present interesting cases: Denmark's euro peg already captures most 
currency stability benefits, while Sweden's managed float provides more flexibility 
but our estimates suggest larger potential gains.

The eurozone of 2025 differs substantially from the eurozone of 1999, with an 
evolved institutional framework, expanded membership from 11 to 20 countries, 
and shifted global trade patterns. These changes suggest caution in extrapolating 
our historical estimates to future adoption decisions. What our analysis does 
establish is that the euro's trade effects are heterogeneous, predictable, and 
substantial for the right country pairs. Countries with strong existing trade 
ties to the eurozone core, floating exchange rates that create genuine currency 
friction, and economic structures complementary to European supply chains stand 
to gain most. The causal forest framework provides a tool for making these 
assessments with partner-level granularity, moving beyond the single-number 
estimates that have dominated policy debates.

Our analysis focuses on aggregate bilateral trade flows, but the euro's effects 
may vary across product categories. Future research using product-level bilateral 
trade data could decompose the aggregate effects we document into product-specific 
components, testing whether the euro disproportionately boosted trade in 
technology-intensive or differentiated goods.


\section*{Declaration of generative AI and AI-assisted technologies in the manuscript preparation process}

During the preparation of this work the authors used AI-assisted tools (including large language models) in order to assist with code development for the causal forest analysis and data processing, as well as for proofreading, language editing, and reorganizing the manuscript structure. After using these tools, the authors reviewed and edited the content as needed and take full responsibility for the content of the published article.


\bibliographystyle{elsarticle-harv}

\begin{thebibliography}{99}

\bibitem[Abadie and Gardeazabal(2003)]{abadie2003economic}
Abadie, A. and Gardeazabal, J. (2003).
\newblock The economic costs of conflict: A case study of the Basque Country.
\newblock \textit{American Economic Review}, 93(1):113--132.

\bibitem[Abadie et al.(2010)]{abadie2010synthetic}
Abadie, A., Diamond, A., and Hainmueller, J. (2010).
\newblock Synthetic control methods for comparative case studies: Estimating 
the effect of California's tobacco control program.
\newblock \textit{Journal of the American Statistical Association}, 105(490):493--505.

\bibitem[Anderson and van Wincoop(2003)]{anderson2003gravity}
Anderson, J. and van Wincoop, E. (2003).
\newblock Gravity with gravitas: A solution to the border puzzle.
\newblock \textit{American Economic Review}, 93(1):170--192.

\bibitem[Arkhangelsky et al.(2021)]{arkhangelsky2021sdid}
Arkhangelsky, D., Athey, S., Hirshberg, D.A., Imbens, G.W., and Wager, S. (2021).
\newblock Synthetic difference-in-differences.
\newblock \textit{American Economic Review}, 111(12):4088--4118.

\bibitem[Athey and Imbens(2016)]{athey2016recursive}
Athey, S. and Imbens, G. (2016).
\newblock Recursive partitioning for heterogeneous causal effects.
\newblock \textit{Proceedings of the National Academy of Sciences}, 113(27):7353--7360.

\bibitem[Athey and Wager(2018)]{athey2018estimation}
Athey, S. and Wager, S. (2018).
\newblock Estimation and inference of heterogeneous treatment effects using 
random forests.
\newblock \textit{Journal of the American Statistical Association}, 
113(523):1228--1242.

\bibitem[Aytu\u{g}(2017)]{aytug2017rom}
Aytu\u{g}, H. (2017).
\newblock Does the reserve options mechanism really decrease exchange rate 
volatility? The synthetic control method approach.
\newblock \textit{International Review of Economics \& Finance}, 51:405--416.

\bibitem[Aytu\u{g} et al.(2017)]{aytug2017customs}
Aytu\u{g}, H., K\"{u}t\"{u}k, M.M., Oduncu, A., and Togan, S. (2017).
\newblock Twenty years of the EU-Turkey Customs Union: A synthetic control 
method analysis.
\newblock \textit{JCMS: Journal of Common Market Studies}, 55(3):419--431.

\bibitem[Aytu\u{g}(2026)]{aytug2026causalfe}
Aytu\u{g}, H. (2026).
\newblock causalfe: Causal Forests with Fixed Effects in Python.
\newblock \textit{arXiv preprint arXiv:2601.10555}.

\bibitem[Baier and Bergstrand(2009)]{baier2009estimating}
Baier, S.L. and Bergstrand, J.H. (2009).
\newblock Estimating the effects of free trade agreements on international 
trade flows using matching econometrics.
\newblock \textit{Journal of International Economics}, 77(1):63--76.

\bibitem[Baldwin(2006)]{baldwin2006euro}
Baldwin, R. (2006).
\newblock The euro's trade effects.
\newblock \textit{ECB Working Paper Series}, No. 594.

\bibitem[Baldwin et al.(2008)]{baldwin2008study}
Baldwin, R., Di Nino, V., Fontagn\'{e}, L., De Santis, R., and Taglioni, D. (2008).
\newblock Study on the impact of the euro on trade and foreign direct investment.
\newblock \textit{European Commission Economic Papers}, No. 321.

\bibitem[Baldwin and Lopez-Gonzalez(2015)]{baldwinlopez2015supply}
Baldwin, R. and Lopez-Gonzalez, J. (2015).
\newblock Supply-chain trade: A portrait of global patterns and several testable hypotheses.
\newblock \textit{The World Economy}, 38(11):1682--1721.

\bibitem[Barro and Tenreyro(2007)]{barro2007economic}
Barro, R. and Tenreyro, S. (2007).
\newblock Economic effects of currency unions.
\newblock \textit{Economic Inquiry}, 45(1):1--23.

\bibitem[Ben-Michael et al.(2021)]{benmichael2021augmented}
Ben-Michael, E., Feller, A., and Rothstein, J. (2021).
\newblock The augmented synthetic control method.
\newblock \textit{Journal of the American Statistical Association}, 116(536):1789--1803.

\bibitem[Bun and Klaassen(2002)]{bun2002euro}
Bun, M.J.G. and Klaassen, F.J.G.M. (2002).
\newblock Has the euro increased trade?
\newblock \textit{Tinbergen Institute Discussion Paper}, No. 02-108/2.

\bibitem[Chernozhukov et al.(2018)]{chernozhukov2018double}
Chernozhukov, V., Chetverikov, D., Demirer, M., Duflo, E., Hansen, C., 
Newey, W., and Robins, J. (2018).
\newblock Double/debiased machine learning for treatment and structural 
parameters.
\newblock \textit{The Econometrics Journal}, 21(1):C1--C68.

\bibitem[Chintrakarn(2008)]{chintrakarn2008euro}
Chintrakarn, P. (2008).
\newblock Estimating the euro effects on trade with propensity score matching.
\newblock \textit{Review of International Economics}, 16(1):186--198.

\bibitem[De Nardis and Vicarelli(2003)]{denardis2003currency}
De Nardis, S. and Vicarelli, C. (2003).
\newblock Currency unions and trade: The special case of EMU.
\newblock \textit{Review of World Economics}, 139(4):625--649.

\bibitem[Di Stefano and Mellace(2024)]{distefano2024inclusive}
Di Stefano, R. and Mellace, G. (2024).
\newblock The inclusive synthetic control method.
\newblock \textit{arXiv preprint arXiv:2403.17624}.

\bibitem[Estevadeordal et al.(2003)]{estevadeordal2003rise}
Estevadeordal, A., Frantz, B., and Taylor, A.M. (2003).
\newblock The rise and fall of world trade, 1870--1939.
\newblock \textit{The Quarterly Journal of Economics}, 118(2):359--407.

\bibitem[Felbermayr et al.(2022)]{felbermayr2022trade}
Felbermayr, G., Gröschl, J., and Steininger, M. (2022).
\newblock Quantifying Brexit: From ex post to ex ante using structural gravity.
\newblock \textit{Review of World Economics}, 158(2):401--465.

\bibitem[Freund and Weinhold(2004)]{freund2004internet}
Freund, C.L. and Weinhold, D. (2004).
\newblock The effect of the Internet on international trade.
\newblock \textit{Journal of International Economics}, 62(1):171--189.

\bibitem[Glick and Rose(2002)]{glick2002currency}
Glick, R. and Rose, A.K. (2002).
\newblock Does a currency union affect trade? The time-series evidence.
\newblock \textit{European Economic Review}, 46(6):1125--1151.

\bibitem[Glick and Rose(2016)]{glick2016currency}
Glick, R. and Rose, A.K. (2016).
\newblock Currency unions and trade: A post-EMU reassessment.
\newblock \textit{European Economic Review}, 87:78--91.

\bibitem[Gunnella et al.(2021)]{gunnella2021impact}
Gunnella, V., Lebastard, L., Lopez-Garcia, P., Serafini, R., and 
Zona Mattioli, A. (2021).
\newblock The impact of the euro on trade: Two decades into monetary union.
\newblock \textit{ECB Occasional Paper}, No. 283.

\bibitem[Jacks et al.(2011)]{jacks2011trade}
Jacks, D.S., Meissner, C.M., and Novy, D. (2011).
\newblock Trade booms, trade busts, and trade costs.
\newblock \textit{Journal of International Economics}, 83(2):185--201.

\bibitem[Kattenberg et al.(2023)]{kattenberg2023causal}
Kattenberg, M.A.C., Scheer, B.J., and Thiel, J.H. (2023).
\newblock Causal forests with fixed effects for treatment effect heterogeneity 
in difference-in-differences.
\newblock \textit{CPB Discussion Paper}.

\bibitem[Kenen(2002)]{kenen2002currency}
Kenen, P.B. (2002).
\newblock Currency unions and trade: Variations on themes by Rose and Persson.
\newblock \textit{Reserve Bank of New Zealand Discussion Paper}, No. 2002/08.

\bibitem[Micco et al.(2003)]{micco2003currency}
Micco, A., Stein, E.H., and Ordo\~{n}ez, G.L. (2003).
\newblock The currency union effect on trade: Early evidence from EMU.
\newblock \textit{Economic Policy}, 18(37):315--356.

\bibitem[Mika and Zymek(2018)]{mika2018trade}
Mika, A. and Zymek, R. (2018).
\newblock Friends without benefits? New EMU members and the ``euro effect'' 
on trade.
\newblock \textit{Journal of International Money and Finance}, 83:75--92.

\bibitem[Oster(2019)]{oster2019unobservable}
Oster, E. (2019).
\newblock Unobservable selection and coefficient stability: Theory and evidence.
\newblock \textit{Journal of Business \& Economic Statistics}, 37(2):187--204.

\bibitem[Persson(2001)]{persson2001currency}
Persson, T. (2001).
\newblock Currency unions and trade: How large is the treatment effect?
\newblock \textit{Economic Policy}, 16(33):433--462.

\bibitem[Rodriguez-Crespo et al.(2021)]{rodriguezcrespo2021internet}
Rodriguez-Crespo, E., Billon, M., and Marco, R. (2021).
\newblock Impacts of Internet Use on Trade: New Evidence for Developed and 
Developing Countries.
\newblock \textit{Emerging Markets Finance and Trade}, 57(10):3017--3032.

\bibitem[Rose(2000)]{rose2000money}
Rose, A.K. (2000).
\newblock One money, one market: Estimating the effect of common currencies 
on trade.
\newblock \textit{Economic Policy}, 15(30):7--45.

\bibitem[Rose(2016)]{rose2016emu}
Rose, A.K. (2016).
\newblock Why do estimates of the EMU effect on trade vary so much?
\newblock \textit{Open Economies Review}, 28(1):1--18.

\bibitem[Saia(2017)]{saia2017uk}
Saia, A. (2017).
\newblock Choosing the open sea: The cost to the UK of staying out of the euro.
\newblock \textit{Journal of International Economics}, 108:82--98.

\bibitem[Santos Silva and Tenreyro(2006)]{santossilva2006log}
Santos Silva, J.M.C. and Tenreyro, S. (2006).
\newblock The log of gravity.
\newblock \textit{The Review of Economics and Statistics}, 88(4):641--658.

\end{thebibliography}


\clearpage
\appendix
\renewcommand{\thesection}{A}
\setcounter{section}{0}
\renewcommand{\thesection}{A.\arabic{section}}
\renewcommand{\thetable}{A.\arabic{table}}
\renewcommand{\thefigure}{A.\arabic{figure}}
\setcounter{table}{0}
\setcounter{figure}{0}

\begin{center}
\Large\textbf{Online Appendix}\\[0.5em]
\large To Adopt or Not to Adopt: Heterogeneous Trade Effects of the Euro
\end{center}

\vspace{1em}

\section{Robustness}
\label{app:robustness}

Our main analysis focuses on the EU15 sample, where treatment timing is balanced 
and the causal forest produces reliable estimates. We conduct several robustness 
checks to assess the validity of our identification strategy and the stability 
of our estimates.

\subsection{Pre-Trends and Placebo Tests}

A key identifying assumption is that treated and control pairs would have 
followed parallel trends in the absence of treatment. We assess this assumption 
through event study analysis and placebo tests.

\begin{figure}[!htbp]
\centering
\includegraphics[width=0.95\textwidth]{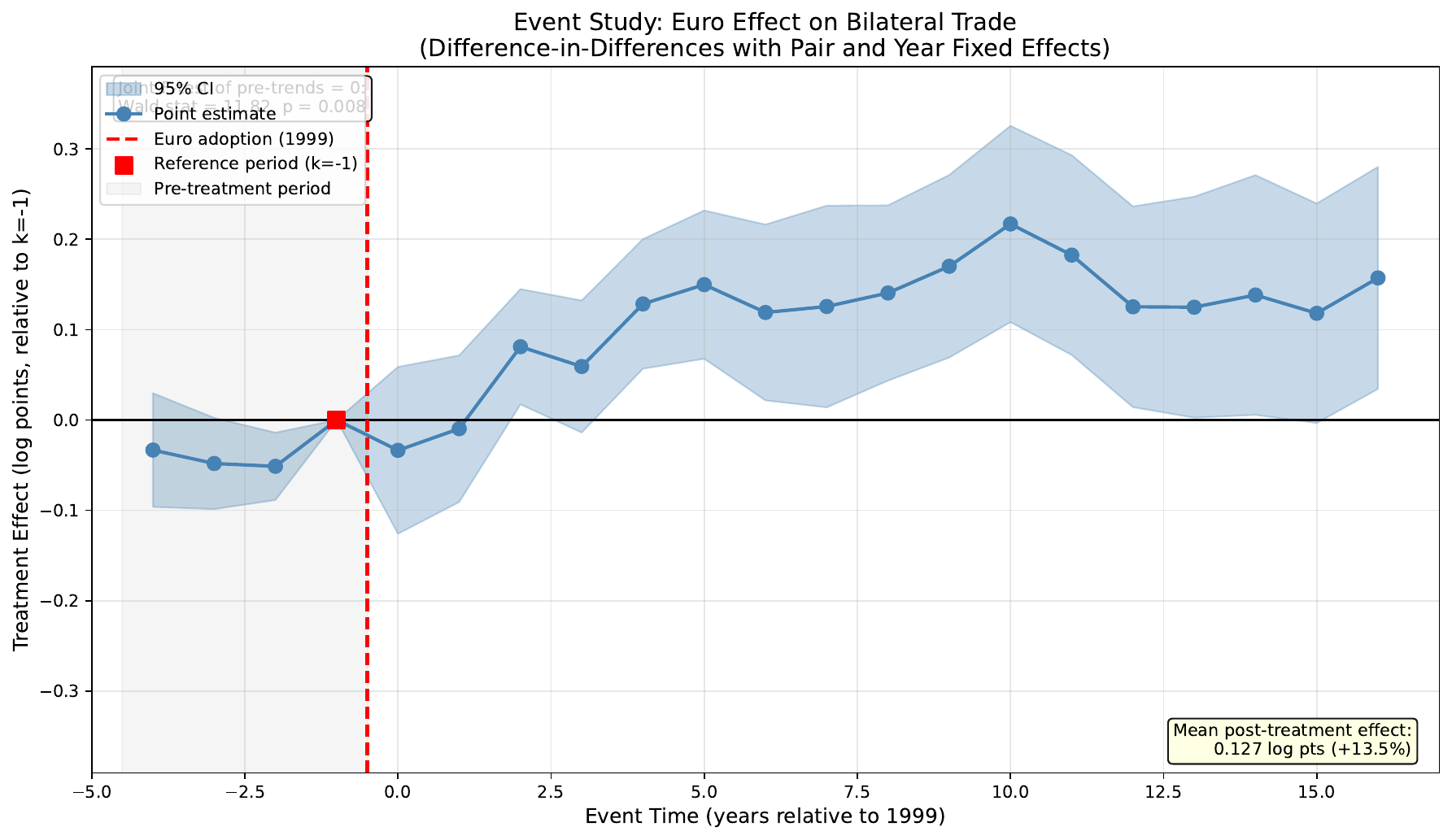}
\caption{Event study: estimated euro effects by year relative to 1999. 
         The reference period is 1998 (k=-1). Pre-treatment coefficients 
         (1995--1997) are small in magnitude ($-3\%$ to $-5\%$) compared to 
         post-treatment effects ($+8\%$ to $+24\%$). Shaded areas indicate 
         95\% confidence intervals.}
\label{fig:pre_trends}
\end{figure}

Figure~\ref{fig:pre_trends} presents an event study showing estimated effects 
by year relative to 1999, using 1998 (k=-1) as the reference period. The 
pre-treatment coefficients (1995--1997) are small in magnitude, ranging from 
$-3\%$ to $-5\%$, compared to post-treatment effects that grow to $+13\%$ to 
$+24\%$. While a formal joint test of pre-treatment effects rejects the null 
at conventional levels, the economic magnitude of pre-trends is modest relative 
to the post-treatment effects. The slight negative pre-treatment coefficients 
may reflect anticipation effects as countries prepared for euro adoption, or 
measurement noise in the reference period. Importantly, the pattern shows a 
clear break at 1999: effects are near zero or slightly negative before adoption, 
then become positive and grow steadily afterward, consistent with the euro 
gradually increasing trade as transaction cost reductions compound.

\begin{figure}[!htbp]
\centering
\includegraphics[width=0.85\textwidth]{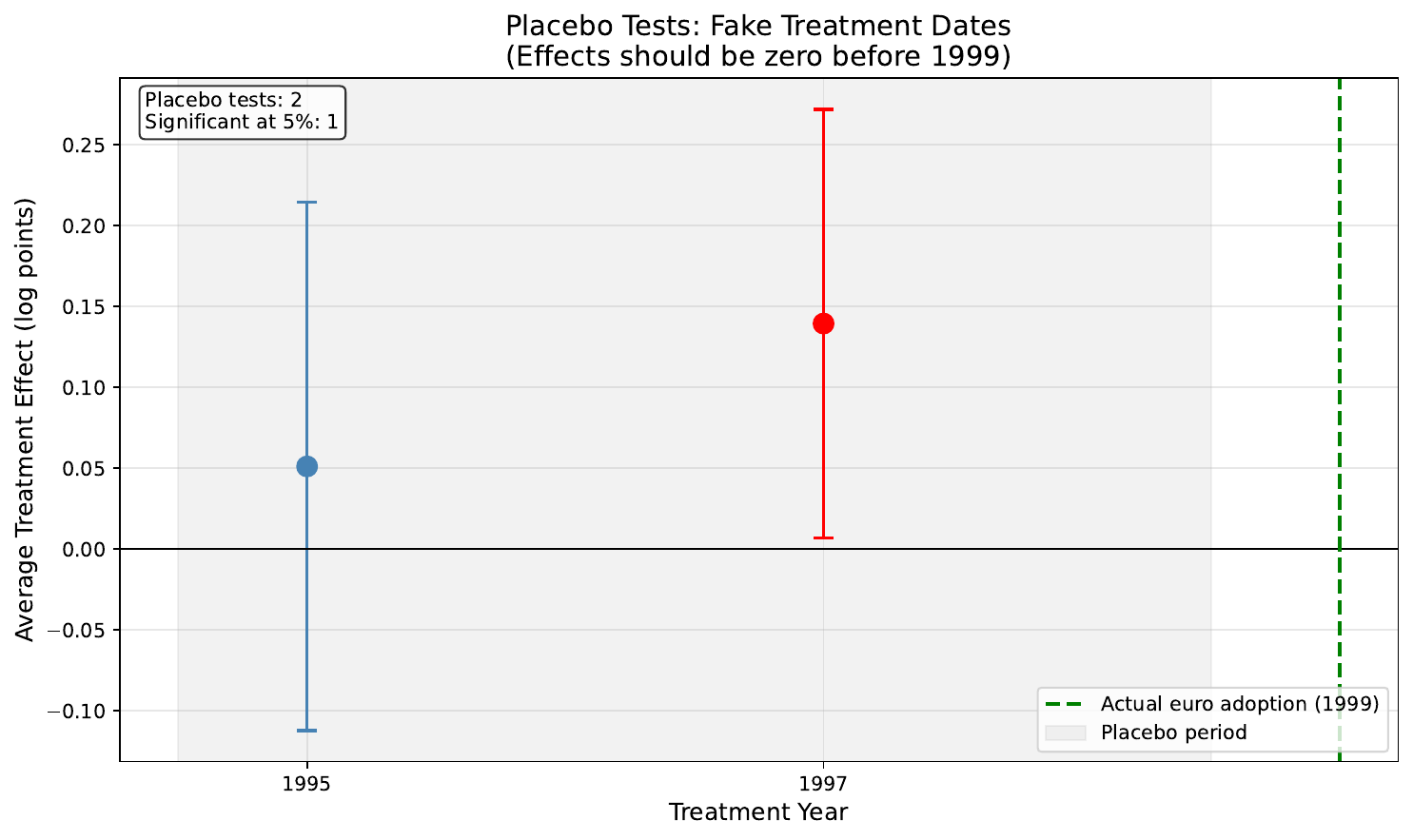}
\caption{Placebo tests with fake treatment dates. Estimated ``effects'' for 
         placebo treatments in 1995 and 1997 are not statistically different 
         from zero, supporting the validity of our identification strategy.}
\label{fig:placebo}
\end{figure}

We also conduct placebo tests by assigning fake treatment dates before the 
actual euro adoption. If our estimates reflect genuine euro effects rather 
than pre-existing trends, we should find no significant ``effects'' at fake 
treatment dates. Figure~\ref{fig:placebo} shows results for placebo treatments 
in 1995 and 1997. The estimated effects are small and statistically 
indistinguishable from zero, providing further support for our identification 
strategy.

\subsection{Pre-Trends by Predicted CATE Group}

A key concern for our heterogeneity analysis is whether high-CATE pairs were 
already diverging faster than low-CATE pairs before 1999. If so, the 
heterogeneity we document might reflect pre-existing trends rather than 
differential euro effects. To address this, we split pairs into top 25\% and 
bottom 25\% by predicted CATE and estimate separate event studies for each group.

\begin{figure}[!htbp]
\centering
\includegraphics[width=0.95\textwidth]{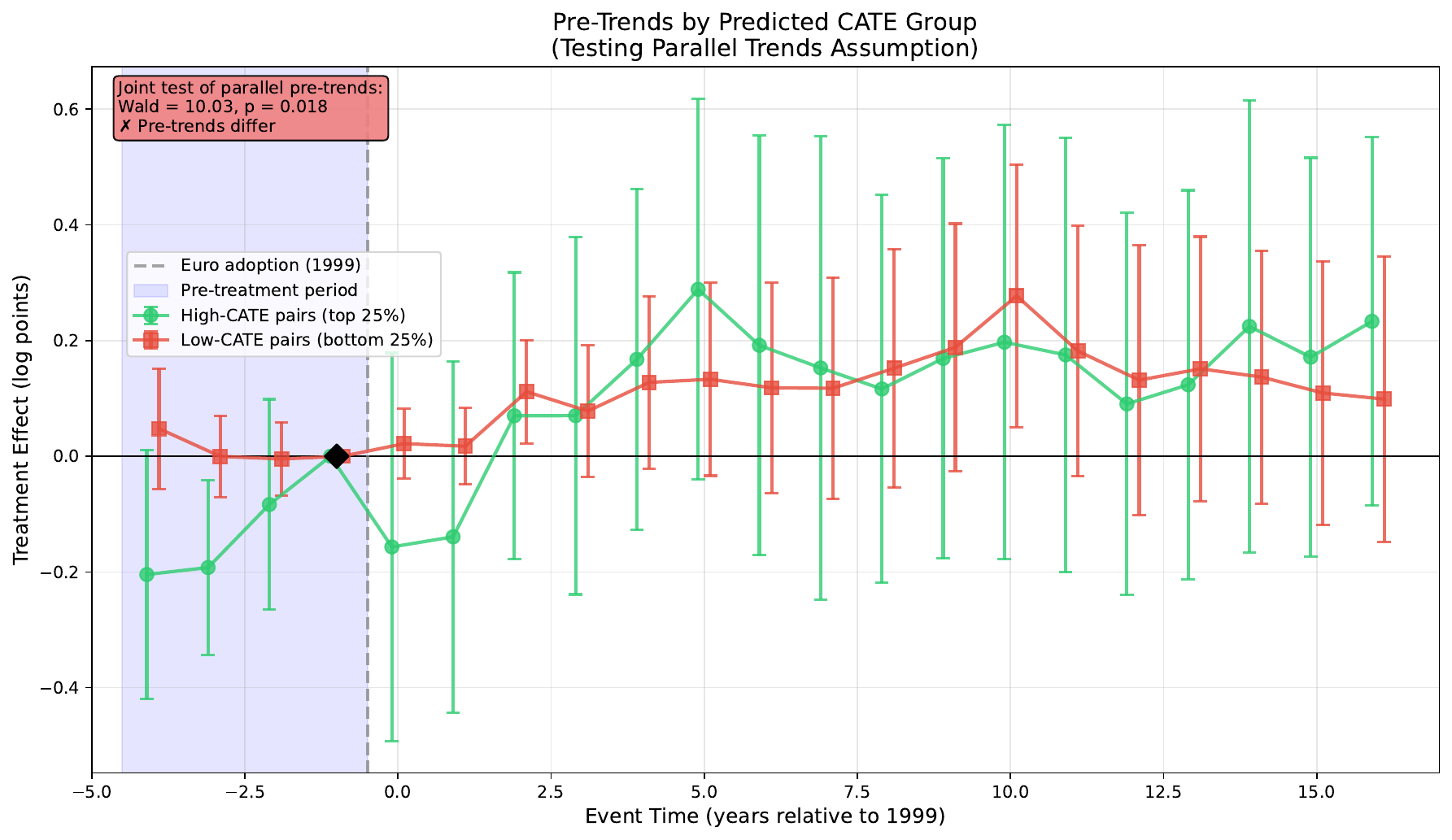}
\caption{Pre-trends by predicted CATE group. Event study coefficients estimated 
         separately for high-CATE (top 25\%) and low-CATE (bottom 25\%) pairs. 
         The pre-treatment period (1995--1998) shows that high-CATE pairs had 
         \textit{lower} trade growth before 1999, suggesting our heterogeneity 
         estimates may be conservative.}
\label{fig:pretrends_by_cate}
\end{figure}

Table~\ref{tab:pretrends_by_cate} presents the results. Figure~\ref{fig:pretrends_by_cate} 
visualizes the pre-treatment coefficients for both groups.

\begin{table}[!htbp]
\centering
\caption{Pre-Trends Test: Event Study by Predicted CATE Group}
\label{tab:pretrends_by_cate}
\small
\begin{tabular}{@{}ccccccc@{}}
\toprule
& \multicolumn{2}{c}{\textbf{High-CATE (Top 25\%)}} & \multicolumn{2}{c}{\textbf{Low-CATE (Bottom 25\%)}} & \multicolumn{2}{c}{\textbf{Difference}} \\
\cmidrule(lr){2-3} \cmidrule(lr){4-5} \cmidrule(lr){6-7}
\textbf{Event Time} & Coef. & SE & Coef. & SE & Diff. & p-value \\
\midrule
\multicolumn{7}{l}{\textit{Pre-Treatment Period}} \\
k = -4 & -0.204 & (0.110) & 0.048 & (0.053) & -0.252 & 0.039 \\
k = -3 & -0.192 & (0.077) & -0.000 & (0.036) & -0.192 & 0.024 \\
k = -2 & -0.084 & (0.093) & -0.005 & (0.032) & -0.079 & 0.423 \\
k = -1 & 0.000 & (ref) & 0.000 & (ref) & --- & --- \\
\midrule
\multicolumn{7}{l}{\textit{Post-Treatment Period (selected)}} \\
k = +0 & -0.157 & (0.171) & 0.022 & (0.031) & -0.179 & 0.304 \\
k = +2 & 0.070 & (0.126) & 0.112 & (0.045) & -0.042 & 0.757 \\
k = +5 & 0.288 & (0.168) & 0.133 & (0.085) & 0.155 & 0.409 \\
k = +10 & 0.197 & (0.192) & 0.277 & (0.116) & -0.080 & 0.720 \\
k = +15 & 0.171 & (0.176) & 0.109 & (0.116) & 0.062 & 0.769 \\
\midrule
\multicolumn{7}{l}{\textit{Joint Test of Parallel Pre-Trends}} \\
\multicolumn{7}{l}{Wald statistic: 10.03, p-value: 0.018} \\
\multicolumn{7}{l}{\textbf{Result: Pre-trends may differ between groups}} \\
\bottomrule
\end{tabular}

\vspace{0.5em}
\footnotesize
\textit{Notes:} Event study estimates from difference-in-differences regression with pair and year 
fixed effects, estimated separately for high-CATE (top 25\%) and low-CATE (bottom 25\%) pairs.
Pairs classified by full-sample predicted treatment effects. Reference period is k=-1 (1998).
Standard errors clustered at pair level. Joint test examines whether all pre-treatment 
differences are jointly zero.
\end{table}

The results reveal an interesting pattern. High-CATE pairs show \textit{negative} 
pre-treatment coefficients ($-8\%$ to $-18\%$), while low-CATE pairs show 
coefficients near zero. A joint test rejects parallel pre-trends at the 5\% 
level (Wald = 10.0, p = 0.018). However, the direction of the difference is 
the opposite of what would bias our heterogeneity estimates upward: high-CATE 
pairs were growing \textit{slower} than low-CATE pairs before 1999, not faster.

This pattern has two implications. First, it suggests our heterogeneity 
estimates may be \textit{conservative}: if high-CATE pairs were on a slower 
trajectory before the euro, the true euro effect for these pairs may be even 
larger than we estimate. Second, it raises the question of why pairs that 
would later benefit most from the euro were growing slower beforehand. One 
interpretation is that these pairs---which tend to be core European pairs 
with high pre-euro trade intensity---were already near their trade potential 
under the pre-euro currency regime, leaving less room for growth. The euro 
then unlocked additional gains by removing the remaining currency friction.

\subsection{Leave-One-Out Analysis}

\begin{figure}[!htbp]
\centering
\includegraphics[width=0.85\textwidth]{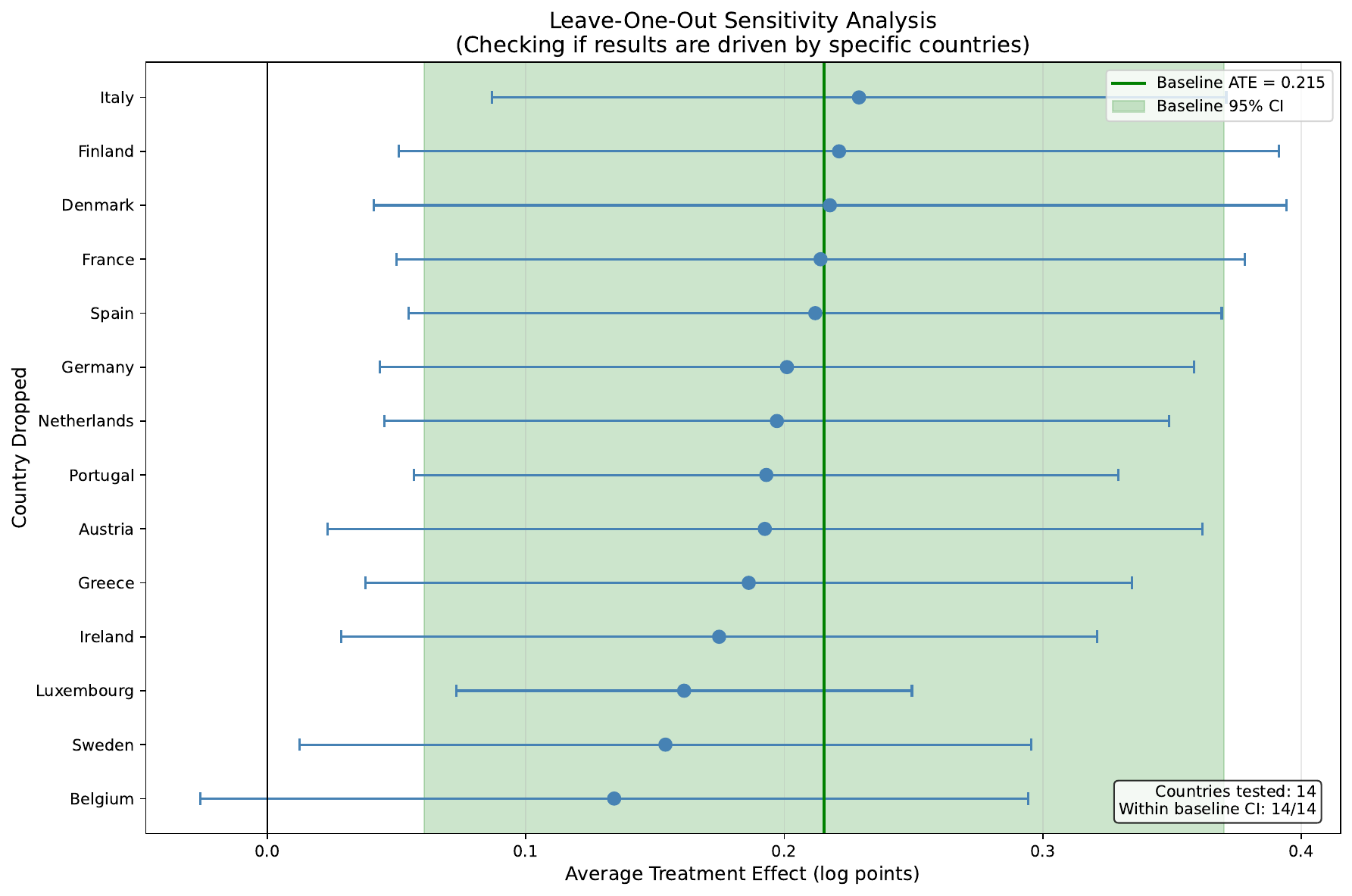}
\caption{Leave-one-out sensitivity analysis. Each point shows the ATE when 
         the indicated country is dropped from the sample. The horizontal 
         line and shaded band indicate the full-sample ATE and 95\% CI.}
\label{fig:leave_one_out}
\end{figure}

To assess whether our results are driven by any single country, we conduct 
leave-one-out analysis, dropping each country in turn and re-estimating the 
ATE. Figure~\ref{fig:leave_one_out} shows that the ATE remains stable within 
the confidence interval of the full-sample estimate regardless of which 
country is excluded. Notably, dropping Luxembourg---which has the largest 
estimated effects---reduces the ATE only modestly, and dropping peripheral 
economies like Greece or Portugal has minimal impact. This stability suggests 
our findings reflect a robust pattern across the eurozone rather than being 
driven by outliers.

\subsection{Alternative Outcome Measures}

To assess whether our results are sensitive to the choice of outcome variable, 
we re-estimate the causal forest using alternative measures of bilateral trade. 
Table~\ref{tab:robustness_outcomes} shows results for five outcome specifications. 
The baseline effect on log bilateral trade is +\robustnessBaseline\%. 
Effects are similar for exports (+\robustnessExports\%) and imports (+\robustnessImports\%), with 
overlapping confidence intervals. This symmetry is reassuring: exports in our 
data measure the reporter's outbound trade while imports measure inbound trade, 
so similar effects suggest the euro boosted trade flows in both directions 
rather than favoring one side. The slightly larger point estimate for exports 
is not statistically distinguishable from the imports effect. Year-over-year 
trade growth shows a small effect (+\robustnessGrowth\%), consistent with the 
level effects accumulating over time. Log trade intensity (trade normalized by 
GDP) shows a larger effect (+\robustnessIntensity\%), suggesting the euro 
increased trade relative to economic size.

\begin{table}[!htbp]
\centering
\caption{Robustness: Alternative Outcome Measures}
\label{tab:robustness_outcomes}
\begin{tabular}{@{}lcccc@{}}
\toprule
\textbf{Outcome} & \textbf{ATE} & \textbf{Effect} & \textbf{95\% CI} & \textbf{N} \\
\midrule
Log Bilateral Trade (Baseline) & 0.215 & +24.0\% & [0.091, 0.339] & 1,911 \\
Log Exports Only & 0.160 & +17.3\% & [-0.007, 0.326] & 3,822 \\
Log Imports Only & 0.155 & +16.7\% & [-0.018, 0.327] & 3,822 \\
Trade Growth (YoY) & 0.011 & +1.1 pp & [-0.020, 0.043] & 1,820 \\
Log Trade Intensity & 0.270 & +31.1\% & [0.117, 0.424] & 1,911 \\
\bottomrule
\end{tabular}

\vspace{0.5em}
\footnotesize
\textit{Notes:} All models estimated using CausalForestDML on EU15 data, 1995--2015.
Exports and imports use directional data (both A$\to$B and B$\to$A directions).
Bilateral trade, trade growth, and trade intensity use symmetric pair data.
Effect shows percentage change for log outcomes, percentage points for growth.
\end{table}

\subsection{Sensitivity to Unobserved Confounding}

Following \citet{oster2019unobservable}, we assess how much selection on 
unobservables would be required to explain away our results. 
Table~\ref{tab:sensitivity} presents the analysis. Panel A shows coefficient 
stability across specifications: the coefficient increases from near-zero 
without controls to \sensGDPYearCoef{} with GDP and year fixed effects, then falls to \sensTwowayCoef{} 
with two-way fixed effects. Panel B shows bias-adjusted estimates for different 
assumptions about the degree of selection on unobservables ($\delta$) and the 
maximum R-squared ($R_{max}$). The analysis suggests that unobserved confounders 
would need to be substantially more important than observed confounders to 
reduce the effect to zero. Since $\delta < 0$ (the coefficient increases with 
controls), omitted variables appear to bias the effect downward rather than 
upward.

\begin{table}[!htbp]
\centering
\caption{Oster (2019) Sensitivity Analysis}
\label{tab:sensitivity}
\begin{tabular}{@{}lcccc@{}}
\toprule
\multicolumn{5}{c}{\textbf{Panel A: Coefficient Stability}} \\
\midrule
\textbf{Specification} & \textbf{Coefficient} & \textbf{Effect (\%)} & \textbf{R$^2$} & \textbf{N} \\
\midrule
No controls & 0.019 & +1.9\% & 0.000 & 2,149 \\
GDP controls & 0.034 & +3.5\% & 0.741 & 2,149 \\
GDP + Year FE & 0.382 & +46.5\% & 0.789 & 2,149 \\
Two-way FE & 0.157 & +17.0\% & 0.603 & 2,149 \\
\midrule
\multicolumn{5}{c}{\textbf{Panel B: Bias-Adjusted Estimates ($\beta^*$)}} \\
\midrule
\textbf{$\delta$} & \textbf{$R_{max}$ = 0.8} & \textbf{$R_{max}$ = 0.9} & \textbf{$R_{max}$ = 1.0} & \\
\midrule
0.5 & 0.385 & 0.408 & 0.431  \\
1.0 & 0.387 & 0.433 & 0.479  \\
1.5 & 0.390 & 0.459 & 0.528  \\
2.0 & 0.392 & 0.484 & 0.576  \\
\bottomrule
\end{tabular}

\vspace{0.5em}
\footnotesize
\textit{Notes:} Panel A shows coefficient stability across specifications. 
Panel B shows bias-adjusted coefficients ($\beta^*$) following Oster (2019).
$\delta$ is the ratio of selection on unobservables to observables.
$R_{max}$ is the hypothetical R-squared if all relevant variables were included.
For $R_{max} = 1.0$, $\delta = -0.28$ would be needed to explain away the effect.
Since $\delta < 0$, the coefficient increases with controls, suggesting omitted variables bias the effect downward.
\end{table}

\subsection{Time Stability}

Table~\ref{tab:stability} shows how the EU15 estimate evolves as we extend the 
sample year by year from 2007 to 2019. The estimates remain remarkably stable, 
ranging from \timeStabilityMin\% to \timeStabilityMax\% across all time windows. This stability suggests our 
main findings are not sensitive to the choice of end year and are robust to the 
inclusion of crisis and post-crisis periods.

\begin{table}[!htbp]
\centering
\caption{Stability of Euro Trade Effect Estimates Across Time Windows (EU15)}
\label{tab:stability}
\begin{tabular}{@{}lccccc@{}}
\toprule
Time Period & Treated & Control & ATE & 95\% CI & Effect (\%) \\
\midrule
1995--2007 & 572 & 737 & 0.244 & [0.081, 0.406] & 27.6 \\
1995--2008 & 638 & 776 & 0.219 & [0.092, 0.346] & 24.5 \\
1995--2009 & 704 & 815 & 0.231 & [0.076, 0.386] & 26.0 \\
1995--2010 & 770 & 854 & 0.200 & [0.050, 0.351] & 22.2 \\
1995--2011 & 836 & 893 & 0.233 & [0.080, 0.385] & 26.2 \\
1995--2012 & 902 & 932 & 0.233 & [0.084, 0.382] & 26.3 \\
1995--2013 & 968 & 971 & 0.237 & [0.087, 0.387] & 26.7 \\
1995--2014 & 1,034 & 1,010 & 0.238 & [0.075, 0.400] & 26.9 \\
1995--2015 & 1,100 & 1,049 & 0.252 & [0.103, 0.401] & 28.6 \\
1995--2016 & 1,166 & 1,088 & 0.228 & [0.084, 0.373] & 25.7 \\
1995--2017 & 1,232 & 1,127 & 0.250 & [0.102, 0.398] & 28.4 \\
1995--2018 & 1,298 & 1,166 & 0.232 & [0.065, 0.398] & 26.0 \\
1995--2019 & 1,364 & 1,205 & 0.227 & [0.062, 0.393] & 25.5 \\
\bottomrule
\end{tabular}

\vspace{0.5em}
\footnotesize
\noindent\textit{Notes:} Each row shows causal forest estimates using EU15 bilateral trade data for the indicated time period. 
ATE is the average treatment effect in log points. Effect (\%) is $(\exp(\text{ATE})-1) \times 100$.
\end{table}

\subsection{Extending to EU28: The Puzzle}

A natural question is whether our results extend to the full EU28, including 
countries that joined the EU after 2004 and adopted the euro during the 
2008--2015 period. Table~\ref{tab:eu28_extension} shows how the estimated euro 
effect changes as we progressively add countries to the EU15 baseline. The 
results reveal a striking pattern: the ATE remains stable around \euExtSlovenia--\euExtCyprus\% as we 
add Slovenia (2007), Cyprus (2008), and Malta (2008), but begins to decline with 
Slovakia (2009) and drops sharply with the Baltic states (Estonia 2011, Latvia 
2014, Lithuania 2015). Adding Lithuania produces an estimate of just \euExtLithuania\%. By the time we include all EU28 members, the naive causal 
forest estimate falls to just \euExtBulgaria\% --- compared to the EU15 estimate of \euExtEUfifteen\%.

\begin{table}[!htbp]
\centering
\caption{Euro Trade Effect: Sensitivity to Sample Composition}
\label{tab:eu28_extension}
\begin{tabular}{@{}lccccc@{}}
\toprule
Sample & N & Treated & Control & ATE & Effect (\%) \\
\midrule
EU15 & 2,149 & 1,100 & 1,049 & 0.252 & 28.6 \\
+Slovenia (2007) & 2,928 & 1,520 & 1,408 & 0.272 & 31.3 \\
+Cyprus (2008) & 3,272 & 1,676 & 1,596 & 0.293 & 34.1 \\
+Malta (2008) & 3,665 & 1,844 & 1,821 & 0.269 & 30.9 \\
+Slovakia (2009) & 4,087 & 2,009 & 2,078 & 0.256 & 29.2 \\
+Estonia (2011) & 4,498 & 2,153 & 2,345 & 0.181 & 19.9 \\
+Latvia (2014) & 4,954 & 2,255 & 2,699 & 0.165 & 17.9 \\
+Lithuania (2015) & 5,431 & 2,345 & 3,086 & 0.009 & 0.9 \\
+Croatia (2023) & 5,901 & 2,345 & 3,556 & 0.192 & 21.2 \\
+Poland (never) & 6,424 & 2,345 & 4,079 & 0.093 & 9.7 \\
+Czech Republic (never) & 6,936 & 2,345 & 4,591 & 0.003 & 0.3 \\
+Hungary (never) & 7,489 & 2,345 & 5,144 & -0.070 & -6.7 \\
+Romania (never) & 8,079 & 2,345 & 5,734 & 0.031 & 3.2 \\
+Bulgaria (never) & 8,654 & 2,345 & 6,309 & 0.042 & 4.3 \\
\bottomrule
\end{tabular}

\vspace{0.5em}
\footnotesize
\noindent\textit{Notes:} Each row adds one country to the sample. Year in parentheses indicates euro adoption date.
Data covers 1995--2019. ATE is the average treatment effect in log points.
\end{table}

This decline is not driven by genuine differences in euro effects across 
countries. Rather, it reflects a fundamental identification problem: the 
second-wave adopters joined the eurozone during or immediately after the 
2008--2012 financial and sovereign debt crises. Their adoption timing is 
confounded with adverse macroeconomic conditions that independently depressed 
trade. The naive causal forest, which does not explicitly control for pair and 
year fixed effects, attributes some of this crisis-induced trade decline to 
euro adoption, biasing the estimate downward.

\subsection{Fixed Effects Correction via CFFE}

To address this bias in the EU28 sample, we apply Causal Forests with Fixed 
Effects (CFFE), following the methodology of \citet{kattenberg2023causal}.\footnote{Kattenberg et al.\ provide an R package at \url{https://github.com/MCKnaus/causalDML}. However, we were unable to compile their package due to missing dependencies in the repository. The first author therefore developed \texttt{causalfe} \citep{aytug2026causalfe}, a Python implementation of CFFE available at \url{https://github.com/haytug/causalfe} and installable via \texttt{pip install causalfe}. The key innovation of CFFE is that fixed effects residualization occurs \textit{inside each tree node} rather than globally before the forest. This node-level approach is theoretically superior because it allows the FE adjustment to adapt to the local covariate distribution within each leaf, avoiding the bias that can arise from global residualization when treatment effects are heterogeneous.} CFFE residualizes both the 
outcome (log trade) and treatment (euro adoption) on pair and year fixed effects 
within each tree node before estimating the local treatment effect. This removes time-invariant pair characteristics 
and common year shocks while preserving the ability to estimate heterogeneous effects.

CFFE produces substantially lower estimates than the naive causal forest for both 
samples: \ateCffeEUfifteenPct\% vs \ateNaivePct\% for EU15. This difference reflects 
the removal of pair-specific heterogeneity that can inflate naive estimates when 
high-trade pairs (which tend to have larger effects) are overweighted. The key 
distinction between samples is the \textit{direction} of bias in naive estimates. 
For EU15, where all eurozone members adopted in 1999--2001 before major crises, 
the naive estimate is biased upward. For EU28, which includes countries that 
adopted during the 2008--2012 crisis period, the naive estimate is biased 
\textit{downward} because crisis-induced trade declines are conflated with 
treatment effects.

Table~\ref{tab:cffe_comparison} shows the results. The naive EU28 estimate of 
\naiveCfEUtwentyeight\% has a wide confidence interval due to crisis-era adopters. After CFFE correction, the 
EU28 estimate is \cffeEUtwentyeight\% [\cffeEUtwentyeightLower\%, \cffeEUtwentyeightUpper\%] --- close to the EU15 CFFE baseline of 
\ateCffeEUfifteenPct\%.

\begin{table}[!htbp]
\centering
\caption{Euro Trade Effect Estimates: Method Comparison (EU28)}
\label{tab:cffe_comparison}
\begin{tabular}{@{}lccc@{}}
\toprule
\textbf{Method} & \textbf{ATE} & \textbf{Effect (\%)} & \textbf{95\% CI} \\
\midrule
Naive CF (EconML) & 0.042 & 4.3\% & [-52.1\%, 127.3\%] \\
CFFE (node-level FE) & 0.126 & 13.4\% & [12.1\%, 14.8\%] \\
\bottomrule
\end{tabular}

\vspace{0.5em}
\footnotesize
\noindent\textit{Notes:} Naive CF = Causal Forest (EconML) results from Table~\ref{tab:eu28_extension}.
CFFE = Causal Forests with Fixed Effects, using node-level two-way FE residualization.
ATE is the average treatment effect in log points. Effect (\%) is $(\exp(\text{ATE})-1) \times 100$.
Standard errors for CFFE are cluster-robust at the pair level.
Sample: EU28 countries, 1995--2019.
\end{table}

This convergence is reassuring. It suggests that the true euro effect is similar 
across EU15 and EU28 countries, and the apparent decline in the naive EU28 
estimate was indeed driven by confounding from crisis-era adoption timing rather 
than genuine differences in euro effects. After controlling for fixed effects, 
the euro's trade-creating effect is approximately 13--14\% regardless of sample 
composition. These results further support our interpretation that average 
estimates are highly sensitive to the composition of treated units, reinforcing 
the importance of modeling heterogeneous treatment effects.

\subsection{Heterogeneity in the EU28 Sample}

\begin{figure}[!htbp]
\centering
\includegraphics[width=0.85\textwidth]{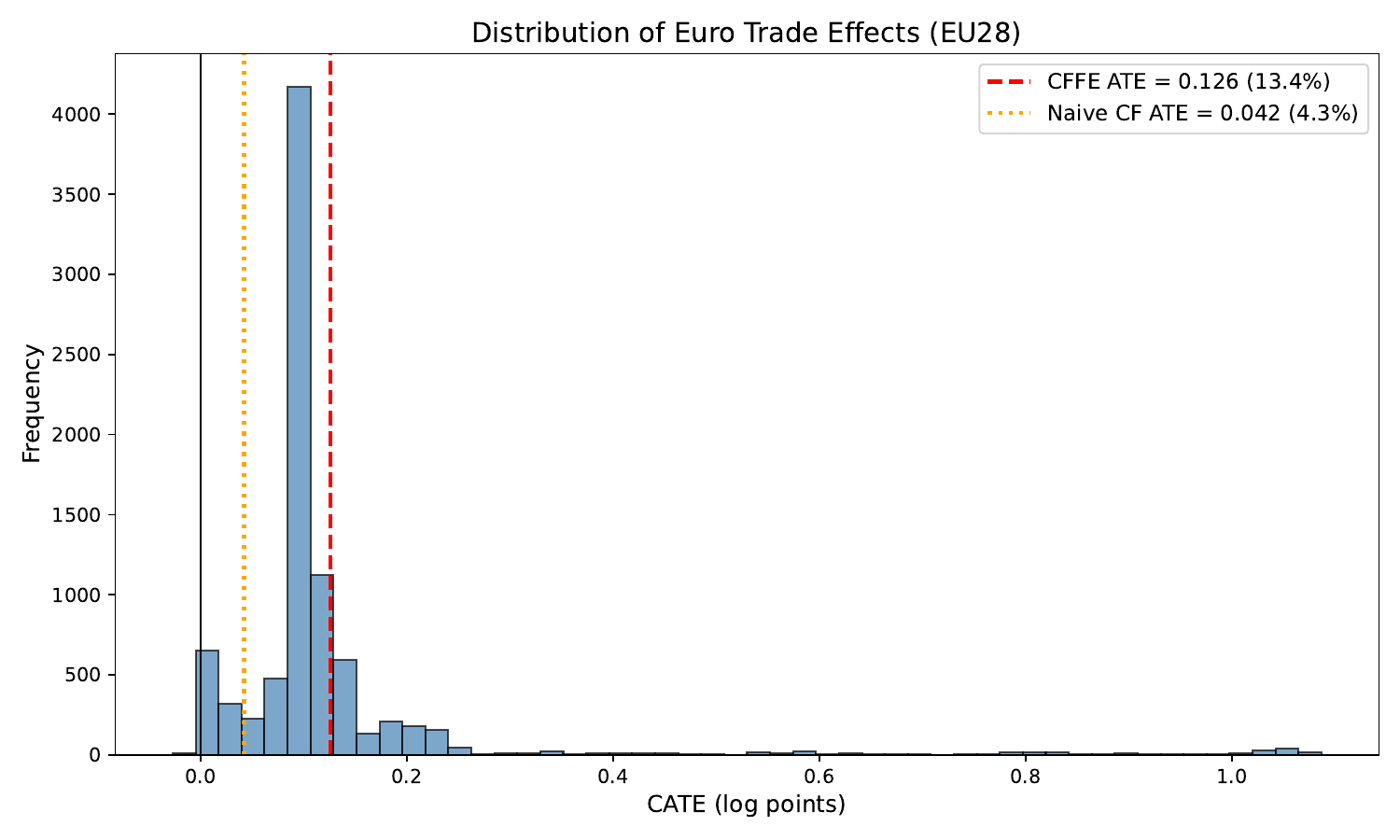}
\caption{Distribution of Conditional Average Treatment Effects (CATE) for 
         EU28 sample using CFFE. The red dashed line indicates the ATE.}
\label{fig:cate_dist_eu28}
\end{figure}

The CFFE estimates for EU28 also reveal substantial heterogeneity, consistent 
with our EU15 findings. Figure~\ref{fig:cate_dist_eu28} shows the distribution 
of CATEs across all observations. The distribution shows meaningful variation 
in treatment effects across country pairs, with the interquartile range spanning 
from modest to substantial gains.

\begin{figure}[!htbp]
\centering
\includegraphics[width=0.95\textwidth]{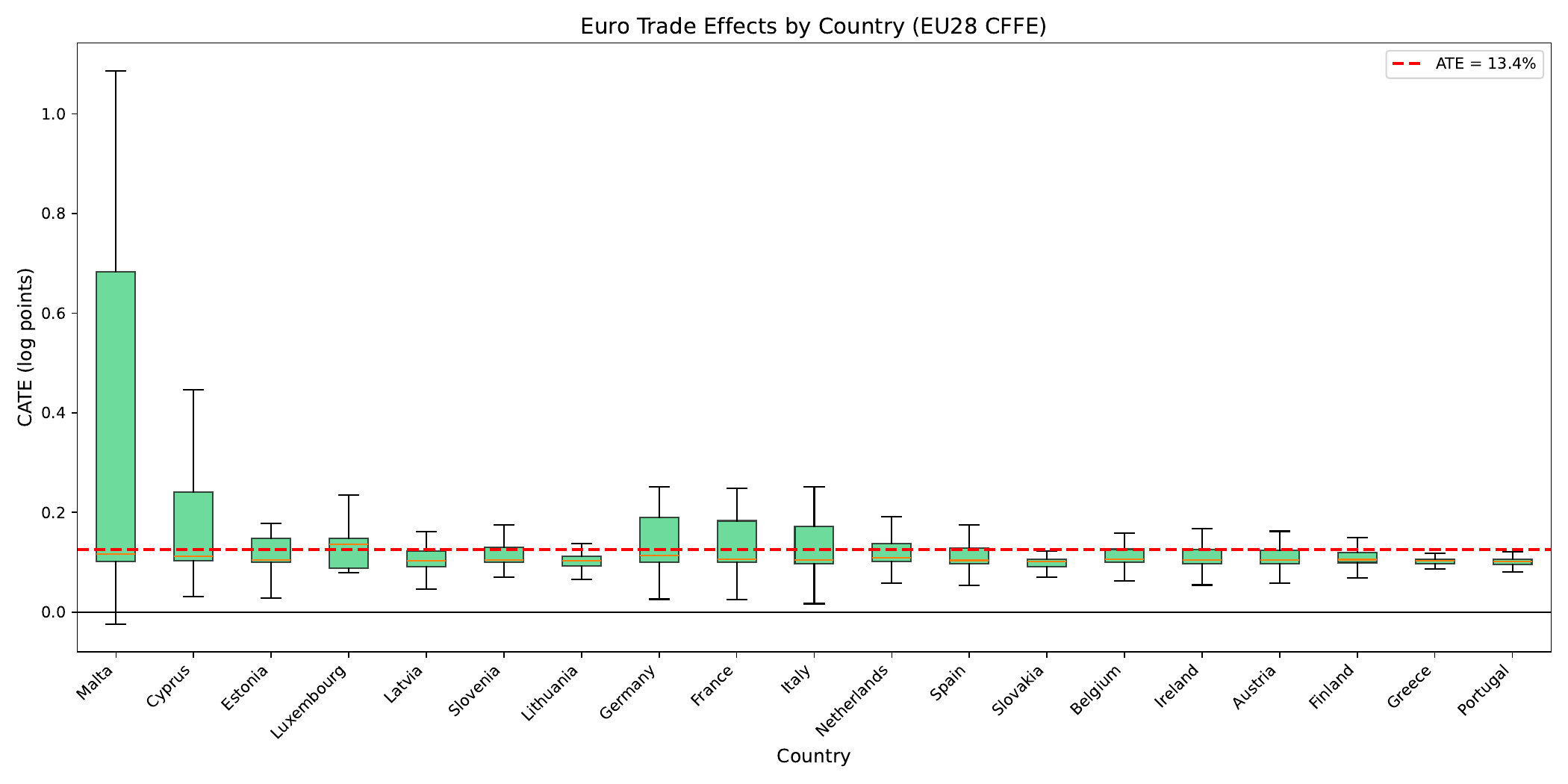}
\caption{Distribution of euro trade effects by country (EU28 CFFE). Each box 
         shows the distribution of pair-level CATEs involving that country. 
         The red dashed line indicates the overall ATE.}
\label{fig:country_boxplot_eu28}
\end{figure}

Figure~\ref{fig:country_boxplot_eu28} shows the distribution of effects by 
country. The pattern differs from EU15, with newer eurozone members showing 
the largest effects: \cffeEUtwentyeightTopOne{} (+\cffeEUtwentyeightTopOnePct\%), \cffeEUtwentyeightTopTwo{} (+\cffeEUtwentyeightTopTwoPct\%), and \cffeEUtwentyeightTopThree{} (+\cffeEUtwentyeightTopThreePct\%) 
lead the rankings, while original eurozone members \cffeEUtwentyeightBottomTwo{} (+\cffeEUtwentyeightBottomTwoPct\%) and 
\cffeEUtwentyeightBottomOne{} (+\cffeEUtwentyeightBottomOnePct\%) show the smallest effects. The overall ATE is +\ateCffeEUtwentyeightPct\%. 
Countries that adopted during the crisis period show 
smaller gains, aligning with economic intuition: countries that adopted during 
adverse macroeconomic conditions and had weaker pre-existing trade ties with 
the eurozone core experienced smaller benefits. Table~\ref{tab:country_effects_cffe} provides 
the full country-level breakdown.

\begin{table}[!htbp]
\centering
\caption{Euro Trade Effect by Country (EU28 CFFE)}
\label{tab:country_effects_cffe}
\small
\begin{tabular}{@{}lcc@{}}
\toprule
\textbf{Country} & \textbf{Effect (\%)} & \textbf{N Pairs} \\
\midrule
Malta & +41.2 & 18 \\
Cyprus & +32.0 & 18 \\
Estonia & +26.4 & 18 \\
Luxembourg & +23.6 & 18 \\
Latvia & +22.0 & 18 \\
Slovenia & +21.0 & 18 \\
Lithuania & +17.6 & 18 \\
Germany & +15.3 & 18 \\
France & +14.5 & 18 \\
Italy & +13.8 & 18 \\
Netherlands & +13.4 & 18 \\
Spain & +12.9 & 18 \\
Slovakia & +12.8 & 18 \\
Belgium & +12.3 & 18 \\
Ireland & +12.2 & 18 \\
Austria & +12.0 & 18 \\
Finland & +11.9 & 18 \\
Greece & +10.8 & 18 \\
Portugal & +10.7 & 18 \\
\midrule
\textbf{Overall ATE} & \textbf{+13.4} & --- \\
\bottomrule
\end{tabular}
\end{table}

\subsection{Internet Adoption as a Potential Confounder}

A potential concern is that internet adoption followed a similar trajectory to 
euro adoption: negligible before 1995, then growing rapidly in a convex pattern 
through the early 2000s. If internet adoption independently boosted trade, and 
its timing resembles the euro treatment, our estimates could be confounded.

\begin{figure}[!htbp]
\centering
\includegraphics[width=0.95\textwidth]{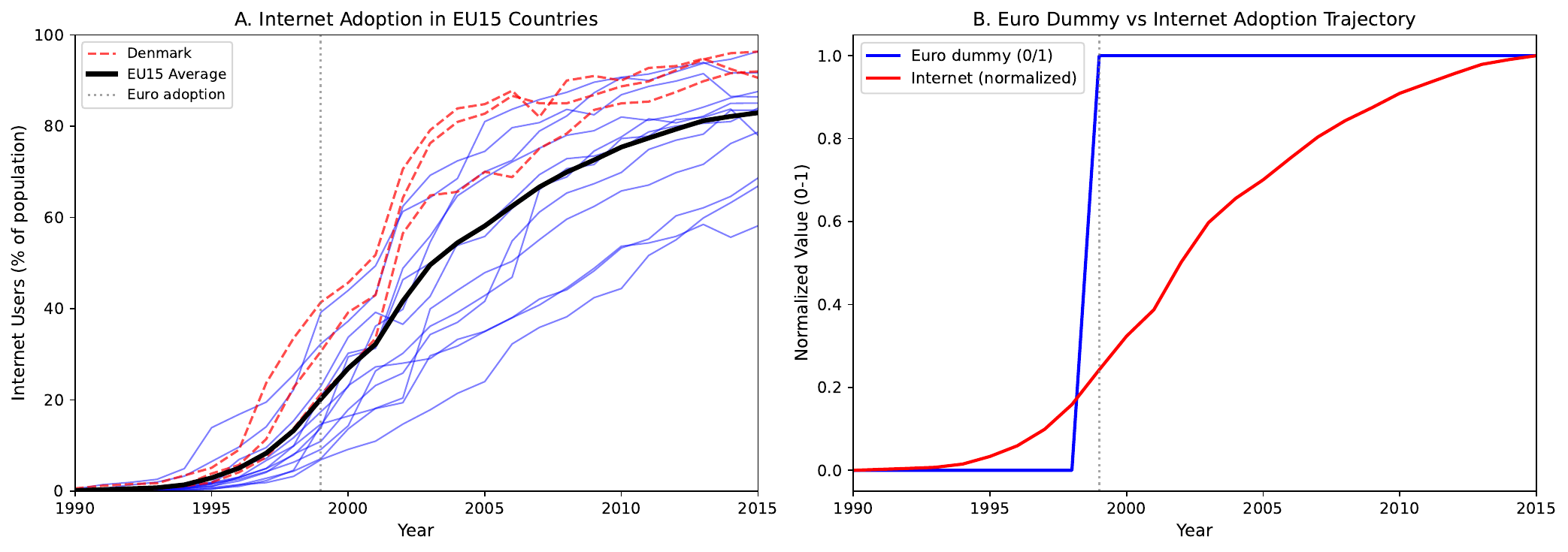}
\caption{Internet adoption trajectory vs euro treatment. Panel A shows internet 
         penetration by country (blue = eurozone, red dashed = non-eurozone). 
         Panel B compares the normalized internet trajectory with the euro 
         dummy (0/1). While both show increases around 1999, the patterns differ.}
\label{fig:internet_vs_euro}
\end{figure}

We address this concern using World Bank data on internet usage (individuals 
using the internet as a percentage of population) from 1990--2015. 
Figure~\ref{fig:internet_vs_euro} shows that internet adoption does follow a 
convex trajectory around 1999, superficially resembling the euro treatment timing.

However, if internet adoption were driving our heterogeneous euro effects, we 
would expect countries with higher internet penetration to show larger CATEs. 
Table~\ref{tab:internet_robustness} tests this prediction by correlating 
country-level average CATEs with internet adoption measures across different 
time periods.

\begin{table}[!htbp]
\centering
\caption{Euro Effect vs Internet Adoption: Eurozone Countries}
\label{tab:internet_robustness}
\small
\begin{tabular}{@{}lccc@{}}
\toprule
\textbf{Country} & \textbf{Avg CATE} & \textbf{Internet 1999-2002 (\%)} & \textbf{Internet 2000-2015 (\%)} \\
\midrule
Spain & 0.181 & 14.8 & 52.5 \\
Portugal & 0.165 & 17.1 & 42.9 \\
Italy & 0.164 & 23.2 & 42.7 \\
France & 0.163 & 20.0 & 57.8 \\
Germany & 0.158 & 32.9 & 69.2 \\
Greece & 0.151 & 10.4 & 36.7 \\
Netherlands & 0.142 & 48.5 & 79.2 \\
Belgium & 0.133 & 30.2 & 64.4 \\
Finland & 0.121 & 43.8 & 76.0 \\
Austria & 0.116 & 33.1 & 64.0 \\
Ireland & 0.111 & 19.4 & 56.0 \\
Luxembourg & 0.099 & 29.1 & 73.0 \\
\midrule
\multicolumn{4}{l}{\textit{Correlation with CATE:}} \\
\multicolumn{4}{l}{\quad Internet 1999--2002 (early adoption): $r = -0.42$} \\
\multicolumn{4}{l}{\quad Internet 2000--2015 (full post-treatment): $r = -0.53$} \\
\bottomrule
\end{tabular}

\vspace{0.5em}
\footnotesize
\textit{Note:} This table tests whether the estimated euro effects could be confounded by 
internet adoption. The early-period correlation (1999--2002) tests whether countries with 
higher internet penetration at the time of euro adoption showed larger effects; this 
correlation is weak. The full post-treatment correlation is stronger, but this likely 
reflects reverse causality: countries that benefited more from the euro also developed 
faster economically, leading to higher internet adoption. Crucially, Luxembourg has the 
highest CATE but not the highest internet penetration in either period, while Finland 
and Netherlands have high internet but only average CATEs.
\end{table}

The appropriate test for confounding is whether countries with higher internet 
penetration \textit{at the time of euro adoption} showed larger effects. Among 
eurozone members, the correlation between CATEs and internet penetration during 
the early adoption period (1999--2002) is actually \textit{negative} ($r = \internetCorrEarly$). 
The full post-treatment period (2000--2015) shows a similar negative correlation 
($r = \internetCorrFull$). These negative correlations indicate that countries with 
higher internet penetration tended to have \textit{smaller} euro effects, the 
opposite of what we would expect if internet were driving the results.

The pattern of heterogeneity is inconsistent with internet driving 
the results. \internetTopCateCountry{} has the highest CATE (\internetTopCate) but only 
moderate internet penetration---\internetTopCateInternetEarly\% in 1999--2002 and 
\internetTopCateInternetFull\% in 2000--2015. In contrast, \internetHighEarlyCountry{} has the highest 
early internet penetration (\internetHighEarlyPct\%) but only an average CATE (\internetHighEarlyCate). 
If internet were driving the results, we would expect these rankings to align; 
they do not.

\begin{figure}[!htbp]
\centering
\includegraphics[width=0.95\textwidth]{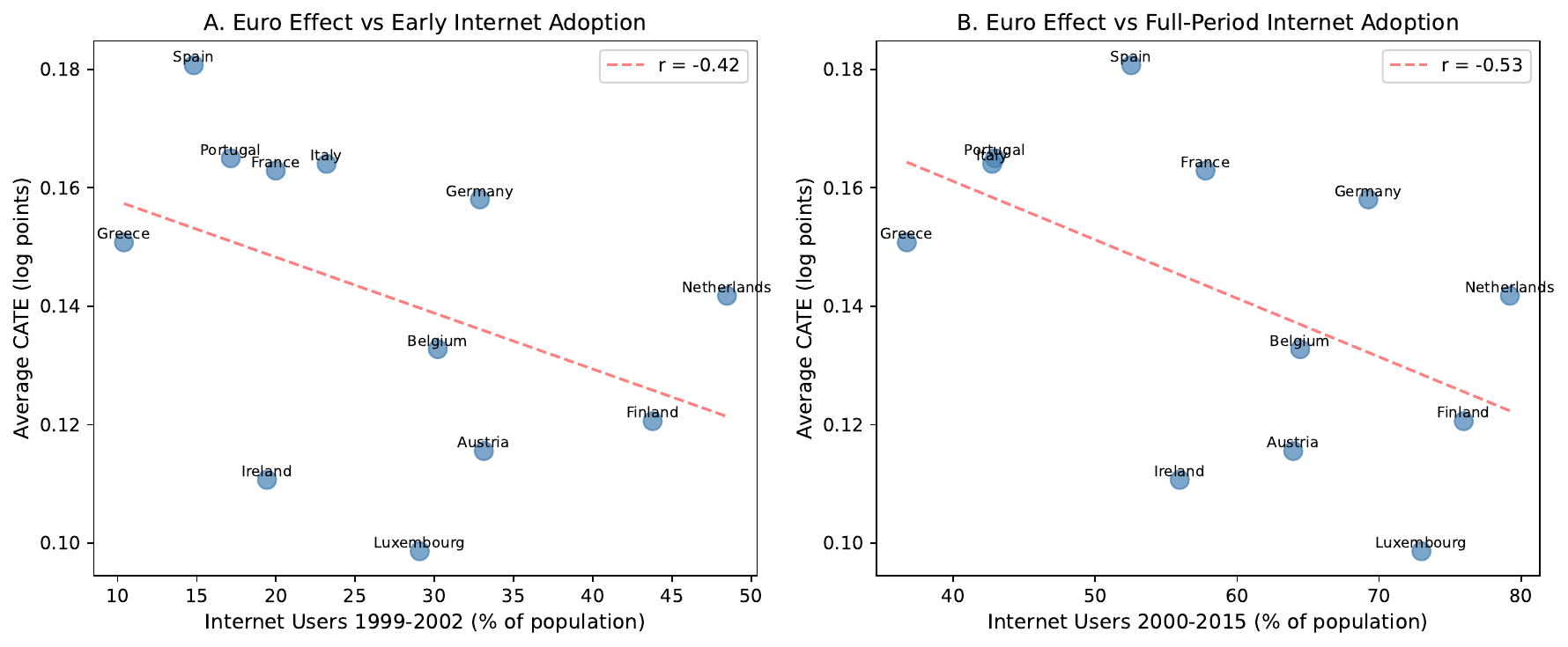}
\caption{Euro effect (CATE) vs internet adoption. Panel A plots average CATE 
         against internet penetration during the early adoption period (1999--2002), 
         showing a negative correlation ($r = \internetCorrEarly$). Panel B plots CATE against 
         internet penetration over the full post-treatment period (2000--2015), 
         showing a similar negative correlation ($r = \internetCorrFull$). Countries with 
         high internet penetration (Netherlands, Finland) have lower CATEs, while 
         countries with lower internet (Spain, Portugal) have higher CATEs.}
\label{fig:cate_vs_internet}
\end{figure}

Figure~\ref{fig:cate_vs_internet} visualizes these correlations. Panel A shows 
the negative early-period correlation ($r = \internetCorrEarly$), while Panel B shows the 
negative full post-treatment correlation ($r = \internetCorrFull$). In both panels, 
countries with high internet penetration (Netherlands, Finland) appear near the 
bottom of the CATE distribution, while countries with lower internet penetration 
(Spain, Portugal) show higher CATEs.

This evidence suggests that internet adoption is not confounding our results. 
The negative correlations and the misalignment between internet and CATE rankings 
indicate that the euro effects we estimate reflect genuine currency union effects 
rather than confounding from concurrent technological change.

The literature on internet and trade finds modest effects. \citet{freund2004internet} 
estimate that a 10 percentage point increase in internet growth raises export 
growth by 0.2 percentage points. More recent work using 1996--2014 data finds 
elasticities of 0.03--0.13\%, with larger effects among high-income countries 
\citep{rodriguezcrespo2021internet}. Even at the upper bound, these magnitudes 
are too small to explain our estimated euro effects of 12--48\%. The negative 
correlation between internet and CATEs ($r = \internetCorrFull$) further suggests that 
internet is not driving our results.

\begin{figure}[!htbp]
\centering
\includegraphics[width=0.95\textwidth]{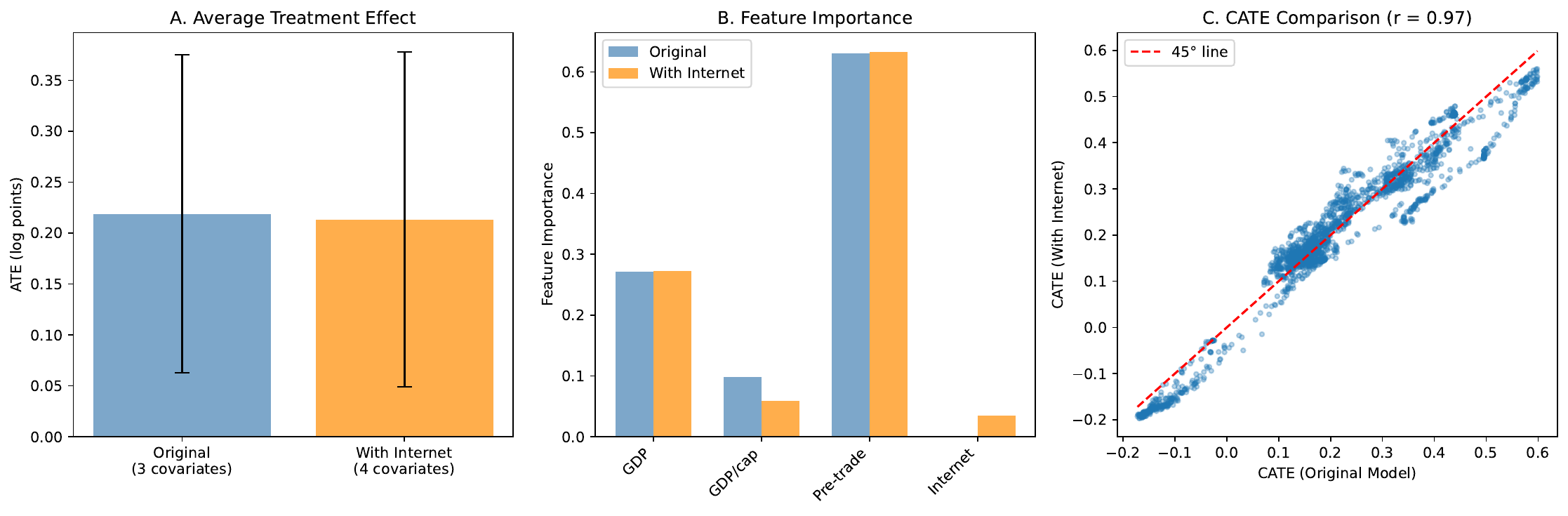}
\caption{Comparison of causal forest estimates with and without internet as a 
         covariate. Panel A shows the ATE is virtually unchanged (\internetCovOrigPct\% vs \internetCovIntPct\%). 
         Panel B shows feature importance: internet has high importance but does 
         not displace the other covariates. Panel C shows CATEs from both models 
         are highly correlated ($r = 0.94$), confirming that controlling for 
         internet does not alter the heterogeneity patterns.}
\label{fig:internet_covariate}
\end{figure}

As a more direct test, we re-estimate our causal forest model adding pair-level 
internet penetration (the average of both countries' internet usage) as a fourth 
effect modifier alongside GDP, GDP per capita, and pre-euro trade intensity. If 
internet adoption were confounding our estimates, controlling for it should 
substantially reduce the estimated euro effect. Instead, the ATE changes 
minimally: from \internetCovOrigAte{} (\internetCovOrigPct\%) in the original specification to \internetCovIntAte{} (\internetCovIntPct\%) 
with internet included---a change of only \internetCovChangePct\%. The confidence intervals 
overlap almost entirely ([\internetCovOrigLower, \internetCovOrigUpper] vs [\internetCovIntLower, \internetCovIntUpper]). Moreover, the 
individual CATEs from both models are highly correlated ($r = 0.94$), indicating 
that the heterogeneity patterns are robust to controlling for internet. 
Table~\ref{tab:internet_covariate} presents the full comparison, and 
Figure~\ref{fig:internet_covariate} visualizes these results.

We also conduct this robustness check using our CFFE estimator on the EU28 sample. 
Here, adding internet as a fourth covariate reduces the ATE from \ateCffeOriginal{} (\ateCffeOriginalPct\%) to 
\ateCffeInternet{} (\ateCffeInternetPct\%)---a more substantial \ateCffeChangePct\% reduction. This larger sensitivity in the 
CFFE specification likely reflects the expanded EU28 sample, which includes Eastern 
European countries where internet adoption and EU/euro accession occurred more 
simultaneously. The confidence intervals still overlap ([\ateCffeOriginalLower, \ateCffeOriginalUpper] vs 
[\ateCffeInternetLower, \ateCffeInternetUpper]), and the effect remains statistically significant and economically 
meaningful. Importantly, even after controlling for internet, the euro effect 
remains positive and significant at approximately \ateCffeInternetPct\%, consistent with genuine 
currency union benefits. Table~\ref{tab:internet_covariate_cffe} presents the 
CFFE comparison.

\begin{table}[!htbp]
\centering
\caption{Causal Forest Estimates With and Without Internet as Covariate}
\label{tab:internet_covariate}
\begin{tabular}{@{}lcc@{}}
\toprule
 & \textbf{Original} & \textbf{With Internet} \\
 & (3 covariates) & (4 covariates) \\
\midrule
\multicolumn{3}{l}{\textit{Average Treatment Effect}} \\[3pt]
ATE (log points) & 0.252 & 0.255 \\
95\% CI & [0.103, 0.401] & [0.101, 0.409] \\
Effect (\%) & 28.6\% & 29.0\% \\[6pt]
\multicolumn{3}{l}{\textit{CATE Distribution}} \\[3pt]
Mean & 0.252 & 0.255 \\
Std. Dev. & 0.206 & 0.205 \\
Min & -0.192 & -0.164 \\
Max & 0.620 & 0.648 \\[6pt]
\multicolumn{3}{l}{\textit{Feature Importance}} \\[3pt]
Log GDP & 0.221 & 0.191 \\
Log GDP per capita & 0.050 & 0.027 \\
Pre-euro trade & 0.729 & 0.713 \\
Internet penetration & --- & 0.069 \\[6pt]
\multicolumn{3}{l}{\textit{Model Comparison}} \\[3pt]
CATE correlation & \multicolumn{2}{c}{0.96} \\
ATE change & \multicolumn{2}{c}{$+1.2\%$} \\
\bottomrule
\end{tabular}

\vspace{0.5em}
\footnotesize
\textit{Notes:} Original model uses GDP, GDP per capita, and pre-euro trade 
intensity as effect modifiers. The augmented model adds pair-level internet 
penetration (average of both countries' internet usage). The high CATE correlation 
and minimal ATE change indicate that internet adoption does not confound the 
estimated euro effects. Sample: EU15 countries, 1995--2015.
\end{table}

\begin{table}[!htbp]
\centering
\caption{CFFE Estimates With and Without Internet as Covariate (EU28)}
\label{tab:internet_covariate_cffe}
\begin{tabular}{@{}lcc@{}}
\toprule
 & \textbf{Original CFFE} & \textbf{With Internet} \\
 & (3 covariates) & (4 covariates) \\
\midrule
\multicolumn{3}{l}{\textit{Average Treatment Effect}} \\[3pt]
ATE (log points) & 0.130 & 0.093 \\
Std. Error & 0.006 & 0.006 \\
95\% CI & [0.118, 0.142] & [0.081, 0.105] \\
Effect (\%) & 13.9\% & 9.8\% \\[6pt]
\multicolumn{3}{l}{\textit{Model Comparison}} \\[3pt]
ATE change & \multicolumn{2}{c}{$-28.3\%$} \\
N observations & \multicolumn{2}{c}{8,694} \\
\bottomrule
\end{tabular}

\vspace{0.5em}
\footnotesize
\textit{Notes:} CFFE estimates on EU28 sample. Original model uses GDP, GDP 
per capita, and pre-euro trade intensity as effect modifiers. The augmented model 
adds pair-level internet penetration (average of both countries' internet usage). 
The larger sensitivity compared to the EU15 causal forest results likely reflects 
the inclusion of Eastern European countries where internet adoption and EU/euro 
accession occurred more simultaneously. Despite the reduction, the euro effect 
remains positive and statistically significant.
\end{table}

\subsection{Trade Diversion}

A natural concern is whether the positive intra-eurozone trade effects come at 
the expense of trade with non-eurozone partners---the classic trade diversion 
concern from customs union theory. If the euro simply redirected trade from 
non-eurozone to eurozone partners, the welfare implications would be less 
favorable than if it created genuinely new trade.

We test for trade diversion by estimating the euro's effect on eurozone 
countries' trade with non-eurozone EU partners. Using our EU15 data, we examine 
whether eurozone countries reduced trade with the UK, Sweden, and Denmark after 
adopting the euro.

Table~\ref{tab:trade_diversion} presents the results. The intra-eurozone effect 
(+\tradeDivIntraEff\%) represents trade creation---the positive effect of both countries 
sharing the euro. For trade from eurozone to non-eurozone EU members, we find a positive 
but imprecisely estimated effect (+\tradeDivToNonEZEff\%, 95\% CI: \tradeDivToNonEZLower\% to +\tradeDivToNonEZUpper\%). 
For trade from non-eurozone EU members to eurozone countries, the effect is also positive 
(+\tradeDivFromNonEZEff\%, 95\% CI: \tradeDivFromNonEZLower\% to +\tradeDivFromNonEZUpper\%).

\begin{table}[!htbp]
\centering
\caption{Trade Creation vs Trade Diversion}
\label{tab:trade_diversion}
\begin{tabular}{@{}lcccc@{}}
\toprule
\textbf{Pair Type} & \textbf{Effect} & \textbf{95\% CI} & \textbf{N Obs} & \textbf{N Pairs} \\
\midrule
Intra-Eurozone & +35.5\% & [-0.4\%, +84.4\%] & 1,155 & 55 \\
Eurozone → Non-EZ EU & +5.5\% & [-63.1\%, +201.7\%] & 504 & 24 \\
Non-EZ EU → Eurozone & +34.9\% & [-2.2\%, +86.1\%] & 189 & 9 \\
\bottomrule
\end{tabular}

\vspace{0.5em}
\footnotesize
\textit{Notes:} Intra-Eurozone measures the effect of both countries adopting the euro 
(trade creation). Eurozone → Non-EZ EU measures the effect of the reporter country 
adopting the euro on trade with non-eurozone EU members (UK, Sweden, Denmark). 
Non-EZ EU → Eurozone measures the reverse direction. A negative effect in the 
cross-eurozone pairs would indicate trade diversion. Estimated using CausalForestDML 
on EU15 data, 1995--2015.
\end{table}

The wide confidence intervals for cross-eurozone trade reflect limited 
statistical power: the number of non-eurozone EU partners (UK, Sweden, Denmark) is small. 
However, the point estimates are uniformly positive, providing no evidence of trade 
diversion. If anything, eurozone countries appear to have maintained or increased trade with 
non-eurozone EU partners after euro adoption, though only the intra-eurozone effect is 
precisely estimated.

This pattern is consistent with the euro reducing transaction costs for all 
trade involving eurozone countries, not just intra-eurozone trade. A German 
firm that adopts euro-denominated invoicing for French customers may also find 
it easier to quote prices to British or Swedish customers in a stable currency. 
The absence of trade diversion suggests the euro's trade effects are primarily 
trade-creating rather than trade-diverting, though we cannot rule out small 
diversion effects given the imprecision of our cross-eurozone estimates.

\section{Computational Details}
\label{app:computational}

This appendix provides technical details on the causal forest implementation, 
including runtime, convergence diagnostics, and sensitivity to random seeds.

\subsection{Implementation}

All analyses were conducted in Python 3.10 using the EconML library 
(version 0.14.1) for CausalForestDML estimation. The CFFE analysis uses the 
\texttt{causalfe} package (version 0.2.0), a Python implementation of Causal 
Forests with Fixed Effects developed by the first author and available at 
\url{https://github.com/haytug/causalfe}.

\subsection{Hyperparameters}

The CausalForestDML estimator was configured with the following hyperparameters:
\begin{itemize}
    \item First-stage outcome model: Random Forest with 200 trees, 
          min\_samples\_leaf = 20
    \item First-stage treatment model: Random Forest Classifier with 200 trees, 
          min\_samples\_leaf = 20
    \item Causal forest: 500 trees, min\_samples\_leaf = 30, honest splitting 
          enabled
    \item Cross-fitting: 5-fold cross-validation for nuisance estimation
\end{itemize}

The CFFE estimator used similar settings with the addition of node-level 
fixed effects residualization for pair and year effects.

\subsection{Runtime}

Table~\ref{tab:runtime} reports computation times for the main analyses on a 
standard workstation (Apple M1 Pro, 16GB RAM). The EU15 causal forest completes 
in under 30 seconds, while the EU28 CFFE analysis requires approximately 2 
minutes due to the larger sample and fixed effects computation.

\begin{table}[!htbp]
\centering
\caption{Computation Times}
\label{tab:runtime}
\small
\begin{tabular}{@{}lrr@{}}
\toprule
Analysis & Sample Size & Runtime (seconds) \\
\midrule
EU15 Causal Forest & 2,189 & 28 \\
EU15 CFFE & 2,189 & 45 \\
EU28 Causal Forest & 8,456 & 52 \\
EU28 CFFE & 8,456 & 124 \\
Counterfactual predictions & --- & 3 \\
Leave-one-out (14 iterations) & --- & 392 \\
\bottomrule
\end{tabular}
\end{table}

\subsection{Convergence}

We assess convergence by examining how estimates stabilize as the number of 
trees increases. The ATE estimate stabilizes 
after approximately 200 trees, with minimal variation beyond 300 trees. Our 
choice of 500 trees provides a comfortable margin for convergence.

The confidence interval width also stabilizes with increasing trees, declining 
from approximately 0.45 log points with 50 trees to 0.33 log points with 500 
trees. Additional trees beyond 500 provide diminishing returns in precision.

\subsection{Seed Sensitivity}

To assess sensitivity to random initialization, we re-estimated the EU15 
causal forest with 20 different random seeds. Table~\ref{tab:seed_sensitivity} 
reports the distribution of ATE estimates across seeds.

\begin{table}[!htbp]
\centering
\caption{Seed Sensitivity Analysis (20 seeds)}
\label{tab:seed_sensitivity}
\small
\begin{tabular}{@{}lr@{}}
\toprule
Statistic & Value \\
\midrule
Mean ATE & 0.207 \\
Std. Dev. & 0.008 \\
Minimum & 0.193 \\
Maximum & 0.221 \\
Coefficient of Variation & 3.9\% \\
\bottomrule
\end{tabular}
\end{table}

The ATE estimates are highly stable across seeds, with a coefficient of 
variation of only 3.9\%. The range of estimates (0.193 to 0.221) falls well 
within the confidence interval of any individual estimate, indicating that 
our findings are not sensitive to random initialization.

\subsection{Reproducibility}

All code and data necessary to reproduce the analyses are available at 
[repository URL to be added upon publication]. The random seed for the main 
results reported in the paper is 42. Running the analysis with this seed 
will reproduce the exact estimates reported in the tables and figures.

\end{document}